\documentclass[]{aa}%

\usepackage{graphicx}
\usepackage{txfonts}
\usepackage{natbib}
\usepackage{verbatim}
\usepackage{multirow}
\usepackage[T1]{fontenc}
\usepackage{tikz}
\usepackage{soul}
\usepackage{morefloats}

\newcommand{\MJ}{$M_{\mbox{\footnotesize{J} }}$}

\newcommand{\Me}{M$_{\oplus}$}

\newcommand{\kms}{\,km\,s$^{-1}$}
\newcommand{\ms}{\,m\,s$^{-1}$}
\newcommand{\cms}{\,cm\,s$^{-1}$}

\newcommand{\logR}{$\log{(R'_{\rm HK})}$}

\newcommand{\prob}[2]{p(#1|#2)}
\newcommand{\prior}[2]{\pi(#1|#2)}
\newcommand{\like}{\mathcal{L}(\boldsymbol{\theta}_i)}
\newcommand{\teta}{\boldsymbol{\theta}}

\newcommand{\M}{\citetalias{mayor2009}}

\newcommand{\T}{\citetalias{tuomi2013b}}

\newcommand{\Perr}{\citetalias{perrakis2014}}

\newcommand{\CJ}{\citetalias{chibjeliazkov2001}}

\defcitealias{mayor2009}{M09}
\defcitealias{tuomi2013b}{T13}
\defcitealias{perrakis2014}{P14}
\defcitealias{chibjeliazkov2001}{CJ01}

\usepackage{color}

\begin{document}

   \title{The  HARPS search for southern extra-solar planets. }
          \subtitle{XXXVIII. Bayesian re-analysis of three systems. New super-Earths, unconfirmed signals, and magnetic cycles.
        \thanks{Based on observations made with the {\footnotesize HARPS} instrument on the ESO 3.6 m telescope at La Silla Observatory under the GTO programme ID 072.C-0488, and its continuation programmes ID 183.C-0972, 091.C-0936, and 192.C-0852.}\fnmsep
        \thanks{Tables 3, 6, and 10 are only available in electronic form at the CDS via anonymous ftp to cdsarc.u-strasbg.fr (130.79.128.5) or via http://cdsweb.u-strasbg.fr/cgi-bin/qcat?J/A+A/}
        }

\titlerunning{Bayesian re-analysis of three HARPS systems.}
          
\author{R.~F.~D\'iaz\inst{1}
        \and D. S\'egransan\inst{1}
        \and S.~Udry\inst{1}
        \and C.~Lovis\inst{1}
        \and F.~Pepe\inst{1} 
        \and X.~Dumusque\inst{2, 1} 
        \and M.~Marmier\inst{1}         
        \and R.~Alonso\inst{3,4}
        \and W.~Benz\inst{5} 
        \and F.~Bouchy\inst{1,6}  
        \and A.~Coffinet\inst{1}
        \and A.~Collier Cameron\inst{7}
        \and M.~Deleuil\inst{6}
        \and P.~Figueira\inst{8}
        \and M.~Gillon\inst{9}
        \and G.~Lo Curto\inst{10}
        \and M.~Mayor\inst{1} 
        \and C.~Mordasini\inst{5}
        \and F.~Motalebi\inst{1}
        \and C.~Moutou\inst{6, 11}
        \and D.~Pollacco\inst{12}
        \and E.~Pompei\inst{10}
        \and D.~Queloz\inst{1, 13} 
        \and N.~Santos\inst{8, 14}  
        \and A.~Wyttenbach\inst{1}
      } 
     
   \offprints{Rodrigo.Diaz@unige.ch}

   \institute{Observatoire astronomique de l'Universit\'e de Gen\`eve, 
                   51 ch. des Maillettes, CH-1290 Versoix, Switzerland
          \and
          Harvard-Smithsonian Center for Astrophysics, 60 Garden Street, Cambridge, Massachusetts 02138, USA
          \and
          Instituto de Astrof\'isica de Canarias, 38205, La Laguna, Tenerife, Spain
          \and
          Dpto. de Astrof\'isica, Universidad de La Laguna, 38206, La Laguna, Tenerife, Spain
          \and
          Physikalisches Institut, Universitat Bern, Silderstrasse 5, CH-3012 Bern, Switzerland
          \and
          Aix Marseille Universit\'e, CNRS, LAM (Laboratoire d'Astrophysique de Marseille) UMR 7326, 13388, Marseille, France
          \and
          School of Physics and Astronomy, University of St Andrews, North Haugh, St Andrews, Fife KY16 9SS
          \and
          Instituto de Astrof\'isica e Ci\^encias do Espa\c{c}o, Universidade do Porto, CAUP, Rua das Estrelas, 4150-762 Porto, Portugal
          \and
          Institut d'Astrophysique et de G\'eophysique, Universit\'e de Li\`ege, All\'ee du 6 Ao\^ut 17, B\^at. B5C, 4000, Li\`ege, Belgium
          \and
          European Southern Observatory, Alonso de Cordova 3107, Vitacura, Casilla 19001, Santiago 19, Chile
          \and 
         Canada France Hawaii Telescope Corporation, Kamuela, 96743, USA
         \and
         Department of Physics, University of Warwick, Coventry, CV4 7AL, UK
         \and
          Cavendish Laboratory, J J Thomson Avenue, Cambridge CB3 0HE, UK          \and 
          Departamento de F\'isica e Astronomia, Faculdade de Ci\^encias, Universidade do Porto, Rua do Campo Alegre, 4169-007 Porto, Portugal
          }
   
  \abstract{We present the analysis of the entire HARPS observations of three stars that host planetary systems: HD1461, HD40307, and HD204313. The data set spans eight years and contains more than 200 nightly averaged velocity measurements for each star. This means that it is sensitive to both long-period and low-mass planets and also to the effects induced by stellar activity cycles. \\ We modelled the data using Keplerian functions that correspond to planetary candidates and included the short- and long-term effects of magnetic activity. A Bayesian approach was taken both for the data modelling, which allowed us to include information from activity proxies such as \logR\ in the velocity modelling, and for the model selection, which permitted determining the number of significant signals in the system. The Bayesian model comparison overcomes the limitations inherent to the traditional periodogram analysis. We report an additional super-Earth planet in the HD1461 system. Four out of the six planets previously reported for HD40307 are confirmed and characterised. We discuss the remaining two proposed signals. In particular, we show that when the systematic uncertainty associated with the techniques for estimating model probabilities are taken into account, the current data are not conclusive concerning the existence of the habitable-zone candidate HD40307 g. \\ We also fully characterise the Neptune-mass planet that orbits HD204313 in 34.9 days.}
   
 \keywords{
  Techniques: radial velocities --
  Methods: data analysis -- Methods: statistical --
 Stars: individual: \object{HD1461},  \object{HD40307} , \object{HD204313} -- Stars: activity
}
\maketitle

\section{Introduction}
The detailed architecture of multiplanet systems is a key observable to constrain formation and evolution theories. Observations made
over many years with very stable instruments are necessary to fully unveil the structure of planetary systems. Detecting companions on long-period orbits and low-mass planets at shorter periods usually requires many dozens of radial velocity measurements \citep[see e.g.][]{mayor2009, pepe2011}, especially when the candidates are found in multi-planet systems, as is common \citep{mayor2011, lissauer2012, lissauer2014}. These data sets tend to span many years.

The HARPS search for extrasolar planets in the southern hemisphere \citep{mayor2003} has recently celebrated its tenth anniversary. The most inactive stars in the solar neighbourhood have been monitored continuously for over a decade, producing data sets with over two hundred measurements. These are expected to permit an in-depth exploration of the planetary systems around them. However, even for these very weakly active stars, the presence of magnetic cycles complicates the detection of low-mass companions \citep{santos2010, lovis2011b, dumusque2011c}.  For this reason, additional observables are routinely obtained from the HARPS spectra: activity proxies based on the line fluxes (mainly the proxies based on  the Ca II H\&K lines, but recently also H$\alpha$ and Na I D), the mean line bisector span, its full-width at half-maximum (FWHM), etc. All of these can help identify activity cycles and ultimately correct for their effect on the radial velocity time series.

Which analysis method is best applied on these data sets has been the subject of some debate. While the classical frequentist approach of studying the velocity periodograms is known to have drawbacks \citep[e.g.][]{sellke2001, lovis2011, tuomi2012}, the alternative Bayesian model comparison has led to different numbers of signals reported for the same system. For the star Gl667C, for example, for which the periodogram analysis revealed two planetary companions \citep{delfosse2013}, different groups have reported a number of planets ranging from two \citep{ferozhobson2014} to six or even seven \citep{gregory2012, angladaescude2013} when they used Bayesian methods. Moreover, stellar activity can mimic planetary signals, and the nature of the detected signals is frequently difficult to identify \citep[see the case of Gl581:][]{udry2007, robertson2014} and not always agreed upon \citep{angladaescude2014, robertson2015, angladaescude2015}.

The radial velocity signal produced by stellar activity can be separated into two types of effects: the short-term effect produced by active regions (spots and plages) that rotate in and out of view as the star revolves, and the long-term effect associated with changes in the global activity level of cyclic stars \citep[e.g.][]{baliunas95, hall2007, hall2009, santos2010, isaacsonfischer2010, gomesdasilva2012, robertson2013}. The short-term modulation is produced by the difference in flux and convective blueshift of active regions with respect to the surrounding photosphere \citep[e.g.][see also \citealp{boisse2012, dumusque2014}]{saardonahue97, hatzes2002, desort2007}. This creates a radial velocity signal whose frequency power is concentrated on the stellar rotational period and its harmonics \citep{boisse2011}. This short-term signal strongly depends on the detailed configuration of the active regions and is therefore difficult to model precisely because no clear correlation with activity proxies is systematically seen. On the other hand, long-term activity variations are related to global changes in the convective pattern of the star \citep{lindegrendravins2003, meunier2010, meunierlagrange2013} that are produced by a change in the typical number of active regions. The effect is therefore less sensitive to the details of individual active regions, and a clear correlation is seen with activity proxies such as \logR\ \citep{noyes84} and the width and asymmetry of the mean spectral line \citep{dravins1982, santos2010, lovis2011b, dumusque2011c}.  

Here we analyse the radial velocity data of HD1461, HD40307, and HD204313; the data include more than ten years of HARPS observations. All three stars have been reported to host at least one planet. They all exhibit long-term variability in their activity levels, as would be produced by magnetic cycles, and their effect on the radial velocity data are clearly detected. However, a full
cycle is only observed for HD1461. We employed the traditional periodogram analysis to identify potential periodic signals and used a Bayesian model comparison to asses their significance. Taking the discussion from the previous paragraph into consideration, we take a different approach to model each of the activity effects: the short-term variability is described using a non-deterministic model that does not require a detailed description of the activity signal, while the long-term variability is modelled assuming a simple relation between the activity effect on the radial velocities and the activity proxy \logR. Effects produced by stellar pulsations and granulation have timescales of minutes and can be efficiently averaged out \citep{dumusque2011a} or treated as additional Gaussian noise. 

The paper is organised as follows: Sect.~\ref{sect.obs} briefly presents the data used in the analysis, Sect.~\ref{sect.stellarparams} reports the results from the spectroscopic analysis of the three target stars and their main characteristics, Sect.~\ref{sect.dataanalysis} describes the models employed, including the model for the short- and long-term activity effects. In this section we also detail the technique used to compare and select the models, we discuss the algorithm for sampling from the model posterior distribution, and we present the choice of parameter priors. In the following three sections we describe the results for each system. Finally,
we discuss the results and present our conclusions in Sect. 8.

\section{Observations and data reduction \label{sect.obs}}
All three targets were observed as part of the Guaranteed Time Observations programme to search for southern extrasolar planets and its continuation high-precision HARPS programmes. The observations were reduced using the HARPS pipeline (version 3.5), and the stellar radial velocities were obtained through a weighted cross-correlation with a numerical mask \citep{baranne1996, pepe2002}. The FWHM and bisector span of the peak in the cross-correlation function (CCF) were also measured for each spectrum, as well as the activity proxy based on the Ca II H and K lines, \logR, calibrated as described in \citet{lovis2011b}. 

The number of measurements and basic characteristics of the observations studied here are presented in Table~\ref{table.obslog}, and the nightly averaged radial velocity measurements are given in Tables~\ref{table.rvHD1461}, \ref{table.rvHD40307}, and \ref{table.rvHD204313}, available online. The HARPS observations of HD204313 started around three years later than for the other two stars because this target was regularly monitored by the CORALIE instrument on the 1.2
m Swiss telescope at La Silla. For the analysis of HD204313 we also included 104 RV measurements by CORALIE, 56 of which were obtained after the instrument upgrade performed in 2007 \citep{segransan2010}. 

\begin{table*}
\center
\caption{Basic characteristics of the HARPS observations of the three target stars. $N$ is the total number of spectra, and $N_\text{nights}$ is the total number of nights on which the target was observed. The average signal-to-noise ratio (<S/N>) is computed over the nightly average spectra at 550 nm. \label{table.obslog}}
\begin{tabular}{l  cc  ccc  c }
\hline
\hline
& & & \multicolumn{2}{c}{Dates} &\\
Target & $N$ & $N_\text{nights}$ & start & end & time span [yr] & <S/N>$_{550\,\mathrm{nm}}$    \\
\hline
HD1461    & 448 & 249 & 2003-09-16 & 2013-11-28 & 10.2 & 277\\
HD40307  & 441 & 226 & 2003-10-29 & 2014-04-05 & 10.4 & 246\\
HD204313&  96  &  95  & 2006-05-05 & 2014-10-17 &   8.5 & 151\\
\end{tabular}
\end{table*}

\section{Stellar characteristics \label{sect.stellarparams}}

\begin{table*}[t!]
\center
\caption{\label{table.stellarparams}
Observed and inferred stellar parameters.}
\begin{tabular}{llccc}
\hline\hline
Parameters    		&   			&\object{HD~1461}	&\object{HD~40307}&\object{HD~204313} \\
\hline                                   
Sp. T.$^{(1)}$  	  	&     			&G0V   		   & K3V   &	G5V  \\
V$^{(1)}$              	&     			&6.47  		   & 7.17  	&   	7.99	\\
$B-V^{(1)}$           	&     			&0.674  		    & 0.935 &   	0.697\\
$\pi^{(2)}$            	&[mas]		&$43.02\pm0.51$ & $76.95\pm0.37$ & $21.11\pm0.62$\\
\hline
 $T_{\rm eff}^{(3)}$		& [K] 	&$5765 \pm 18$	&$4977 \pm 59$	&$5776 \pm 22$    \\
$[Fe/H]^{(3)}$		&[dex]		&$+0.19 \pm 0.01$	&$-0.31 \pm 0.03$	&$0.18 \pm 0.02$\\
$\log{(g)}^{(3)}$		& [cgs]		&$4.38 \pm 0.03$	&$4.47 \pm 0.16$	&$4.38 \pm 0.02$\\
$M_{\star}^{(3)}$		&[M$_{\odot}$]	& 1.02 & 0.77 & 1.02	 \\
\logR$^{(4)}$		&			& $-5.021\pm0.013$ & $-4.940\pm0.058$ &$-5.024\pm0.019$\\
$P_{rot}^{(5)}$		&[days]		& $30.2\pm3.5$	& $47.9\pm6.4$	& $34.1\pm3.7$ \\
$P_{rot}^{(6)}$		&[days]		& --				& 37.4			& -- \\

$v\sin{(i_\star)}$	&[km\,s$^{-1}$]	&-&1.61& 2.4\\
\end{tabular}
\tablebib{
(1) As listed in ~\citet{hipparcos}; (2)~\citet{vanleeuwen2007}; (3)~\citet{sousa2008}; (4)~This work: mean and standard deviation.; (5)~Estimated from \logR\ using the relationship by \citet{mamajekhillenbrand2008}. The reported uncertainty is the quadratic addition of the individual errors propagated from the \logR\ uncertainty and the deviation in the values produced by the change in the activity level.; (6) Measured in the \logR\, FWHM, and/or bisector time series.
}
\end{table*}

The atmospheric parameters for the three stars studied here have been obtained by \citet{sousa2008} based on HARPS spectra. Although in some cases now a larger number of spectra are available, the gain in signal-to-noise ratio (S/N) and therefore in the precision of the parameters, is surely limited. We therefore decided to use the parameters as reported in \citet{sousa2008}, which are listed in Table~\ref{table.stellarparams}. We note that the reported uncertainties for the atmospheric parameters do not consider potential systematic errors and may therefore be underestimated. 

The stellar mass is given without uncertainty because the statistical error bar is certainly plagued with systematic errors (the choice of the stellar tracks, the physics used to compute the track, etc.). To compute the masses and semi-major axes of the detected companions, we conservatively fixed the uncertainty of the stellar mass to 10\%. For all companions reported in this article, except for HD1461~c and HD40307~f, the contribution of the uncertainty in the stellar mass to the companion mass is larger than the contribution of the uncertainty in the orbital parameters. This illustrates the importance of improving our knowledge of the fundamental stellar characteristics, and the relevance of space missions such as Gaia and PLATO.

All three stars are magnetically quiet, with a mean \logR\ below -4.9, but they all exhibit magnetic variability on timescales similar to the duration of the HARPS observations. These variations are reminiscent of the solar activity cycle and are described in more detail in the following sections. Despite their relative brightness and solar-like characteristics, these stars have not been systematically included in the southern surveys of stellar activity targeting solar-like stars \citep[e.g.][]{henry1996, cincunegui2007a, mauas2012}. To the best of our knowledge, the only mention of one of the target stars in southern surveys appears in \citet{arriagada2011}. Based on eight observations of HD204313 from the Magellan planet search programme, \citet{arriagada2011} computed \logR=-5.0. Long-term observations of the magnetic activity level of HD1461 exist from northern surveys \citep{hall2009},
however. These observations show that the low level of activity of HD1461 was maintained for at least 15 years and seem to confirm the cyclic behaviour detected in the HARPS data (see Sect.~\ref{sect.HD1461}).

The mean activity level is used to estimate the rotational period of the targets using the relation reported by \citet{mamajekhillenbrand2008}. This estimate is also reported in Table~\ref{table.stellarparams}.

\section{Data analysis \label{sect.dataanalysis}}

\subsection{Description of the models \label{sect.modeldescription}}

The stellar radial velocity (RV) variations are described using a physical model $M_n$ consisting of $n$ Keplerian curves representing potential planetary companions and activity signals with timescales of the order of the rotational period, plus an additional signal produced by long-term stellar activity effects, $a(t)$, which could take different forms depending on the knowledge we have on the stellar activity cycle (its period, etc.). The Keplerian functions plus the long-term activity signal constitute the deterministic part of the model ($m$). To this we add a statistical noise component $\epsilon$ that represents the stellar activity "jitter" -- that is, the short-term activity-induced variability that is not correctly modelled by a Keplerian curve-- and all remaining systematic errors not considered in the reported uncertainties. As mentioned in the introduction, the short-term activity signal depends on the details of the active regions visible at a given time on the stellar disk, their positions, sizes, and evolutions. This signal is therefore very hard to model using a single deterministic model, which is why we decided to add to it a statistical (non-deterministic) { component that we call the stellar jitter}. Therefore, the RV prediction for model $M_n$ at time $t_i$ can be written as 
\begin{equation}
m_i + \epsilon_i = \sum_{j = 1}^n k_j(t_i) + a(t_i) + \epsilon_i\;\;,\label{eq.model}
\end{equation}
where $k_j$ is the Keplerian curve of companion $j$. The Keplerian curves were parametrised using their period, amplitude, eccentricity, argument of periastron $\omega$, and mean longitude at epoch $L_0$. We describe the statistical model for the stellar jitter in detail below, but we note that we explicitly added the subindex $i$ to the noise component of the model $\epsilon_i$ to indicate
that it can potentially depend on time. Additionally, the data errors are assumed to be uncorrelated and normally distributed.

Furthermore, assuming that the error term $\epsilon$ is distributed as a zero-centred normal, with standard deviation $\sigma_J(t_i) = \sigma_{Ji}$ (potentially a function of time), the likelihood function of our model takes the form \citep[][Sect. 4.8]{gregory}
\begin{equation}
\mathcal{L} = \prod_i \frac{1}{\sqrt{2\pi}\sqrt{\sigma_i^2 + \sigma_{Ji}^2}} \exp{\left[-\frac{(v_i - m_i)^2}{2 (\sigma_i^2 + \sigma_{Ji}^2)}\right]}\; ,
\label{eq.likelihood}
\end{equation}
where $(v_i, \sigma_i)$ is the velocity measurement and its associated uncertainty at time $t_i$.

\subsubsection*{Stellar jitter \label{sect.jitter}}

\begin{figure}
\center
\input{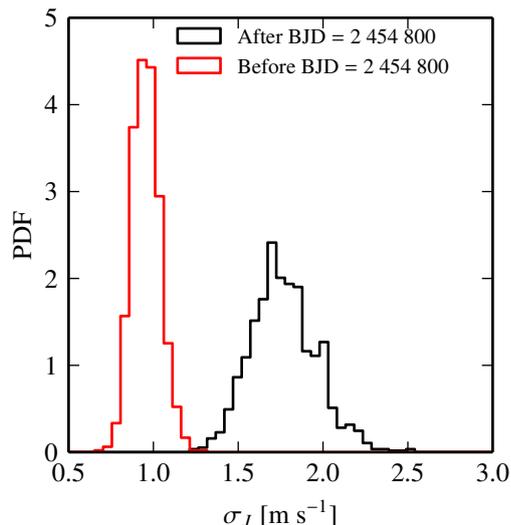}
\caption{Stellar jitter for the constant-jitter model, fitted to the RV data of HD40307 before and after BJD=2'454'800, which have different mean levels of activity (see Sect.~\ref{sect.HD40307} for details).\label{fig.HD40307_jitter}}
\end{figure}

The stellar jitter is included in our model as an additional, statistical error in the model prediction. We note that this is different from the approach taken by other authors \citep[e.g.][]{tuomi2013b, tuomi2013}, who constructed a deterministic model of the stellar activity at short timescales.  In this analysis, we make the strong assumption that the additional noise is uncorrelated and normally distributed. { We note, however, that this term appears in addition to any potential rotational signal modelled by a Keplerian curve. It aims at accounting for the parts of the rotational activity signal that are not represented by the deterministic model. In that sense, the white-noise assumption is probably less dramatic than if we were to use it to model the entire rotational activity signal}. Two simple models of the amplitude of the stellar jitter were explored. 

In the first one, the added jitter term has a constant amplitude $\sigma_J$ for all times. In this case, the global value of $\sigma_J$ is the sole parameter of the statistical model. However, we know beforehand that this model does not correctly describe the data because it is known that the dispersion in RV time series is larger for more active stars. To illustrate this, the constant jitter model was fitted separately to the RV measurements of HD40307 (see Sect.~\ref{sect.HD40307}) obtained before and after BJD=2,454,800. As these data sets have different stellar activity levels, it is no surprise to find a clear difference in the distribution of the parameter $\sigma_J$ (Fig.~\ref{fig.HD40307_jitter}). Therefore, we decided to use a second model of the stellar activity jitter, in which the standard deviation of the noise component $\epsilon$ increases linearly with \logR. The dependency on \logR\ is motivated by the fact that the scatter in the RV measurements increases when the \logR\ activity proxy does, but the linearity is an additional assumption of the model that needs to be tested. The second model was parametrised using the jitter level when \logR$ = -5.0$ and the slope of the dependence of the jitter amplitude with \logR. The jitter level when \logR$ = -5.0$ (the base-level jitter) represents any RV effect that might exist for solar-type stars with such a low level of magnetic activity, such as the granulation noise \citep[e.g.][]{dumusque2011a} and the undetected instrumental systematics changing from one night to another, which do not appear anywhere else in our model. The model requires an extra parameter and therefore suffers the Occam penalty described in Sect.~\ref{sect.modelcomp}. However, it is preferred by the available data for all three systems studied here, and we therefore only consider the evolving jitter model in the analyses presented below.

\subsubsection*{Stellar activity cycles}
All three stars analysed here exhibit long-term activity variations reminiscent of the solar activity cycle. We claim that the effect on the RV time series is clearly detected and include it in the model in the form of the term $a(t)$. The functional form of $a(t)$ is not fixed \emph{\textup{a priori}}, but  is instead
taken from a fit to the \logR\ time series. We tried a number of different models to fit the \logR\ time series (see Sect.~\ref{sect.HD1461}), but in all cases, the "shape" of the \logR\ time series was used to model the long-term variations seen in the RV data. 

To transform the variations in \logR\ into variations in RV, a scaling constant $\alpha$ is included in the model. Previous studies exploring the effect of magnetic cycles on RV data have parametrised an equivalent scaling constant as a function of the effective temperature and the metallicity of the star \citep{lovis2011b}. However, the dispersion around the fit is considerable, and we therefore decided not to include a prior for parameter $\alpha$. We instead compared in each case the obtained result with the expected value based on the $T_\mathrm{eff}$ - $\mathrm{[Fe/H]}$ parametrisation.

We note that by modelling the effect of the activity cycle in this way, we are assuming a linear relation between the variations in the \logR\ proxy and the long-term activity-induced RV. This is a different assumption from the one used for the jitter model, which states that the amplitude of the additional Gaussian noise scales linearly with \logR.

\subsection{Posterior sampling}
To estimate the model parameter credible regions, we obtained samples from the posterior distribution using the Markov chain Monte Carlo (MCMC) algorithm described in \citet{diaz2014} with normal proposal distributions for all parameters. The algorithm uses an adaptive principal component analysis to efficiently
sample densities with non-linear correlations.

To increase the efficiency of the MCMC algorithm, the starting point for the chain was chosen using the genetic algorithm (GA) implemented in the \emph{yorbit} package \citep{segransan2011}. This drastically reduces the burn-in period and guarantees that the entire parameter space has been explored. We nevertheless launched a number of independent chains to explore the possibility of multi-modal posterior distributions. The chains were combined after thinning using their autocorrelation length. 

The posterior of the stellar mass, which is not constrained in our model, was assumed to be a normal distribution centred at the value reported in Table~\ref{table.stellarparams}, with a width equivalent to 10\% of this value. A randomly drawn sample from this distribution was coupled to the MCMC sample of the remaining model parameters to obtain the posteriors of model parameters such as the planet minimum masses or the semi-major axes.

\subsection{Choice of priors \label{sect.priors40307}}
The priors of the model parameters are presented in detail in the Appendix for each system. In general, the only parameters with informative priors are the orbital eccentricity and the parameters of the activity signal $a(t)$. For the eccentricity we chose a Beta distribution, as advocated by \citet{kipping2013b}, who derived the shape parameters that best match a sample of around 400 RV-measured orbital eccentricities ($a = 0.867$, $b = 3.03$). The priors for the long-term activity signal were chosen as normal distributions around the least-squares fit to the \logR\ time series, neglecting the covariances between the fit parameters (see Tables~\ref{table.priors1461}, \ref{table.40307priors}, and \ref{table.priors204313}). This is the practical way in which we incorporated the information present in the \logR\ time series to our model.

For the remaining parameters we used uninformative priors (i.e. uniform or Jeffreys). The limits chosen for each parameter are shown in the tables in the Appendix.

\subsection{Bayesian model comparison \label{sect.modelcomp}}

One of the aims of our analysis is to establish the number of periodic signals present in a given RV data set, independently of their nature. Traditionally, this is addressed by studying the periodogram of the RV time series and by estimating the significance of the highest peak found. To do this, a series of synthetic datasets are obtained by reshuffling or permuting the original data points. The periodogram is computed on each newly created data set and the power of the highest peak is recorded. The histogram of the maximum peak powers is used to estimate the $p$-value as a function of power level. This $p$-value is estimated under the null hypothesis --in this case no (further) signal-- since no real signal is expected in the reshuffled data sets. If the $p$-value of the highest peak in the original histogram is lower than a predefined threshold, the best-fit Keplerian signal at the peak frequency is subtracted from the data and the periodogram analysis is again performed on the velocity residuals. This process is repeated until no further peaks appear above the threshold.  Finally, a global fit including all detected frequencies is performed.

This technique has the advantage of being computationally inexpensive and is expected to produce the correct number of significant signals if { the threshold $p$-value is chosen to be low enough and provided} the removed signals are well constrained. However, it has { two main limitations: a) the interpretation of the $p$-value as a false-alarm probability is in general incorrect and leads to an overestimation of the evidence against the null hypothesis \citep{sellke2001}, and} b) the uncertainties of the signals subtracted from the data are not taken into account when computing the statistical significance of any potential remaining signal \citep[see e.g.][]{lovis2011, tuomi2012, hatzes2013, baluev2013}. Therefore, when dealing with signals with amplitudes below $\sim$~1 or 2 \ms, which are similar to the activity signals and to the uncertainty of the individual observations, { it is not advisable to conclude on the} significance of the signals { based on the $p$-values obtained from the periodogram}. In these cases we resorted to the more rigorous technique of Bayesian model comparison. { We note, however, that the periodogram was used throughout the analysis to identify possible periodicities in the data, and when the associated $p$-value was low enough (typically below 0.1\%), more sophisticated analyses were not deemed necessary to declare the signal significant.}

Bayesian statistics permits, unlike the frequentist approach, computing the probability ($p$) of any logical proposition, where the probability is understood to be a degree of plausibility for that proposition. In this framework, comparing two models ($M_1$ and $M_2$) in the face of a given data set $D$ and some information $I$ can be made rigorously by computing the ratio of their posterior probabilities, known as the \emph{\textup{odds ratio}}:
\begin{equation}
O_{1,2} = \frac{\prob{M_1}{D, I}}{\prob{M_2}{D, I}} = \frac{\prob{M_1}{I}}{\prob{M_2}{I}}\cdot\frac{\prob{D}{M_1, I}}{\prob{D}{M_2, I}}\;\;,
\label{eq.oddsratio}
\end{equation}
where the first term on the right-hand side is called the \emph{\textup{prior odds}} and is independent of the data, and the second term is the \emph{\textup{Bayes factor}} and encodes all the support the data give to one model over the other.

The Bayesian approach to model comparison treats models with different numbers of parameters and non-nested models. The Bayes factor has a built-in mechanism that penalises models according to the number of free parameters they have \citep[known as Occam's factor, see][Sect. 3.5]{gregory}. We note that when there is no prior preference for any model ($\prob{M_1}{I} / \prob{M_2}{I} = 1$), the Bayes factor is directly the odds ratio. To compute the Bayes factor, the \emph{\textup{evidence}} or \emph{\textup{marginal likelihood}} of each model are needed, defined as the weighted average of the model likelihood ($\prob{D}{\teta_i, M_i, I} = \like$) over the prior parameter space\footnote{The prior distribution was assumed to be normalised to unity.}:
\begin{equation}
\mathcal{E}_i = \prob{D}{M_i, I} = \int{\prior{\teta_i}{M_i, I}\cdot\like\cdot\mathrm{d}\teta_i}\;\;,
\label{eq.evidence}
\end{equation} 
where $\teta_i$ denotes the parameter vector of model $i$, and $\prior{\teta_i}{M_i, I}$ is the parameter prior distribution. In multi-dimensional parameter spaces, such as those associated with models of multi-planet systems, the integral of Eq.~\ref{eq.evidence} is often intractable and has to be estimated numerically. Moreover, the basic Monte Carlo integration estimate consisting of obtaining the mean value of $\like$ over a sample from the prior density is expected to fail for high-dimensional problems if the likelihood is concentrated relative to the prior distribution because most elements from the prior sample will have very low likelihood values.

A considerable number of methods for estimating the evidence exist in the literature (see \citeauthor{frielwyse2012} \citeyear{frielwyse2012} for a recent review and \citeauthor{kassraftery1995} \citeyear{kassraftery1995}). Estimating the evidence is difficult in multi-dimensional spaces, and different techniques can lead to very different results \citep[see e.g.][]{gregory2007}. We therefore decided to use three different methods and compare their results. { All of them rely on} posterior distribution samples { and are therefore relatively fast to compute because they use the sample obtained with the} MCMC algorithm described above. { In some cases, further samples are needed from known distributions from which they can be drawn in a straightforward manner}.

\begin{itemize}
\item The \citet{chibjeliazkov2001} estimator (hereafter \CJ) is based on the fact that the marginal likelihood is the normalising constant of the posterior density. The method requires estimating the posterior density at a single point in parameter space $\mathbf{\theta^\star}$. To do this, a sample from the posterior density and from the proposal density used to produce the trial steps in the MCMC algorithm are needed. { The method is straightforward and relatively fast, but can run into problems for multi-modal posterior distributions \citep{frielwyse2012}.} In this study, the sample from the proposal distribution was obtained by approximating the proposal density by a multivariate normal with covariance equal to the covariance of the posterior sample. The uncertainty was estimated by repeatedly sampling from the proposal density and using different subsets of the posterior sample. { A weakness of this method as implemented here is the approximation of the proposal density. Moreover, computing the likelihood on the sample from this distribution is the most computationally expensive step in the process, which limits the sample size that can be drawn. Additionally, some of the draws from the proposal distribution fall outside the prior domain, reducing the effective sample size further.}

\item The \citet{perrakis2014} estimator (hereafter \Perr) is based on the \emph{\textup{importance sampling}} technique. Importance sampling improves the efficiency of Monte Carlo integration { of a function over a given distribution} using samples from a \emph{\textup{different}} distribution, known as the importance sampling density \citep[see e.g.][]{geweke1989, kassraftery1995}.This technique can readily be employed to estimate the integral in Eq.~\ref{eq.evidence}. \Perr\ proposed using the product of the marginal posterior densities of the model parameters as importance sampling density, which yields the estimator
\begin{equation}
\hat{\mathcal{E}_i} = N^{-1} \sum_{n=1}^N \frac{\mathcal{L}(\teta^{(n)})\prior{\teta^{(n)}}{M_i, I}}{\prod_{j=1}^{q_i} p_j(\teta^{(n)}|D, I)}\;\;,
\label{eq.perrakisestimator}
\end{equation}
where the $p_j$, with $j = 1, ... , q_i$, are the marginal posterior densities of the model parameters, $q_i$ is the number of parameters in model $i$, and $\teta^{(n)}$, with $n = 1, ..., N$, are the parameter vectors sampled from the marginal posterior densities.

We here obtained the sample from the marginal posterior densities by reshuffling the $N$ elements from the joint posterior sample obtained with the MCMC algorithm, so that correlations between parameters are lost, { as suggested by \Perr}. { We note that if the marginal posterior sample is obtained in this way, no further draws are necessary from the posterior distribution, although computing the likelihood in the reshuffled sample is still necessary and is the most time-consuming step in the estimation.} The technique also requires evaluating the marginal posterior probabilities that appear in the denominator of Eq.~\ref{eq.perrakisestimator}. We estimated these densities using the normalised histogram of the MCMC sample for each parameter. { The error produced by this estimation of the marginal posterior distributions is a weak point of our implementation because it increases with the number of parameters as a result of the product in the denominator.} We are currently studying more sophisticated techniques such as the non-parametric kernel density estimation. The uncertainty was estimated by repeatedly reshuffling the joint posterior sample to produce new samples from the marginal posterior distributions.

\item \citet{tuomijones2012} also used importance sampling to estimate the marginal likelihood. The importance sampling function $\mathcal{I}$ is a mixture of posterior distribution samples at different stages of a MCMC:\begin{equation}
\mathcal{I} \propto (1 - \lambda)\,\like\prior{\teta_i}{I} + \lambda\,\mathcal{L}(\teta_{i-h})\prior{\teta_{i-h}}{I}
\label{eq.tpmimportance}
.\end{equation}

The level of mixture ($\lambda$) and the lag between samples ($h$) are two parameters of the method that the authors explored. The result is called a truncated posterior-mixture (TPM) estimate. { This estimator is designed to solve the well-known stability problem of the harmonic mean estimator (HME) \citep{newtonraftery1994, kassraftery1995}, which uses the posterior density as importance sampling density. The HME converges very slowly to the evidence \citep{frielwyse2012, robertwraith2009} and usually produces an estimator with infinite variance \citep{robertwraith2009}. In addition, as the HME is based solely in samples from the posterior, which is typically much more peaked than the prior distribution, it will generally not be very sensitive to changes in the prior. This is documented in \citet{frielwyse2012} and is a clear drawback of the HME because the evidence is known to be extremely sensitive to prior choice.}

The TPM estimate aims at solving the stability problem by using a mixture for the importance sampling density. { This estimator converges} to the HME of \citet{newtonraftery1994} as $\lambda$ tends to zero, and therefore its variance also tends to infinity\footnote{TPM converges in probability to the HME, which implies convergence in distribution (E. Cameron, priv. comm.).}. However, when $\lambda$ is different from zero, the TPM estimate is inconsistent, that is, it does not converge to the evidence as the sample size increases. In addition, the TPM estimate has the very important drawback of inheriting the prior-insensitivity of the HME. It is therefore unable to correctly reproduce the effect of Occam's penalisation found in the Bayes factor. This is documented in the \citet{tuomijones2012} article where TPM is introduced, but is presented as an advantage of the estimator. In summary, for $\lambda=0$, the TPM estimate is equivalent to the problematic HME, and if $\lambda > 0$ the estimator is inconsistent. Therefore we do not expect this estimator to produce reliable results, but we included it for comparison.
\end{itemize}

 \section{HD1461 \label{sect.HD1461}} 
 
\object{HD1461} hosts a super-Earth on a 5.77-day period orbit. \citet{rivera2010} reported its discovery based on 167 radial velocity measurements taken with HIRES on the Keck telescope over 12.8 years. The presence of two additional companions in longer period orbits ($P=446.1$ days and $P=5017$ days) is also discussed by the authors. We analysed 249 nightly averaged HARPS measurements spanning { more than} ten years with a mean internal uncertainty  of 49 \cms, which include photon noise and the error in the wavelength calibration.

\begin{figure}
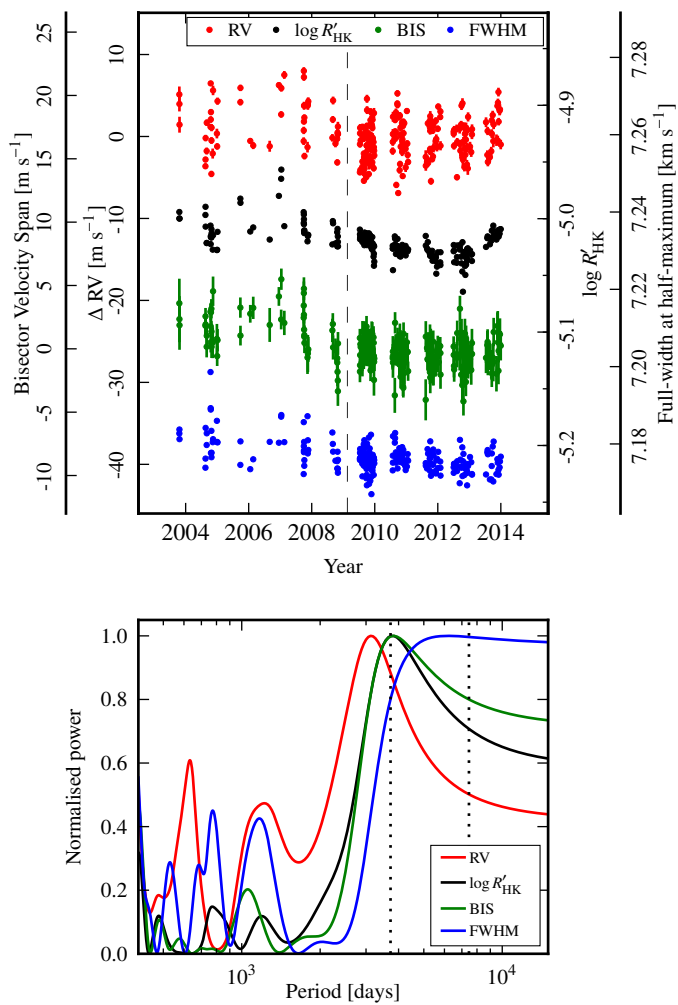

\input{HD1461_activity_timeseries_withHall_fwhmcorr_rv.pgf}
\input{HD1461_activityGLS_short.pgf}
\caption{Top panel: HARPS time series of \object{HD1461}. { The vertical dashed line separates the active (BJD < 2'454'850) from the inactive data set.} Lower panel: GLS at periods over 400 days for the four time series plotted in the top panel. The vertical dotted lines represent the time span of observations and twice this value.}
\label{fig.HD1461obs}
\end{figure}

A preliminary analysis of the HARPS radial velocities produced by the instrument pipeline revealed a periodic one-year oscillation with an amplitude of $\sim 1.4$ \ms. This one-year signal has previously been identified as a systematic effect in HARPS data (Dumusque et al.\ 2015, submitted). Its origin is the manufacturing of the E2V CCD by stitching together ($512\times1024$)-pixel blocks to reach the total detector size ($4096\times2048$ pixels). The spacing between these blocks is not as regular as the spacing between the columns within a block. Such discontinuities are at the moment not taken into account in the HARPS wavelength calibration. Despite the great stability of HARPS, the position of the stellar spectral lines on the detector varies throughout a year due to the changes in the Earth orbital velocity. Depending on the content of the spectrum and the systemic velocity of the star some spectral lines may go through these stitches and produce the observed yearly oscillation. This is the case for HD1461, and we have corrected for this effect by removing the responsible lines from the spectral correlation mask. When this is done, the signal at one year disappears. The average uncertainty in the velocity increases around 13\% due to the smaller number of lines used for the correlation. The velocities reported in Table~\ref{table.rvHD1461} and plotted in Fig.~\ref{fig.HD1461obs} are the corrected version.

\onllongtab{
\begin{longtable}{r r r r r r r}
\caption{HARPS measurements of HD1461. \label{table.rvHD1461}}\\
\hline\hline
\multicolumn{1}{c}{BJD} &	\multicolumn{1}{c}{RV} &	\multicolumn{1}{c}{$\sigma_\text{RV}$} &	\multicolumn{1}{c}{BIS} &	\multicolumn{1}{c}{FWHM} &	 \multicolumn{1}{c}{\logR} &	\multicolumn{1}{c}{$\sigma_\text{\logR}$} \\
-2 450 000  &(\kms) & (\kms) & (\ms) & (\kms) & & \\
\hline
\noalign{\smallskip}
\endfirsthead
\caption{Continued.}\\
\hline
\multicolumn{1}{c}{BJD} &	\multicolumn{1}{c}{RV} &	\multicolumn{1}{c}{$\sigma_\text{RV}$} &	\multicolumn{1}{c}{BIS} &	\multicolumn{1}{c}{FWHM} &	 \multicolumn{1}{c}{\logR} &	\multicolumn{1}{c}{$\sigma_\text{\logR}$} \\
-2 450 000  &(\kms) & (\kms) & (\ms) & (\kms) & & \\
\hline
\noalign{\smallskip}
\endhead
\hline
\endfoot
\hline
\hline
\endlastfoot
52899.7648 &	-10.0528 &	0.0011 &	-11.32 &	7.1703 &	-4.9972 &	0.0040 \\
52900.7412 &	-10.0549 &	0.0010 &	-11.55 &	7.1782 &	-4.9941 &	0.0028 \\
52901.7596 &	-10.0560 &	0.0009 &	-12.80 &	7.1792 &	-4.9998 &	0.0021 \\
52902.7191 &	-10.0585 &	0.0010 &	-13.28 &	7.1767 &	-4.9999 &	0.0026 \\
52903.7775 &	-10.0566 &	0.0010 &	-14.92 &	7.1736 &	-5.0105 &	0.0035 \\
52937.6699 &	-10.0628 &	0.0011 &	-12.44 &	7.1739 &	-5.0338 &	0.0055 \\
52942.6907 &	-10.0588 &	0.0010 &	-15.09 &	7.1700 &	-5.0194 &	0.0036 \\
52946.7119 &	-10.0576 &	0.0011 &	-14.01 &	7.1730 &	-5.0442 &	0.0056 \\
53202.9094 &	-10.0636 &	0.0003 &	-13.32 &	7.1695 &	-4.9967 &	0.0017 \\
53203.8970 &	-10.0628 &	0.0003 &	-12.58 &	7.1719 &	-4.9895 &	0.0015 \\
53204.8831 &	-10.0602 &	0.0004 &	-13.22 &	7.1798 &	-5.0070 &	0.0021 \\
53216.8799 &	-10.0617 &	0.0004 &	-14.98 &	7.1754 &	-5.0215 &	0.0028 \\
53217.8455 &	-10.0583 &	0.0004 &	-14.11 &	7.1784 &	-5.0214 &	0.0027 \\
53262.7950 &	-10.0588 &	0.0003 &	-12.21 &	7.1943 &	-5.0061 &	0.0013 \\
53264.7455 &	-10.0535 &	0.0003 &	-13.37 &	7.1850 &	-5.0224 &	0.0015 \\
53265.7976 &	-10.0579 &	0.0004 &	-14.63 &	7.1846 &	-5.0155 &	0.0022 \\
53267.7273 &	-10.0570 &	0.0004 &	-12.27 &	7.1734 &	-5.0164 &	0.0021 \\
53272.7395 &	-10.0646 &	0.0004 &	-11.98 &	7.1791 &	-5.0166 &	0.0022 \\
53273.7144 &	-10.0605 &	0.0004 &	-13.78 &	7.1723 &	-5.0094 &	0.0022 \\
53274.8129 &	-10.0589 &	0.0004 &	-14.99 &	7.1734 &	-5.0130 &	0.0033 \\
53287.6716 &	-10.0621 &	0.0006 &	-14.66 &	7.1762 &	-5.0273 &	0.0067 \\
53291.6729 &	-10.0544 &	0.0006 &	-10.60 &	7.1770 &	-5.0146 &	0.0062 \\
53336.6522 &	-10.0596 &	0.0006 &	-14.36 &	7.1818 &	-5.0277 &	0.0036 \\
53339.6722 &	-10.0612 &	0.0004 &	-15.72 &	7.1763 &	-5.0274 &	0.0025 \\
53343.6176 &	-10.0557 &	0.0004 &	-14.46 &	7.1762 &	-5.0116 &	0.0025 \\
53607.7291 &	-10.0558 &	0.0004 &	-11.89 &	7.1709 &	-4.9823 &	0.0018 \\
53608.8581 &	-10.0541 &	0.0004 &	-14.10 &	7.1770 &	-4.9860 &	0.0019 \\
53722.5596 &	-10.0605 &	0.0003 &	-12.38 &	7.1701 &	-5.0113 &	0.0016 \\
53757.5439 &	-10.0611 &	0.0004 &	-11.91 &	7.1728 &	-5.0077 &	0.0029 \\
53946.8074 &	-10.0612 &	0.0007 &	-13.26 &	7.1780 &	-5.0184 &	0.0039 \\
54054.6665 &	-10.0537 &	0.0004 &	-11.01 &	7.1781 &	-4.9799 &	0.0014 \\
54080.5985 &	-10.0573 &	0.0004 &	-12.83 &	7.1850 &	-4.9648 &	0.0014 \\
54082.5917 &	-10.0541 &	0.0004 &	-9.65  &	7.1854 &	-4.9567 &	0.0017 \\
54117.5526 &	-10.0525 &	0.0005 &	-13.10 &	7.1786 &	-5.0066 &	0.0031 \\
54340.7554 &	-10.0605 &	0.0005 &	-12.69 &	7.1714 &	-4.9942 &	0.0022 \\
54341.8223 &	-10.0592 &	0.0004 &	-11.73 &	7.1848 &	-5.0037 &	0.0022 \\
54342.7466 &	-10.0520 &	0.0005 &	-10.75 &	7.1786 &	-4.9963 &	0.0018 \\
54343.8505 &	-10.0527 &	0.0004 &	-10.40 &	7.1774 &	-5.0005 &	0.0024 \\
54344.7875 &	-10.0582 &	0.0003 &	-12.46 &	7.1770 &	-5.0049 &	0.0012 \\
54346.8140 &	-10.0562 &	0.0005 &	-13.63 &	7.1797 &	-5.0107 &	0.0030 \\
54348.7762 &	-10.0562 &	0.0004 &	-14.19 &	7.1735 &	-5.0127 &	0.0016 \\
54349.7839 &	-10.0577 &	0.0004 &	-14.06 &	7.1750 &	-5.0103 &	0.0016 \\
54350.8066 &	-10.0624 &	0.0004 &	-14.68 &	7.1774 &	-5.0123 &	0.0015 \\
54385.6991 &	-10.0613 &	0.0006 &	-15.83 &	7.1866 &	-5.0200 &	0.0039 \\
54388.6725 &	-10.0562 &	0.0004 &	-15.25 &	7.1769 &	-5.0141 &	0.0018 \\
54389.6944 &	-10.0560 &	0.0007 &	-15.51 &	7.1803 &	-5.0153 &	0.0047 \\
54394.6644 &	-10.0557 &	0.0005 &	-15.19 &	7.1779 &	-5.0144 &	0.0022 \\
54673.9339 &	-10.0601 &	0.0005 &	-13.69 &	7.1840 &	-5.0227 &	0.0021 \\
54678.8605 &	-10.0579 &	0.0004 &	-13.19 &	7.1811 &	-5.0021 &	0.0019 \\
54682.8337 &	-10.0556 &	0.0005 &	-15.03 &	7.1791 &	-5.0041 &	0.0021 \\
54709.8365 &	-10.0608 &	0.0005 &	-15.25 &	7.1769 &	-5.0218 &	0.0022 \\
54730.7308 &	-10.0603 &	0.0005 &	-15.89 &	7.1769 &	-5.0119 &	0.0028 \\
54731.7092 &	-10.0631 &	0.0004 &	-16.23 &	7.1751 &	-5.0244 &	0.0017 \\
54734.7128 &	-10.0595 &	0.0006 &	-16.96 &	7.1794 &	-5.0090 &	0.0035 \\
54737.6719 &	-10.0597 &	0.0005 &	-17.65 &	7.1750 &	-5.0156 &	0.0028 \\
54738.6912 &	-10.0588 &	0.0006 &	-18.49 &	7.1740 &	-5.0205 &	0.0038 \\
54739.6726 &	-10.0597 &	0.0004 &	-14.72 &	7.1773 &	-5.0244 &	0.0020 \\
54992.9354 &	-10.0643 &	0.0004 &	-14.29 &	7.1795 &	-5.0096 &	0.0014 \\
54993.9396 &	-10.0609 &	0.0004 &	-16.94 &	7.1791 &	-5.0138 &	0.0019 \\
54994.9272 &	-10.0596 &	0.0004 &	-16.44 &	7.1810 &	-5.0181 &	0.0017 \\
54995.9439 &	-10.0589 &	0.0005 &	-15.64 &	7.1819 &	-5.0165 &	0.0022 \\
54998.9394 &	-10.0640 &	0.0004 &	-14.78 &	7.1798 &	-5.0139 &	0.0015 \\
55001.9369 &	-10.0605 &	0.0004 &	-15.00 &	7.1809 &	-5.0139 &	0.0014 \\
55020.8632 &	-10.0637 &	0.0004 &	-16.16 &	7.1783 &	-5.0154 &	0.0016 \\
55021.9281 &	-10.0630 &	0.0005 &	-16.40 &	7.1750 &	-5.0206 &	0.0024 \\
55022.8615 &	-10.0588 &	0.0004 &	-13.79 &	7.1788 &	-5.0196 &	0.0021 \\
55024.8780 &	-10.0600 &	0.0004 &	-15.05 &	7.1813 &	-5.0157 &	0.0020 \\
55036.8884 &	-10.0603 &	0.0004 &	-14.63 &	7.1835 &	-5.0189 &	0.0016 \\
55037.8596 &	-10.0621 &	0.0003 &	-16.04 &	7.1829 &	-5.0177 &	0.0012 \\
55038.8815 &	-10.0654 &	0.0004 &	-14.59 &	7.1780 &	-5.0147 &	0.0013 \\
55039.8664 &	-10.0637 &	0.0004 &	-15.69 &	7.1802 &	-5.0146 &	0.0014 \\
55040.8019 &	-10.0625 &	0.0004 &	-14.69 &	7.1819 &	-5.0178 &	0.0016 \\
55042.8401 &	-10.0633 &	0.0004 &	-14.60 &	7.1785 &	-5.0145 &	0.0014 \\
55043.8864 &	-10.0639 &	0.0005 &	-15.00 &	7.1846 &	-5.0234 &	0.0024 \\
55044.8621 &	-10.0645 &	0.0005 &	-14.68 &	7.1829 &	-5.0180 &	0.0023 \\
55045.8363 &	-10.0626 &	0.0004 &	-14.52 &	7.1780 &	-5.0145 &	0.0018 \\
55046.8111 &	-10.0596 &	0.0004 &	-13.70 &	7.1792 &	-5.0122 &	0.0014 \\
55064.8108 &	-10.0593 &	0.0005 &	-15.04 &	7.1805 &	-5.0158 &	0.0012 \\
55066.8111 &	-10.0630 &	0.0004 &	-14.17 &	7.1788 &	-5.0191 &	0.0018 \\
55067.7864 &	-10.0649 &	0.0004 &	-15.67 &	7.1749 &	-5.0128 &	0.0015 \\
55068.7851 &	-10.0624 &	0.0004 &	-14.48 &	7.1810 &	-5.0166 &	0.0017 \\
55069.8498 &	-10.0619 &	0.0004 &	-13.39 &	7.1773 &	-5.0192 &	0.0017 \\
55070.7744 &	-10.0611 &	0.0004 &	-15.36 &	7.1790 &	-5.0153 &	0.0014 \\
55071.7619 &	-10.0611 &	0.0005 &	-15.55 &	7.1779 &	-5.0186 &	0.0026 \\
55072.8068 &	-10.0629 &	0.0004 &	-14.30 &	7.1758 &	-5.0149 &	0.0019 \\
55073.7583 &	-10.0601 &	0.0005 &	-14.92 &	7.1731 &	-5.0138 &	0.0024 \\
55074.7614 &	-10.0587 &	0.0004 &	-15.26 &	7.1799 &	-5.0155 &	0.0016 \\
55075.7310 &	-10.0554 &	0.0005 &	-15.00 &	7.1832 &	-5.0233 &	0.0028 \\
55076.7557 &	-10.0568 &	0.0005 &	-15.14 &	7.1839 &	-5.0223 &	0.0027 \\
55095.7257 &	-10.0633 &	0.0007 &	-15.04 &	7.1790 &	-5.0160 &	0.0037 \\
55096.7660 &	-10.0646 &	0.0004 &	-15.34 &	7.1857 &	-5.0182 &	0.0015 \\
55097.7286 &	-10.0597 &	0.0004 &	-15.36 &	7.1818 &	-5.0129 &	0.0016 \\
55099.7339 &	-10.0612 &	0.0004 &	-14.85 &	7.1737 &	-5.0135 &	0.0018 \\
55100.6599 &	-10.0605 &	0.0004 &	-14.62 &	7.1812 &	-5.0146 &	0.0017 \\
55104.7958 &	-10.0586 &	0.0004 &	-15.52 &	7.1818 &	-5.0185 &	0.0017 \\
55106.6555 &	-10.0628 &	0.0004 &	-16.19 &	7.1829 &	-5.0156 &	0.0016 \\
55108.7339 &	-10.0632 &	0.0005 &	-14.84 &	7.1788 &	-5.0154 &	0.0020 \\
55110.6892 &	-10.0626 &	0.0004 &	-14.87 &	7.1791 &	-5.0144 &	0.0014 \\
55111.7366 &	-10.0613 &	0.0004 &	-15.39 &	7.1781 &	-5.0171 &	0.0015 \\
55113.6862 &	-10.0624 &	0.0004 &	-16.21 &	7.1746 &	-5.0206 &	0.0018 \\
55115.7297 &	-10.0580 &	0.0006 &	-15.82 &	7.1792 &	-5.0230 &	0.0028 \\
55116.7073 &	-10.0569 &	0.0004 &	-16.85 &	7.1812 &	-5.0193 &	0.0017 \\
55121.7178 &	-10.0598 &	0.0004 &	-14.47 &	7.1866 &	-5.0209 &	0.0016 \\
55122.6418 &	-10.0599 &	0.0004 &	-16.14 &	7.1788 &	-5.0248 &	0.0019 \\
55124.6549 &	-10.0632 &	0.0004 &	-14.58 &	7.1776 &	-5.0210 &	0.0013 \\
55126.6329 &	-10.0616 &	0.0004 &	-15.47 &	7.1712 &	-5.0230 &	0.0020 \\
55128.6423 &	-10.0585 &	0.0004 &	-15.01 &	7.1779 &	-5.0263 &	0.0015 \\
55129.6234 &	-10.0610 &	0.0004 &	-13.76 &	7.1759 &	-5.0239 &	0.0017 \\
55133.6928 &	-10.0587 &	0.0004 &	-14.85 &	7.1804 &	-5.0260 &	0.0018 \\
55135.5894 &	-10.0615 &	0.0003 &	-15.26 &	7.1794 &	-5.0250 &	0.0013 \\
55136.6446 &	-10.0636 &	0.0004 &	-15.57 &	7.1777 &	-5.0253 &	0.0019 \\
55137.5764 &	-10.0631 &	0.0005 &	-13.97 &	7.1805 &	-5.0162 &	0.0012 \\
55138.6778 &	-10.0589 &	0.0004 &	-16.10 &	7.1773 &	-5.0275 &	0.0018 \\
55139.5757 &	-10.0572 &	0.0004 &	-14.72 &	7.1784 &	-5.0232 &	0.0019 \\
55140.6506 &	-10.0579 &	0.0003 &	-15.49 &	7.1784 &	-5.0217 &	0.0013 \\
55141.5705 &	-10.0603 &	0.0003 &	-15.19 &	7.1798 &	-5.0233 &	0.0013 \\
55142.6103 &	-10.0602 &	0.0006 &	-15.45 &	7.1816 &	-5.0244 &	0.0021 \\
55151.6485 &	-10.0607 &	0.0004 &	-14.24 &	7.1809 &	-5.0335 &	0.0022 \\
55153.6136 &	-10.0639 &	0.0004 &	-14.67 &	7.1764 &	-5.0234 &	0.0017 \\
55155.5681 &	-10.0613 &	0.0004 &	-15.71 &	7.1809 &	-5.0309 &	0.0017 \\
55156.6218 &	-10.0571 &	0.0006 &	-17.60 &	7.1802 &	-5.0415 &	0.0038 \\
55158.5914 &	-10.0617 &	0.0004 &	-16.17 &	7.1804 &	-5.0244 &	0.0015 \\
55160.5989 &	-10.0623 &	0.0005 &	-15.91 &	7.1795 &	-5.0379 &	0.0026 \\
55372.9052 &	-10.0618 &	0.0004 &	-15.89 &	7.1795 &	-5.0239 &	0.0027 \\
55373.9121 &	-10.0606 &	0.0005 &	-15.73 &	7.1833 &	-5.0263 &	0.0033 \\
55374.9123 &	-10.0597 &	0.0006 &	-15.36 &	7.1815 &	-5.0264 &	0.0047 \\
55375.8759 &	-10.0576 &	0.0006 &	-16.23 &	7.1883 &	-5.0454 &	0.0054 \\
55396.8763 &	-10.0596 &	0.0007 &	-18.82 &	7.1868 &	-5.0156 &	0.0042 \\
55399.8573 &	-10.0577 &	0.0004 &	-13.09 &	7.1880 &	-5.0176 &	0.0020 \\
55400.9019 &	-10.0580 &	0.0004 &	-15.15 &	7.1894 &	-5.0225 &	0.0018 \\
55402.8284 &	-10.0593 &	0.0010 &	-14.49 &	7.1815 &	-5.0282 &	0.0074 \\
55408.9006 &	-10.0602 &	0.0005 &	-14.95 &	7.1844 &	-5.0254 &	0.0024 \\
55410.7636 &	-10.0571 &	0.0005 &	-14.97 &	7.1821 &	-5.0311 &	0.0029 \\
55412.8554 &	-10.0595 &	0.0004 &	-15.57 &	7.1818 &	-5.0256 &	0.0025 \\
55414.8089 &	-10.0659 &	0.0004 &	-15.64 &	7.1829 &	-5.0244 &	0.0017 \\
55426.8851 &	-10.0559 &	0.0003 &	-14.96 &	7.1841 &	-5.0242 &	0.0017 \\
55427.8178 &	-10.0547 &	0.0004 &	-15.67 &	7.1832 &	-5.0241 &	0.0019 \\
55428.7950 &	-10.0561 &	0.0004 &	-14.30 &	7.1804 &	-5.0178 &	0.0021 \\
55436.7634 &	-10.0669 &	0.0003 &	-14.65 &	7.1838 &	-5.0288 &	0.0014 \\
55450.8136 &	-10.0575 &	0.0005 &	-16.38 &	7.1859 &	-5.0268 &	0.0019 \\
55453.7049 &	-10.0641 &	0.0004 &	-16.31 &	7.1838 &	-5.0329 &	0.0020 \\
55456.7195 &	-10.0574 &	0.0004 &	-14.03 &	7.1828 &	-5.0269 &	0.0023 \\
55458.6711 &	-10.0633 &	0.0004 &	-15.96 &	7.1842 &	-5.0267 &	0.0020 \\
55460.6775 &	-10.0638 &	0.0010 &	-11.26 &	7.1756 &	-5.0261 &	0.0075 \\
55463.8266 &	-10.0589 &	0.0004 &	-14.55 &	7.1824 &	-5.0334 &	0.0025 \\
55464.7892 &	-10.0598 &	0.0005 &	-17.35 &	7.1848 &	-5.0270 &	0.0025 \\
55465.8127 &	-10.0612 &	0.0005 &	-16.16 &	7.1841 &	-5.0309 &	0.0033 \\
55480.6524 &	-10.0583 &	0.0004 &	-14.66 &	7.1857 &	-5.0254 &	0.0018 \\
55482.7088 &	-10.0619 &	0.0005 &	-15.30 &	7.1864 &	-5.0289 &	0.0025 \\
55483.6138 &	-10.0605 &	0.0004 &	-14.85 &	7.1837 &	-5.0231 &	0.0021 \\
55485.6512 &	-10.0603 &	0.0004 &	-16.43 &	7.1844 &	-5.0248 &	0.0015 \\
55488.6668 &	-10.0648 &	0.0003 &	-17.14 &	7.1848 &	-5.0258 &	0.0014 \\
55492.6846 &	-10.0587 &	0.0004 &	-18.11 &	7.1824 &	-5.0302 &	0.0016 \\
55495.7177 &	-10.0582 &	0.0004 &	-18.23 &	7.1859 &	-5.0287 &	0.0016 \\
55498.6598 &	-10.0596 &	0.0004 &	-13.96 &	7.1830 &	-5.0284 &	0.0022 \\
55507.6103 &	-10.0584 &	0.0003 &	-16.12 &	7.1826 &	-5.0259 &	0.0012 \\
55510.5892 &	-10.0612 &	0.0005 &	-17.05 &	7.1827 &	-5.0249 &	0.0024 \\
55515.5958 &	-10.0614 &	0.0004 &	-15.35 &	7.1821 &	-5.0304 &	0.0021 \\
55522.5982 &	-10.0615 &	0.0005 &	-15.77 &	7.1818 &	-5.0240 &	0.0025 \\
55523.5937 &	-10.0638 &	0.0006 &	-15.36 &	7.1809 &	-5.0324 &	0.0033 \\
55537.5798 &	-10.0569 &	0.0004 &	-14.63 &	7.1819 &	-5.0271 &	0.0016 \\
55542.5796 &	-10.0608 &	0.0004 &	-15.42 &	7.1837 &	-5.0269 &	0.0019 \\
55546.5428 &	-10.0622 &	0.0004 &	-15.38 &	7.1852 &	-5.0287 &	0.0019 \\
55551.5554 &	-10.0631 &	0.0004 &	-16.91 &	7.1836 &	-5.0271 &	0.0017 \\
55754.9082 &	-10.0625 &	0.0008 &	-19.17 &	7.1849 &	-5.0152 &	0.0052 \\
55755.8568 &	-10.0635 &	0.0006 &	-14.18 &	7.1843 &	-5.0237 &	0.0036 \\
55769.8918 &	-10.0631 &	0.0003 &	-15.08 &	7.1837 &	-5.0262 &	0.0014 \\
55776.8745 &	-10.0628 &	0.0004 &	-15.95 &	7.1888 &	-5.0276 &	0.0024 \\
55779.7536 &	-10.0584 &	0.0004 &	-16.01 &	7.1843 &	-5.0206 &	0.0021 \\
55782.8311 &	-10.0632 &	0.0005 &	-14.75 &	7.1877 &	-5.0274 &	0.0022 \\
55802.8763 &	-10.0584 &	0.0005 &	-15.93 &	7.1866 &	-5.0306 &	0.0024 \\
55803.7241 &	-10.0589 &	0.0005 &	-16.99 &	7.1866 &	-5.0301 &	0.0023 \\
55805.7579 &	-10.0618 &	0.0004 &	-14.10 &	7.1835 &	-5.0309 &	0.0022 \\
55806.8365 &	-10.0615 &	0.0004 &	-16.27 &	7.1855 &	-5.0296 &	0.0016 \\
55807.8209 &	-10.0593 &	0.0005 &	-16.83 &	7.1877 &	-5.0325 &	0.0022 \\
55808.8181 &	-10.0575 &	0.0006 &	-17.41 &	7.1911 &	-5.0297 &	0.0044 \\
55809.8349 &	-10.0579 &	0.0004 &	-16.79 &	7.1882 &	-5.0328 &	0.0023 \\
55814.7942 &	-10.0588 &	0.0004 &	-14.44 &	7.1871 &	-5.0314 &	0.0025 \\
55816.7379 &	-10.0654 &	0.0004 &	-16.02 &	7.1823 &	-5.0300 &	0.0022 \\
55829.7459 &	-10.0628 &	0.0003 &	-17.18 &	7.1830 &	-5.0283 &	0.0014 \\
55834.7439 &	-10.0593 &	0.0004 &	-15.31 &	7.1853 &	-5.0371 &	0.0016 \\
55836.7294 &	-10.0584 &	0.0003 &	-16.24 &	7.1882 &	-5.0345 &	0.0015 \\
55839.7307 &	-10.0624 &	0.0004 &	-17.17 &	7.1855 &	-5.0341 &	0.0022 \\
55842.7296 &	-10.0587 &	0.0004 &	-14.58 &	7.1894 &	-5.0359 &	0.0023 \\
55845.7200 &	-10.0614 &	0.0006 &	-16.93 &	7.1879 &	-5.0359 &	0.0030 \\
55870.6343 &	-10.0606 &	0.0004 &	-16.61 &	7.1841 &	-5.0324 &	0.0016 \\
55872.6140 &	-10.0581 &	0.0004 &	-15.86 &	7.1837 &	-5.0376 &	0.0021 \\
55873.6070 &	-10.0604 &	0.0004 &	-16.39 &	7.1835 &	-5.0342 &	0.0018 \\
55878.6219 &	-10.0603 &	0.0006 &	-15.64 &	7.1869 &	-5.0485 &	0.0033 \\
55886.5508 &	-10.0613 &	0.0005 &	-15.80 &	7.1874 &	-5.0423 &	0.0027 \\
55895.5885 &	-10.0590 &	0.0004 &	-14.91 &	7.1851 &	-5.0297 &	0.0018 \\
55900.6088 &	-10.0563 &	0.0004 &	-16.24 &	7.1894 &	-5.0404 &	0.0019 \\
55923.5369 &	-10.0576 &	0.0004 &	-14.93 &	7.1871 &	-5.0385 &	0.0016 \\
55925.5468 &	-10.0569 &	0.0004 &	-15.48 &	7.1855 &	-5.0409 &	0.0022 \\
55927.5335 &	-10.0611 &	0.0006 &	-17.18 &	7.1835 &	-5.0296 &	0.0033 \\
56079.9290 &	-10.0594 &	0.0004 &	-14.96 &	7.1873 &	-5.0287 &	0.0018 \\
56080.9230 &	-10.0596 &	0.0005 &	-15.63 &	7.1859 &	-5.0232 &	0.0023 \\
56083.9101 &	-10.0611 &	0.0004 &	-16.50 &	7.1882 &	-5.0282 &	0.0023 \\
56085.9233 &	-10.0592 &	0.0005 &	-16.78 &	7.1883 &	-5.0384 &	0.0026 \\
56115.9074 &	-10.0590 &	0.0005 &	-15.56 &	7.1880 &	-5.0389 &	0.0025 \\
56117.8874 &	-10.0649 &	0.0004 &	-15.85 &	7.1868 &	-5.0338 &	0.0016 \\
56120.8855 &	-10.0607 &	0.0005 &	-13.90 &	7.1894 &	-5.0236 &	0.0025 \\
56149.8137 &	-10.0590 &	0.0006 &	-15.69 &	7.1840 &	-5.0272 &	0.0036 \\
56152.8194 &	-10.0609 &	0.0005 &	-11.98 &	7.1952 &	-5.0265 &	0.0027 \\
56157.9041 &	-10.0617 &	0.0007 &	-14.34 &	7.1902 &	-5.0230 &	0.0043 \\
56166.7512 &	-10.0585 &	0.0004 &	-17.19 &	7.1885 &	-5.0335 &	0.0021 \\
56171.6906 &	-10.0562 &	0.0005 &	-13.86 &	7.1883 &	-5.0394 &	0.0024 \\
56180.8032 &	-10.0625 &	0.0006 &	-18.01 &	7.1936 &	-5.0466 &	0.0033 \\
56182.7775 &	-10.0559 &	0.0006 &	-15.09 &	7.1862 &	-5.0207 &	0.0027 \\
56187.7602 &	-10.0628 &	0.0008 &	-15.58 &	7.1901 &	-5.0644 &	0.0049 \\
56193.7585 &	-10.0606 &	0.0006 &	-19.29 &	7.1921 &	-5.0298 &	0.0027 \\
56202.7348 &	-10.0589 &	0.0006 &	-14.94 &	7.1887 &	-5.0299 &	0.0031 \\
56204.7208 &	-10.0637 &	0.0005 &	-17.17 &	7.1867 &	-5.0294 &	0.0019 \\
56205.7432 &	-10.0614 &	0.0005 &	-15.93 &	7.1888 &	-5.0320 &	0.0020 \\
56210.6728 &	-10.0597 &	0.0006 &	-13.95 &	7.1891 &	-5.0417 &	0.0027 \\
56215.6102 &	-10.0622 &	0.0007 &	-16.64 &	7.1887 &	-5.0496 &	0.0034 \\
56221.7004 &	-10.0604 &	0.0006 &	-17.76 &	7.1884 &	-5.0397 &	0.0022 \\
56223.6517 &	-10.0589 &	0.0006 &	-14.61 &	7.1887 &	-5.0406 &	0.0021 \\
56225.6782 &	-10.0593 &	0.0005 &	-16.27 &	7.1899 &	-5.0346 &	0.0018 \\
56234.5541 &	-10.0634 &	0.0006 &	-16.07 &	7.1882 &	-5.0352 &	0.0019 \\
56236.5865 &	-10.0592 &	0.0006 &	-15.37 &	7.1845 &	-5.0267 &	0.0022 \\
56238.5792 &	-10.0613 &	0.0006 &	-15.21 &	7.1877 &	-5.0365 &	0.0019 \\
56244.5625 &	-10.0639 &	0.0008 &	-15.79 &	7.1899 &	-5.0261 &	0.0042 \\
56247.5825 &	-10.0576 &	0.0006 &	-16.37 &	7.1896 &	-5.0304 &	0.0017 \\
56269.5533 &	-10.0612 &	0.0006 &	-15.96 &	7.1922 &	-5.0280 &	0.0019 \\
56283.5311 &	-10.0619 &	0.0006 &	-16.82 &	7.1910 &	-5.0318 &	0.0019 \\
56291.5477 &	-10.0597 &	0.0006 &	-14.69 &	7.1880 &	-5.0373 &	0.0020 \\
56292.5461 &	-10.0591 &	0.0006 &	-15.55 &	7.1879 &	-5.0312 &	0.0025 \\
56454.9248 &	-10.0582 &	0.0006 &	-15.86 &	7.1899 &	-5.0249 &	0.0025 \\
56469.9145 &	-10.0622 &	0.0006 &	-14.85 &	7.1959 &	-5.0171 &	0.0024 \\
56472.9400 &	-10.0593 &	0.0005 &	-14.60 &	7.1965 &	-5.0156 &	0.0023 \\
56475.9386 &	-10.0631 &	0.0005 &	-16.17 &	7.1954 &	-5.0172 &	0.0025 \\
56498.8045 &	-10.0625 &	0.0005 &	-15.74 &	7.1892 &	-5.0226 &	0.0025 \\
56515.8296 &	-10.0618 &	0.0005 &	-16.42 &	7.1894 &	-5.0195 &	0.0022 \\
56520.8584 &	-10.0580 &	0.0006 &	-16.42 &	7.1929 &	-5.0170 &	0.0018 \\
56521.8030 &	-10.0596 &	0.0005 &	-17.08 &	7.1962 &	-5.0106 &	0.0023 \\
56531.7904 &	-10.0574 &	0.0005 &	-15.83 &	7.1981 &	-5.0100 &	0.0021 \\
56539.7337 &	-10.0605 &	0.0008 &	-16.19 &	7.1892 &	-5.0061 &	0.0034 \\
56543.7076 &	-10.0585 &	0.0006 &	-13.82 &	7.1996 &	-5.0146 &	0.0027 \\
56575.6465 &	-10.0604 &	0.0007 &	-17.11 &	7.1916 &	-5.0184 &	0.0030 \\
56586.7081 &	-10.0563 &	0.0005 &	-13.34 &	7.1956 &	-5.0133 &	0.0026 \\
56590.7000 &	-10.0605 &	0.0006 &	-13.34 &	7.1947 &	-5.0129 &	0.0027 \\
56592.6631 &	-10.0559 &	0.0005 &	-14.63 &	7.1955 &	-5.0102 &	0.0021 \\
56599.5062 &	-10.0546 &	0.0005 &	-16.88 &	7.1930 &	-5.0142 &	0.0024 \\
56601.6179 &	-10.0582 &	0.0005 &	-15.30 &	7.1918 &	-5.0158 &	0.0015 \\
56612.5370 &	-10.0568 &	0.0006 &	-14.77 &	7.1938 &	-5.0132 &	0.0020 \\
56616.6271 &	-10.0567 &	0.0006 &	-14.00 &	7.1954 &	-5.0094 &	0.0025 \\
56625.6194 &	-10.0610 &	0.0006 &	-14.90 &	7.1967 &	-5.0092 &	0.0022 \\
\end{longtable}
}

In Fig.~\ref{fig.HD1461obs} we plot time series of the RV, the \logR\ activity proxy and two spectral line measurements (FWHM and bisector velocity span) that can also be affected by activity. A similar long-term evolution of all four observables is clearly visible, indicating the presence of a magnetic activity cycle (Sect.~\ref{sect.HD1461cycle}). The top panel of Fig.~\ref{fig.HD1461omc} presents the generalized Lomb-Scargle periodograms \citep[GLS; ][]{zechmeisterkurster2009} of the RV data. The periodogram is dominated by a signal with a period of 5.77 days, compatible to the planet candidate reported by \citet{rivera2010}. The amplitude of $2.37\pm0.20$ \ms\ also agrees with \citet{rivera2010} and corresponds to a minimum mass of around 6.4 \Me. In the remaining panels of Fig.~\ref{fig.HD1461omc} the GLS of the residuals around models with increasing number of Keplerian components are shown. Table~\ref{table.HD1461evidences} presents the Bayesian evidence of models with at least three signals (including the activity cycle; see below) and the associated Bayes factors with respect to the three-Keplerian model. { The model probabilities are plotted in Fig.~\ref{fig.1461bayes}.}

\subsection{Two-Keplerian model. A new super-Earth candidate.}

\begin{figure}
\includegraphics[width=\columnwidth]{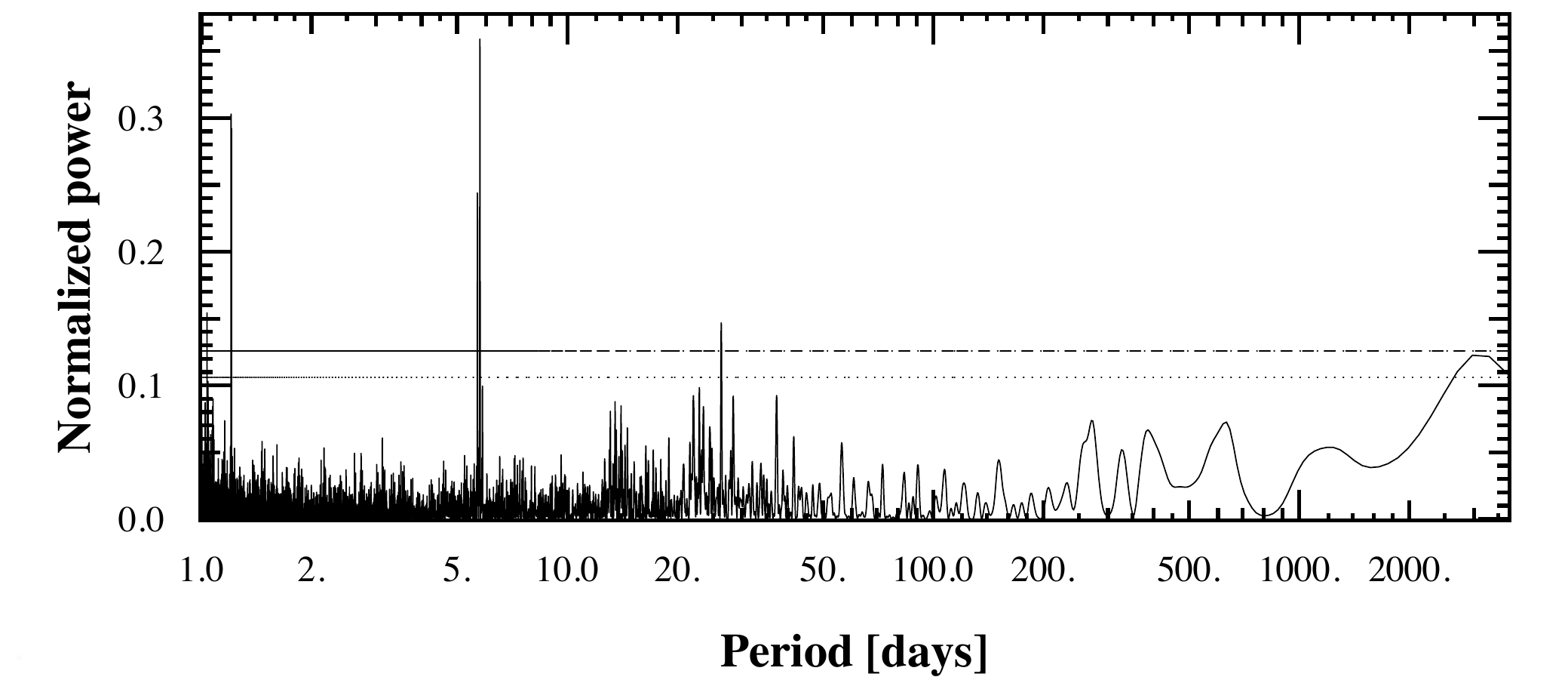}
\includegraphics[width=\columnwidth]{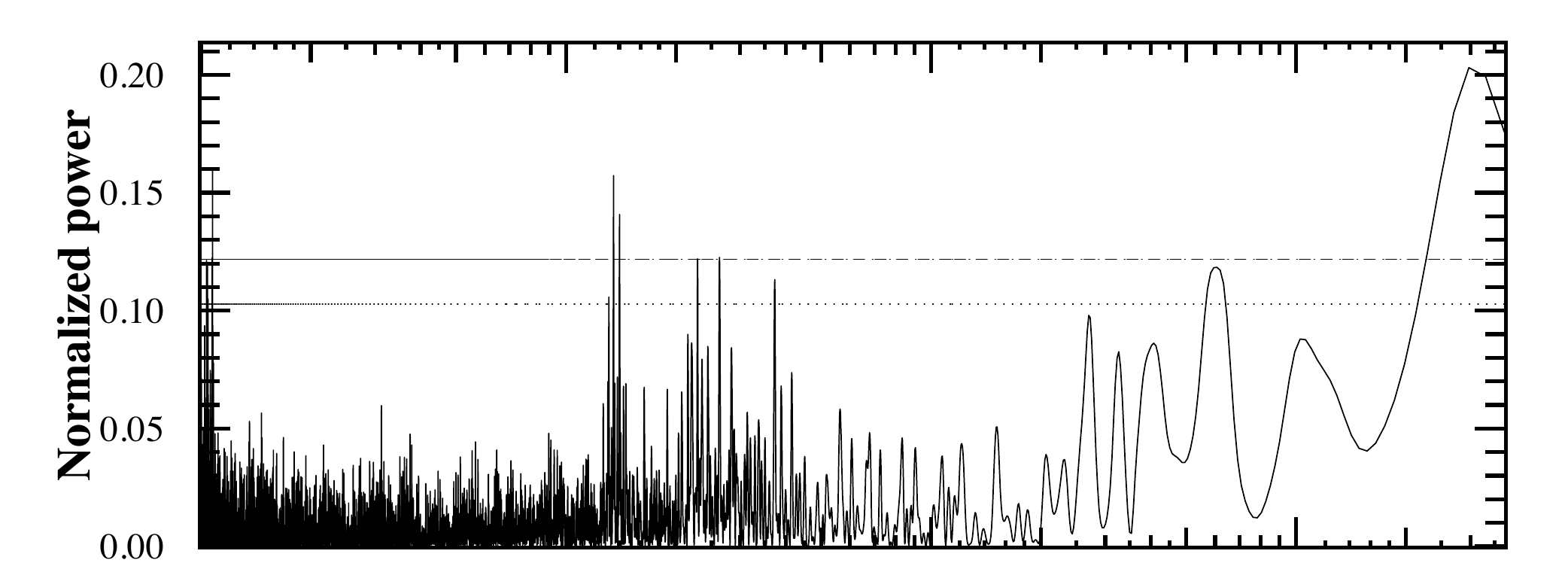}
\includegraphics[width=\columnwidth]{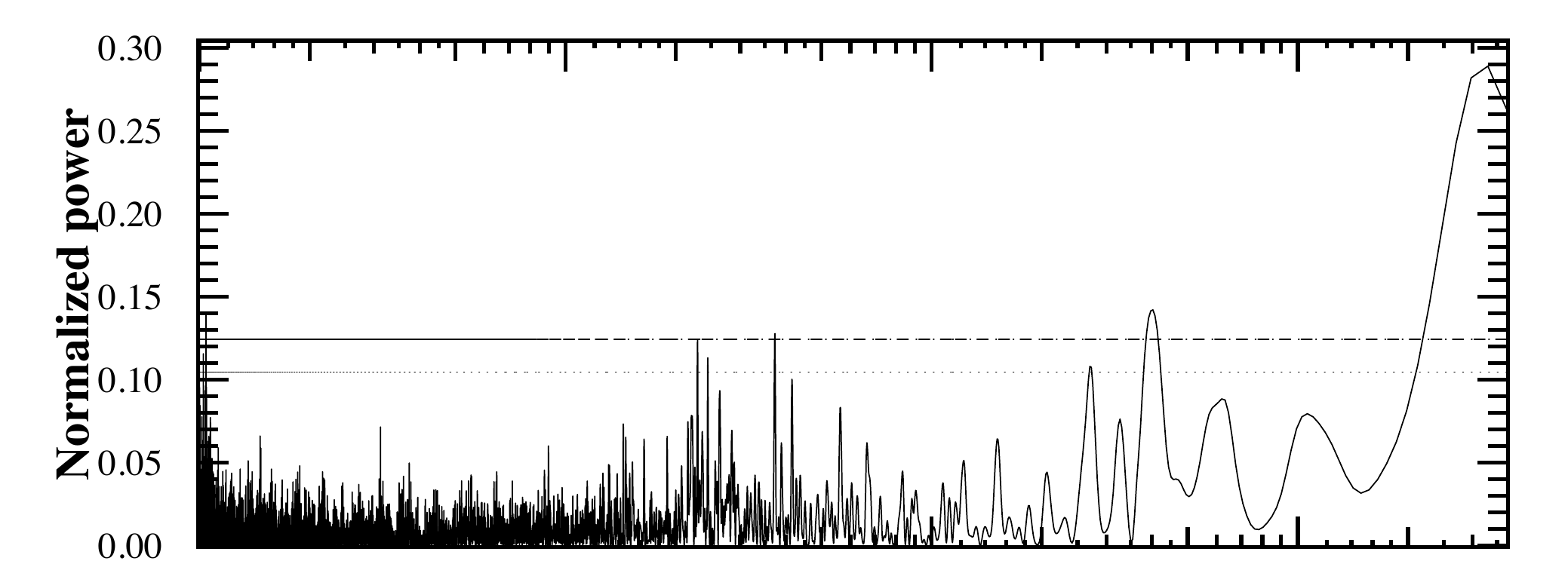}
\includegraphics[width=\columnwidth]{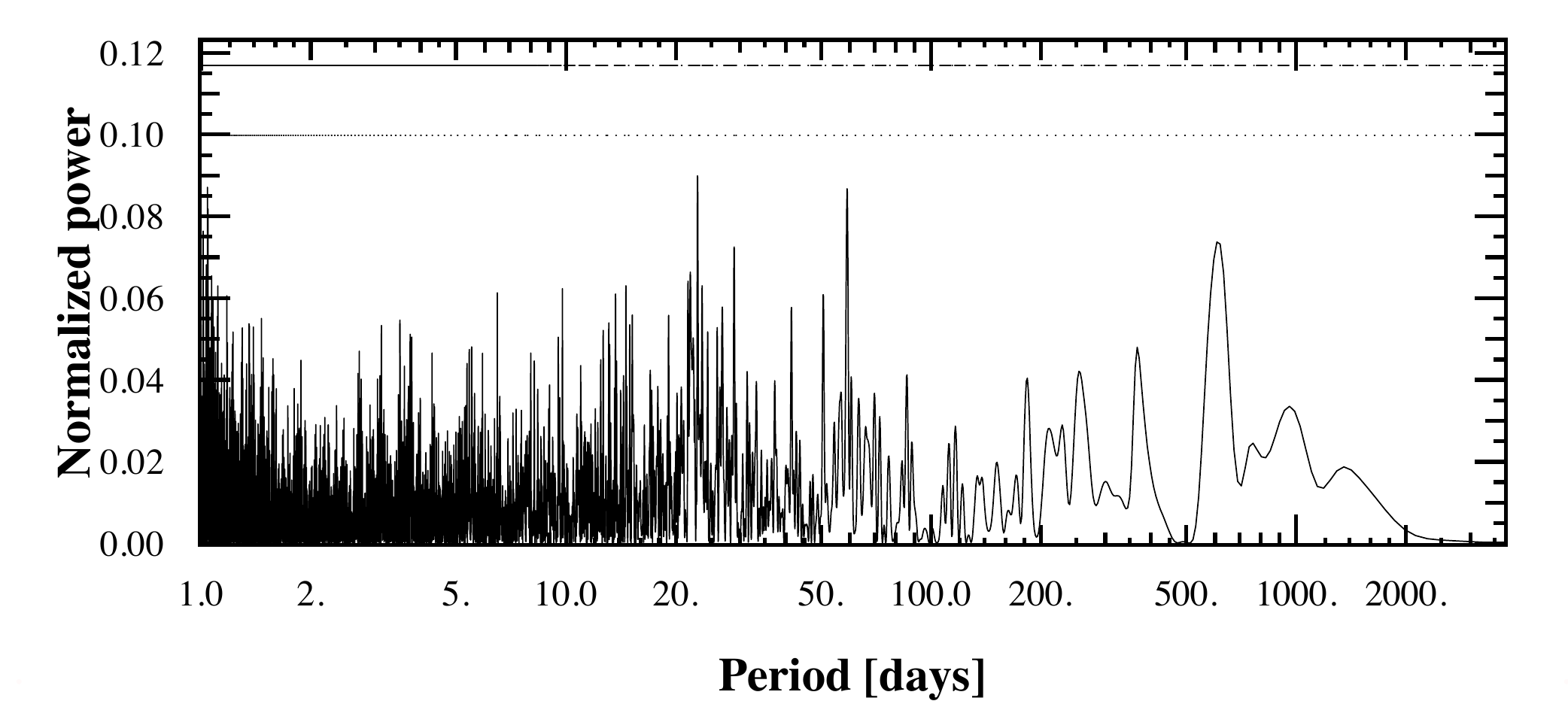}
\caption{Periodograms of the RV data of HD1461 (top panel) and residuals around models with 1, 2, and 3 Keplerians. The horizontal lines are the 10\% and 1\% $p$-value levels.\label{fig.HD1461omc}}
\end{figure}

\begin{figure}[t]
\input{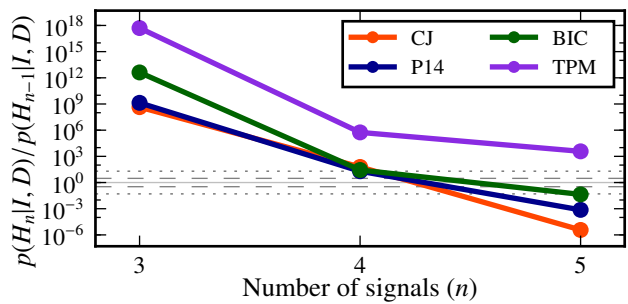}
\caption{HD1461. Odds ratio for models with $n$ Keplerian curves with respect to models with $n-1$ Keplerian curves as a function of model complexity $n$, assuming equal unity prior odds in all cases. The estimates using different techniques are shown and the customary limits for positive ($O_{n+1, n} = 3$) and strong ($O_{n+1, n} = 20$) and their inverses are shown as dashed and dotted lines, respectively. The model with four signals contains the 620-day Keplerian.}
\label{fig.1461bayes}
\end{figure}

The residuals around the one-Keplerian model (Fig.~\ref{fig.HD1461omc}) reveal a new significant signal with $P = 13.5$ days and an amplitude of 1.5 \ms, accompanied by its yearly and seasonal aliases. We employed the technique described by \citet{dawsonfabrycky2010} to identify the peak corresponding to the real signal, but the data were not sufficient to obtain a definitive answer. A long-term signal associated with the magnetic cycle (see below) is also significant. There is no peak in the spectral window function that might indicate that the 13.5-day peak is an alias of the longer activity-induced signal. { On the other hand, the signal period is close to the first harmonic of the estimated rotational period. \citet{boisse2011}, among others, showed that activity-induced signals preferentially appear at the rotational period and its two first harmonics. However}, the signal is recovered with the same period and amplitude if only the last five seasons of observations (BJD > 2'454'850) are considered, when the activity level of HD1461 was at a minimum. This indicates that the signal is coherent over many years, which is not expected from a signal induced by stellar magnetic activity. 

Furthermore, none of the \logR, bisector or FWHM time series show any significant power at this period. The bisector velocity span time series exhibits a dispersion of 1.24 \ms\ and 1.16 \ms, respectively, before and after the degree-three polynomial is used to correct for the effect of the activity cycle (see below). The GLS of the bisector exhibits significant power at 29.2 days with an amplitude of around 60~\cms, most likely caused by the stellar rotational modulation (see Table~\ref{table.stellarparams}). The FWHM time series does not present any significant periodicity when the long-term trend is removed and exhibits a dispersion of only 3 \ms\ over more than ten years. The time series of the \logR\ activity proxy, after correction for the long-term variation interpreted as the activity cycle, still exhibits power at periods $\sim 500$ days. As discussed below, this is surely due to an incorrect modelling of the activity cycle, which introduces aliasing frequencies in the periodogram of the corrected time series.

We conclude that the signal is best explained as produced by an additional planetary companion  to HD1461, with a minimum mass of around 6 \Me\footnote{This companion was previously announced by \citet{mayor2011}.}. The parameters of the new companion are listed in Table \ref{table.params1461}.

 \subsection{Activity cycle and search for additional signals. \label{sect.HD1461cycle}}

A common long-term evolution is conspicuous in the time series plotted in the upper panel of Fig.~\ref{fig.HD1461obs}. The periodograms in the lower panel of the same figure show peaks at periods of around 3000 days, close to the time span of the observations (Table~\ref{table.obslog}). Changes in the bisector span throughout a solar-like magnetic activity cycle are expected from changes in the convective blueshift pattern \citep[see for example][]{graybaliunas95, gray96b, dumusque2011c, lovis2011b}. These variations have a slightly longer period that the variation observed in the RV time series. The period of the FWHM is not yet constrained. We conclude that the long-period signal seen in the RV time series is produced by a magnetic activity cycle, with a period of $P_\mathrm{cycle} = 9.64 \pm 0.21$ yr, as measured by a Keplerian fit to the \logR\ time series. \citet{hall2009} presented seasonally averaged \logR\ measurements between late 1998 and late 2007. These data agree
well with the trend observed in the HARPS time series and seem to confirm the amplitude of activity variations. On the other hand, the combined data set hints at a longer period and at a shorter active season around the year 2007. A fit of the combined data set gives a period between 16 yr and 18 yr.
 
The RV signature of the activity cycle has a period $P = 9.1 \pm 0.4$ years, an amplitude above 3 \ms, and a significant eccentricity of $e = 0.43 \pm 0.07$. It is the dominant feature in the residuals of the model, including the planets at 5.77 and 13.5 days. This signal has to be corrected for to continue searching for signals in the RV time series and to avoid mistaking an alias of this long-term variation with real periodic signals. For example, the one-year aliases of the signal with the period of the activity cycle are located at 330 and 407 days. The GLS periodogram of the RV residuals shows significant power at these frequencies. On the other hand, the best-fit Keplerian curve to the \logR\ has a slightly different period and eccentricity ($e=0.17\pm0.07$ for \logR) than the one for the RVs, as also seen in the periodograms of Fig.~\ref{fig.HD1461obs}. An incorrect correction for the effect of activity can introduce spurious signals in the data.

We therefore decided to study different functional forms for the activity function $a(t)$ included in our model (Eq.~\ref{eq.model}) and to compare the signals obtained under each method. A signal independent of the correction method intuitively has more support than a signal that is only found for one particular correction method. The activity cycle was included in the model of the RV data in two different manners: 
\begin{itemize}
\item[a)] the \logR\ variations are modelled using a sinusoidal function, and the best-fit parameters are used as Gaussian priors for a fit to the RV time series, with the exception of the sine amplitude, which is free to vary, and 
\item[b)] same as (a), but using a Keplerian function instead of a sinusoid. 
\end{itemize}
Additionally, we tested other methods of removing the activity signal from the RV time series \emph{\textup{a priori}}: 
\begin{itemize}
\item[c)] applying a low-pass filter (cutoff at 100 days) to the \logR\ time series and using the filtered time series to detrend the RV data \citep[see][]{dumusque2012}, and 
\item[d)] running a principal component analysis on the combined \logR\ and RV time series; the corrected RV are constructed by using only the second principal component, which is orthogonal to the direction of the joint variation of RV and \logR. 
\end{itemize}
We note that all the methods used to account for the activity cycle assume a linear relation between the variations observed in the \logR\ proxy and those in the RV data. The alternative of fitting an additional Keplerian curve to the RV data without any prior information on the \logR\ variations was discarded
because it does not fully consider all available information. The search for additional signals was also performed on the RV data obtained after JD = 2'454'850, which correspond to the last five observing seasons and to the period of lesser magnetic activity, according to the \logR\ proxy. These inactive data set contains 191 nightly averaged observations spanning  4.5 yr. The activity cycle is less prominent in these data and appears as a weak drift in the radial velocities.

Additional signals were searched for in the RV data using each of the models of the activity cycle and adding a further Keplerian signal to the model with two planets and the magnetic cycle. We initialised the MCMC algorithm using the best-fit solution of the two-Keplerian model for the parameters of the two super-Earths and randomly drawing parameters from the prior joint density for the third potential planet. We note that although the two known planets were started at fixed points, no informative priors were used for their parameters, and they therefore were able
to change freely if the data required it in the model with three planets. To thoroughly explore the parameter space, we launched 75 chains thus initialised. We list the priors used for each parameter in Table~\ref{table.priors1461}.
 
 As expected, the chains became trapped in the numerous local likelihood maxima associated with different values of the period of the putative additional planet. By comparing the value of $\log(\mathcal{L}\; \pi)$ in each maximum, where $\mathcal{L}$ is the likelihood function and $\pi$ is the prior probability density, those that clearly produced a much poorer fit were discarded. 
 
Two signals were found irrespective of the method employed to model the activity cycle or correct its effect: a signal at 22.9 days with an amplitude of around 75 \cms, and another one around 620 days with an amplitude of 1.2 \ms. We note that their frequencies are not significantly present in the periodogram of the residuals of the three-Keplerian model (Fig.~\ref{fig.HD1461omc}). Table~\ref{table.HD1461evidences} presents the results of the Bayesian model comparison between models with and without these two additional periodicities, using a Keplerian model for the activity cycle with priors based on the \logR\ time series, as explained above.
 
 \subsection{Four-Keplerian model I. The 22.9-day signal.}

The parameters of the 22.9-day period Keplerian are approximately the same for different methods and for the inactive data set. In Table~\ref{table.HD1461evidences} both the \CJ\ and \Perr\ estimators indicate that the improvement in the data fit is not enough to justify the inclusion of the 22.9-day signal. The Bayesian information criterion (BIC)\footnote{The BIC is a very popular estimator based solely on the maximum likelihood of the model and the number of free parameters \citep{schwarz78}. It is therefore extremely simple to compute. According to \citet{kassraftery1995}, minus half the BIC tends to the logarithmic evidence of the model as the size of the data set increases. However, the authors warn that that the relative error is $\mathcal{O}(1)$, meaning that even for large samples the correct value is not achieved.} is inconclusive in this respect. We therefore discarded the possibility that only the 22.9-day signal is present. We tested, on the other hand, the inclusion of \emph{\textup{both}} 22.9-day and 620-day signals, but this model is not favoured by the data, probably due to the larger number of parameters. As discussed above, the TPM estimator overestimates the evidence for all cases and does not incorporate the Occam penalisation correctly, leading to a preference for more complex models, as is clearly seen in Table~\ref{table.HD1461evidences}.

\begin{figure}
\input{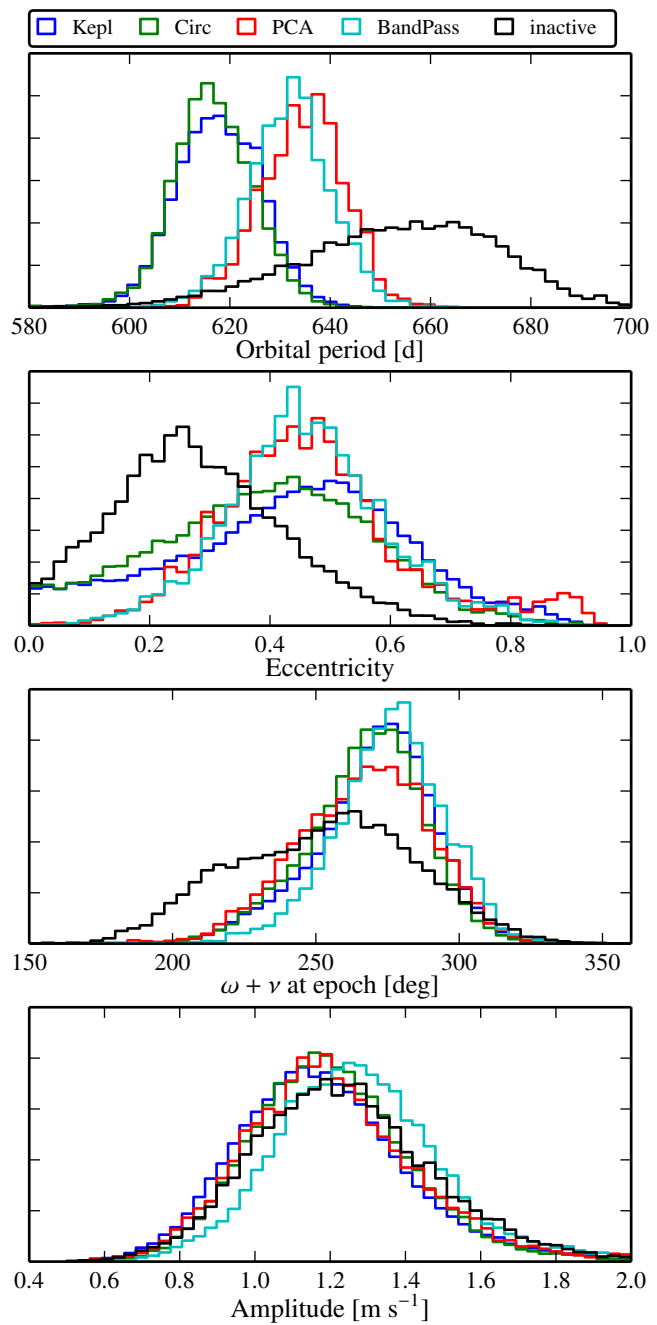}
\caption{Marginal posterior distributions of the orbital period, eccentricity, phase, and signal amplitude for the signal at around 620 days for each method used to account for the RV effect of the activity cycle. Also included are the posterior distributions using only the RVs obtained during the period of lower activity.\label{fig.HD1461posteriors}}
\end{figure}

\begin{table*}
{\tiny
\caption{Model probabilities for HD1461. Estimate of the evidence (marginal likelihood) for models with $n = 3, ...,  5$ Keplerians. The periods of the included signals are listed in the second column. The estimates based on the method of \citet{chibjeliazkov2001} (\CJ) and \citet{perrakis2014} (\Perr) are given in Cols. 3 and 4. For comparison, we also report the estimate obtained based on the \citet{tuomijones2012} TPM estimator and on the Bayesian information criterion (BIC) in Cols. 5 and 6. For clarity, we have subtracted 1100 from each estimation, which corresponds to a change of units in the observed radial velocities. Columns 7 and 8 list the base-10 logarithm of the odds ratio between model $i$ with respect to the three-Keplerian model. Additionally, the posterior estimates of the amplitude of the additional base-level noise and the dispersion of the residuals are given in Cols. 9 and 10.
\label{table.HD1461evidences}}
\begin{tabular}{c p{1.5cm} | c c c c | c c  | c c }
\hline
\hline
        &&\multicolumn{4}{c|}{$\log \prob{D}{M_n, I} - 1100$}                                                    &\multicolumn{2}{c|}{$\prob{M_n}{D, I}/\prob{M_3}{D, I}$}           &$\sigma_{J_n}|_{\log R^\prime_{HK} = -5}$      &$\sigma_{O-C}$\\
$n$             & Periods       [d]                                     & \CJ                     &\Perr                  & TPM                   &BIC            & \CJ                                             &\Perr                          &[ms$^{-1}$]                                                    &[ms$^{-1}$]\\
\hline
3               &\{5.77, 13.5, 3500\}                           &$80.44\pm0.06$ &$81.15\pm0.05$ &132.51                 &93.76  &$1.0$                          &$1.0$                          &$2.20 \pm 0.19$                       &$2.257\pm0.071$\\
\hline
4               &\{5.77, 13.5, 22.9, 3500\}                     &$76.31\pm0.11$ &$79.84\pm0.05$ &143.52                 &93.18  &$0.016\pm0.002$                &$0.274\pm0.016$                 &$2.15 \pm 0.18$                        &$2.242 \pm 0.075$\\
\hline
4               &\{5.77, 13.5, 620, 3500\}                      &$84.57\pm0.04$ &$84.16\pm0.11$ &145.74                 &97.07  &$60.8\pm4.0$                   &$19.6\pm2.4$                   &$1.87 \pm 0.19$                       &$1.965 \pm 0.092$\\
\hline
5               &\{5.77, 13.5, 22.9, 620, 3500\}                &$77.24\pm0.17$ &$76.92\pm0.14$ &153.97                 &93.94  &$(2.25\pm0.30)\times10^{-4}$           &$0.015\pm0.003$                &$1.81 \pm 0.19$       &$1.946 \pm 0.098$\\
\hline
\end{tabular}
}
\end{table*}

\subsection{Four-Keplerian model II. The 620-day period signal}

\begin{figure}
\includegraphics[width=\columnwidth]{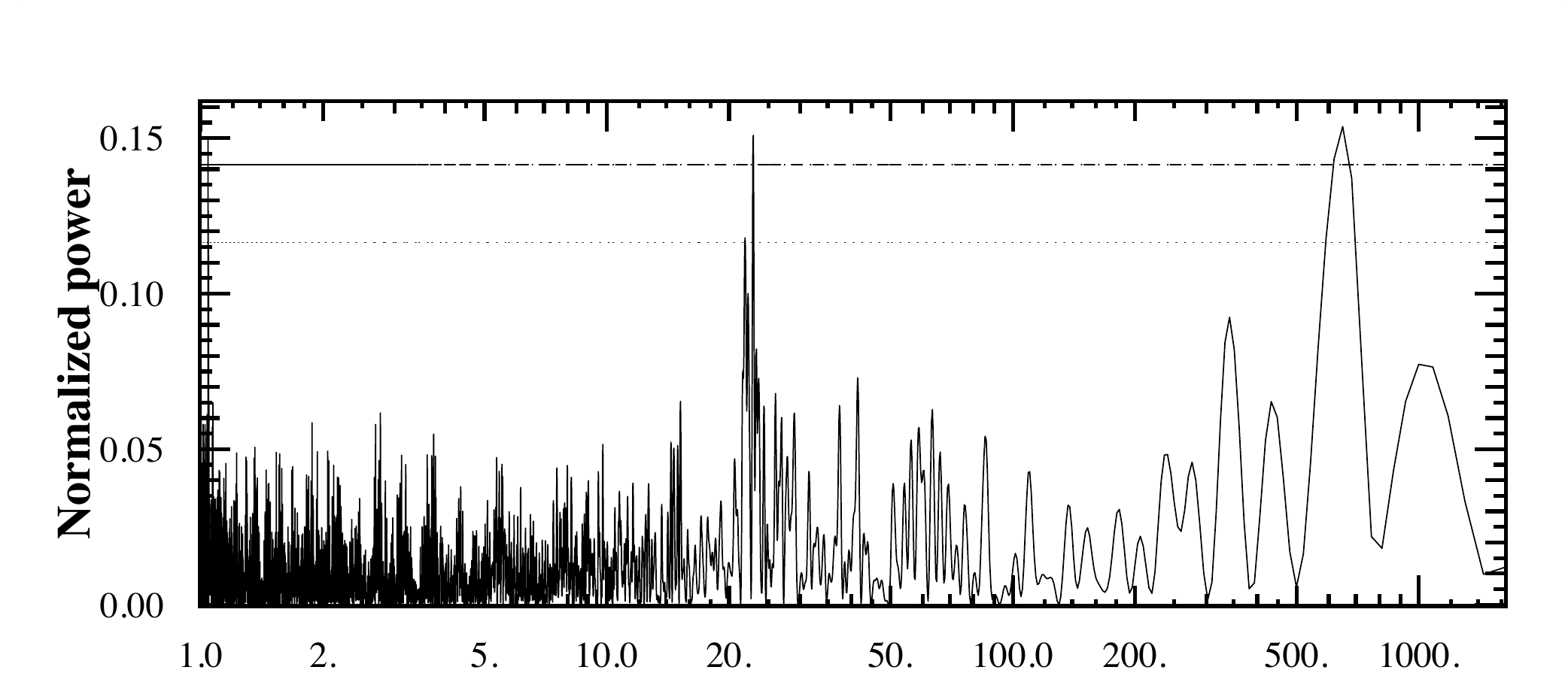}
\includegraphics[width=\columnwidth]{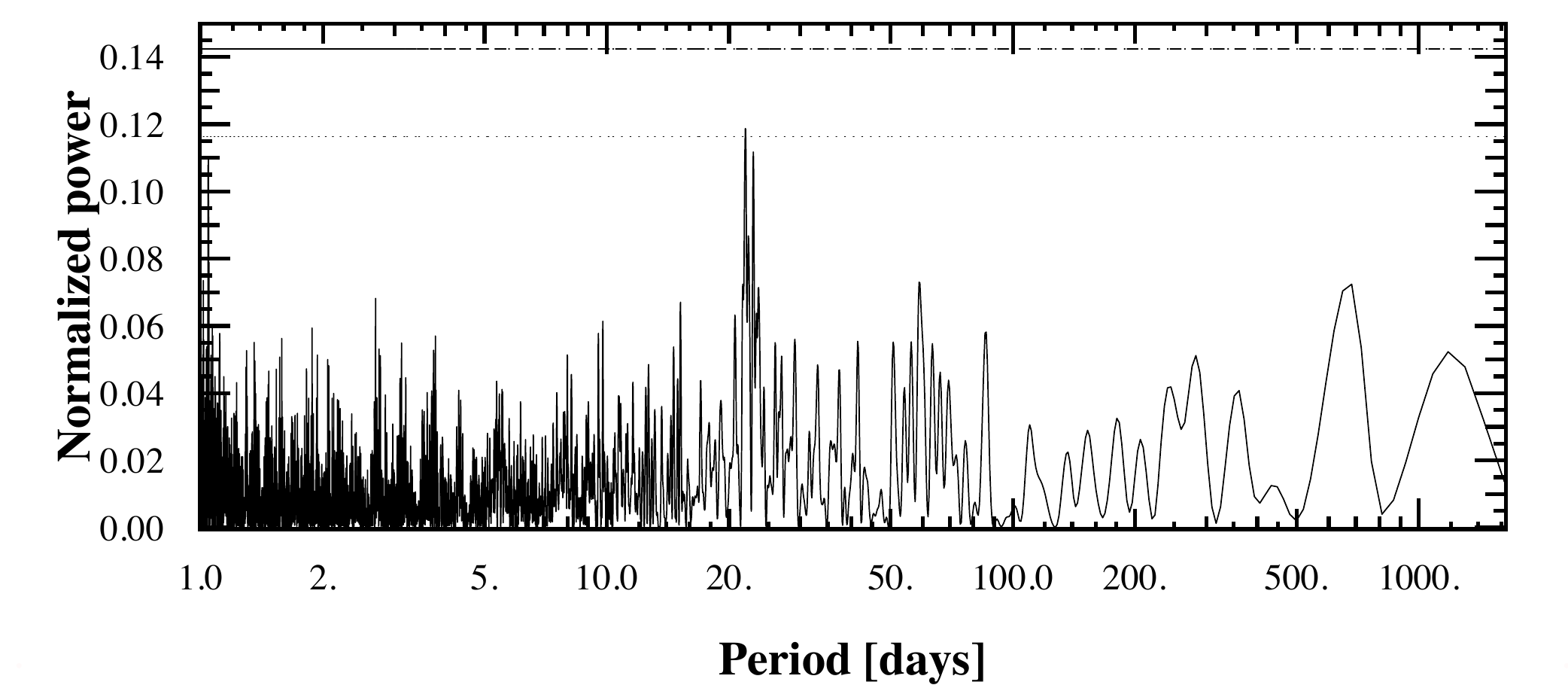}
\caption{HD1461. GLS periodogram of the RV residuals of the two-Keplerian model (top) and two-Keplerian + linear drift (bottom)  for data taken after JD = 2'454'850 (the inactive data set). The two peaks standing out as significant signals in the top panel have periods of 22.9 days and around 650 days. Note that the significance is reduced drastically when the long-term trend caused by the activity cycle is removed, indicating that the observed periodicities are aliases of a long-period signal present in the data.} \label{fig.HD1461inactive_omc_k2}
\end{figure}

Both the \CJ\ and \Perr\ techniques favour the inclusion of a signal at 620 days, with a Bayes factor of 60 and 20, respectively, { which is considered as} strong  evidence \citep{kassraftery1995}. The estimation based on the BIC leads to a  similar conclusion.  

For all models used to describe the effect of the activity cycle of HD1461 on the RV measurements, a signal at around 630 days is seen, albeit its period changes slightly with the method employed. Methods (a) and (b) produce a signal closer to 615 days, while for methods (c) and (d), the signal  is found closer to 640 days (Fig.~\ref{fig.HD1461posteriors}). For all methods, the amplitude is compatible. If this signal is of planetary origin, the minimum mass of the companion would be $M_d \sim 14.5 \pm 1.3$ \Me. No significant power is present at similar periods in the time series of the \logR, the bisector velocity span, or the FWHM, even after subtracting the long-term trend associated with the magnetic cycle. 

If only the inactive data set (BJD > 2'454'850) is considered, the GLS periodogram of the residuals of a two-Keplerian model exhibits significant ($p$-value < 0.01) power at the period of the signal (Fig.~\ref{fig.HD1461inactive_omc_k2}). However, when a linear drift is added to the model to account for the effect of the activity cycle, the amplitude of the peak is strongly reduced (Fig.~\ref{fig.HD1461inactive_omc_k2}), indicating that the periodicity may be an alias of the long-term trend and not a real signal. The 22.9-day signal exhibits the same behaviour. This would explain why the signal is recovered for all the correction methods of the activity cycle, as well as the slight change of the period under different corrections. Since a long-period signal must remain in the data for the alias frequencies to be present, this either means that the correction of the activity cycle is not fully satisfactory with any of the methods or that an additional long-term signal, still not fully sampled, is present in the data. We conclude that although the periodicity at 620-day period is significantly present in the data, its nature is still uncertain and might originate in an incomplete correction of the activity cycle. 

\begin{figure*}
\input{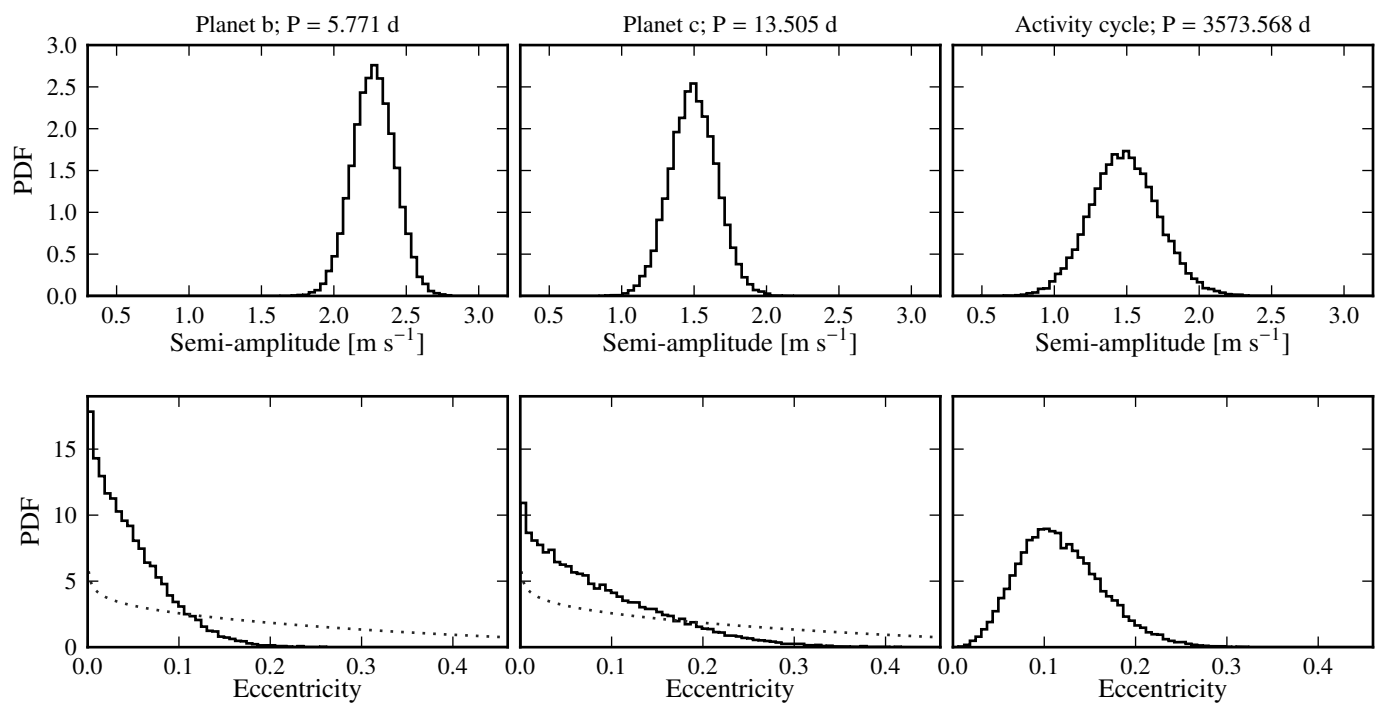}
\caption{Posterior distributions of the amplitude (top row) and eccentricity (bottom row) of the three Keplerian curves used to model the HARPS radial velocities of HD1461. The grey dotted curves represent the eccentricity prior for the planetary signals. To facilitate comparison, the axis scales are the same for the three signals. \label{fig.HD1461_posteriors}}
\end{figure*}

\begin{figure*}
\input{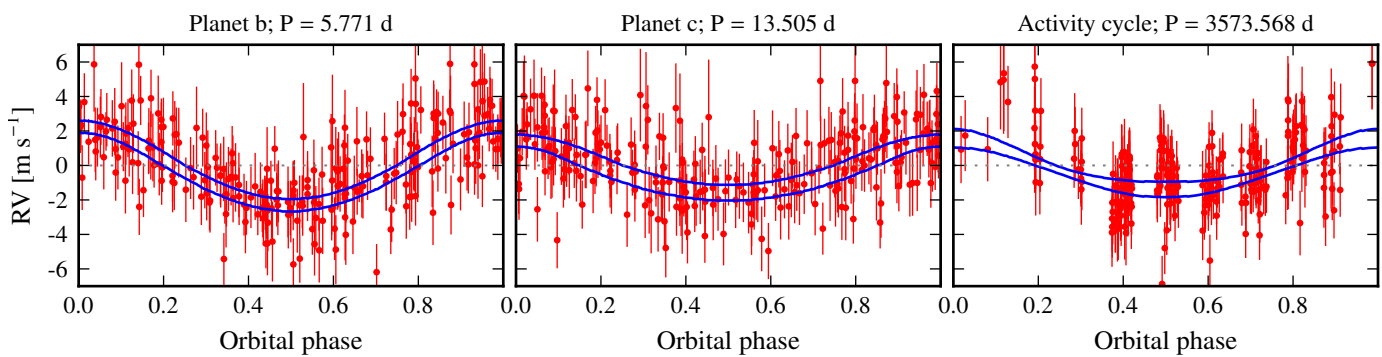}
\caption{Radial velocity data phase-folded to the best-fit period of each of the three Keplerian curves used in the final modelling of HD1461, after subtracting the effect of the remaining signals. The error bars include the additional noise term. The blue lines represent the 95-\% highest density interval (HDI).\label{fig.HD1461_orbits}}
\end{figure*}

\begin{table*}[t]
{\small\center
\caption{Parameter posteriors for the HD1461 system. The epoch is BJD=2,455,155.3854 for planets b and c and BJD=2,455,195.8367 for the magnetic cycle.\label{table.params1461}}            
\begin{tabular}{l l c c c }        
\hline\hline                 
\noalign{\smallskip}

\multicolumn{2}{c}{ Orbital parameters } 		&Planet b	&Planet c &Magnetic cycle\\
\hline
\noalign{\smallskip}
Orbital period, $P^{\bullet}$ 	&[days]	 	&$5.77152 \pm 0.00045$ &$13.5052 \pm 0.0029$	&$3503 \pm 80$\\
RV amplitude, $K^{\bullet}$ 	&[\ms]		&$2.28\pm0.15$ 		&$1.49 \pm 0.17$		&$1.51 \pm 0.26$ \\
Eccentricity, $e$$^{\bullet}$    	& 			&$<0.131; <0.172^\dagger$ &$<0.228; <0.305^\dagger$ &$0.103^{+0.063}_{-0.030}$\\
Argument of periastron, $\omega^{\bullet}$&[deg] & --$^\dagger$	& --$^\dagger$		&$294 \pm 15$\\
$e^{1/2} \cos(\omega)$		&			&				&				&$0.131 \pm 0.092^{\bullet}$\\
$e^{1/2} \sin(\omega)$		&			&				&				&$-0.298 \pm 0.068^{\bullet}$\\
Mean longitude at epoch, $L_0^{\bullet}$ &[deg] &$271.6 \pm 4.1$ 	&$317.9 \pm 6.6$	&$148.6 \pm 6.7$\\
Systemic velocity, $V_0^{\bullet}$		&[\kms]		&\multicolumn{3}{c}{$ -10.05960 \pm 1.7\times10^{-4}$}\\
Semi-major axis of relative orbit, $a$		&[AU] &$0.0634\pm0.0022$ &$0.1117\pm0.0039$	&--\\
Minimum mass, $M \sin i$ 			&[\Me] &$6.44\pm0.61$ 	&$5.59\pm0.73$		&--\\
\noalign{\smallskip}
\multicolumn{2}{c}{Noise model$\ddag$} 		\\
\hline
\noalign{\smallskip}
Additional noise at \logR=-5, $\sigma_J|_{-5.0}^{\bullet}$	&[\ms]		&\multicolumn{3}{c}{$2.13^{+0.25}_{-0.11}$}\\
Slope, $\alpha_J^{\bullet}$						&[\ms/dex]	&\multicolumn{3}{c}{$25.5 \pm 6.4$}\\
Additional noise at <\logR>						&[\ms]		&\multicolumn{3}{c}{$1.67 \pm 0.09$}\\
$\mathrm{rms}(\mathrm{O-C})$				&[\ms]		&\multicolumn{3}{c}{$2.260\pm0.067$}\\
\hline
\hline
\end{tabular}
\tablefoot{

$\bullet$: MCMC jump parameter.

$\dagger$: eccentricity does not differ significantly from zero; the 95\% and 99\% upper limits are reported. The argument of periastron $\omega$ is therefore unconstrained.

$\ddag$: the additional (stellar) noise for measurement $i$ is $\sigma_{Ji} = \sigma_J|_{-5.0} + \alpha_J \cdot (\log{(R'_{\rm HK})}_i + 5.0)$.
}
}
\end{table*}

\subsection{The planetary system around HD1461 \label{sect.1461mcmc}}
Our final model of the RV series includes two Keplerian curves for the known planet candidates at 5.77 and 13.5 days and an additional Keplerian curve to model the activity cycle. The planet signals are independent of the method used to model the activity cycle. For simplicity, we chose the Keplerian model, which also allows us, unlike the filtering and principal components method, to include the uncertainties in the parameters in the error budget of the planet signals.

The resulting posterior distributions for the semi-amplitude and the orbital eccentricity are shown in Fig.~\ref{fig.HD1461_posteriors}, which clearly shows that the three signals have amplitudes significantly different from zero. The covariance between the three semi-amplitudes and the eccentricities is { much} smaller than the variance of each parameter.
In Table \ref{table.params1461} the mode and 68.3\% credible intervals are listed for all MCMC parameters and for a series of derived parameters.  The reflex motion induced by the new companion at the 13.5-day period has an amplitude of $1.49\pm0.17$ \ms, which implies a minimum mass of $5.59\pm0.73$ \Me, in agreement with the values reported by \citet{mayor2011}. The RV amplitude associated with the activity cycle is $1.51\pm0.26$ \ms, which means that the scaling constant $\alpha$ between \logR\ and RV is $74.2\pm12.8$ \ms/dex, in good agreement with the value calibrated as a function of effective temperature and metallicity by \citet{lovis2011b}, which gives 74.5 \ms/dex.

The RV data folded to the best-fit period of each signal are shown in Fig. \ref{fig.HD1461_orbits} after subtracting the effect of the remaining signals. This correction was made by computing the model corresponding to each Keplerian curve for each step of the MCMC, sampled at the data times. The mean value of these models in each data time was subtracted from the observed data. The blue curves in Fig.~\ref{fig.HD1461_orbits} are the 95\% highest density interval (HDI) of the curve sampled in 300 phase points. We computed this in a similar way as \citet{gregory2011}: the period of each signal was sampled at 300 points, and the corresponding RV model was computed for each posterior sample element obtained with the MCMC algorithm. 

{ To study the stability of the system, we performed a numerical integration of the system over half a million years using the Mercury code \citep{chambers99}. Two simulations were run: the first using the minimum masses as the true masses of the companions and coplanar orbits, and the second one increasing the masses by a factor two and including a mutual inclination of ten degrees. The initial eccentricities were set to the 95\% upper confidence level. In both cases the system was stable over the integrated time scale. Additionally, the eccentricities did not increase beyond 0.24 and 0.23 for planet b and c, respectively. The fractional semi-major axis change is smaller than $10^{-6}$ for the outer companion and around $8\times10^{-5}$ for the inner one, which
is similar to the precision of the integrator.}

The residuals of the model with three Keplerians still show significant scatter (2.3 \ms), which forces the additional noise component of the model to be $1.7\pm0.1$ \ms\ for the mean \logR\ value. This is caused partially by the remaining signal originated in an incomplete cycle correction and by other effects that were
not taken into account in our model, such as rotational modulation of the RV data due to stellar spots. It could also be indicative of additional planets in the system. Further observations of this system are needed to fully characterise it.

\citet{rivera2010} reported two potential signals with periods around 450 and 5000 days in their HIRES data set. In the light of the present analysis, their detection might be related to an incompletely sampled magnetic cycle, although more data are needed to reach a firm conclusion on the nature of those suggested periodicities.

\section{HD40307 \label{sect.HD40307}}

HD40307 was reported to host three super-Earth-type planets with orbital periods $P_\mathrm{b} = 4.311$ d, $P_\mathrm{c} = 9.6$ d, and $P_\mathrm{d} = 20.5$ d by \citet[hereafter \M,]{mayor2009} based on 2.4 years of HARPS data. More recently, \citet[][hereafter \T,]{tuomi2013b} analyzed the publicly available HARPS data, which included the \M\ data and observations taken on three additional nights, and claimed the presence of three additional planets in the system, with orbital periods $P_\mathrm{e} = 34.62$ d, $P_\mathrm{f} = 51.76$ d, and $P_\mathrm{g} = 197.8$ d. Planet g would be in the habitable zone of the star. They also detected a periodic signal with $P \sim320$ days that they attributed to magnetic activity effects because its amplitude changes depending on the fraction of the spectrum used to compute the radial velocities. The analysis by \T\ differs  from the one by \M\ mainly in the way the radial velocities are obtained --by template matching instead of mask cross-correlation-- and in that they used the complete HARPS data set instead of the nightly binned velocities, including seven points taken during the commissioning of the instrument. Additionally, \T\ used a moving-average model to take into account the correlation between individual observations taken during a single night, a deterministic model of the short-term activity signal. The analysis presented here includes four additional years of data, for a total of 226 nightly averaged radial velocity measurements taken over eight years. This represents around 70\% more data points than used by \M. We list the data in Table~\ref{table.rvHD40307}.

\onllongtab{
\begin{longtable}{r r r r r r r}
\caption{HARPS measurements of HD40307. \label{table.rvHD40307}}\\
\hline\hline
\multicolumn{1}{c}{BJD} &	\multicolumn{1}{c}{RV} &	\multicolumn{1}{c}{$\sigma_\text{RV}$} &	\multicolumn{1}{c}{BIS} &	\multicolumn{1}{c}{FWHM} &	 \multicolumn{1}{c}{\logR} &	\multicolumn{1}{c}{$\sigma_\text{\logR}$} \\
-2 450 000  &(\kms) & (\kms) & (\ms) & (\kms) & & \\
\hline
\noalign{\smallskip}
\endfirsthead
\caption{Continued.}\\
\hline
\multicolumn{1}{c}{BJD} &	\multicolumn{1}{c}{RV} &	\multicolumn{1}{c}{$\sigma_\text{RV}$} &	\multicolumn{1}{c}{BIS} &	\multicolumn{1}{c}{FWHM} &	 \multicolumn{1}{c}{\logR} &	\multicolumn{1}{c}{$\sigma_\text{\logR}$} \\
-2 450 000  &(\kms) & (\kms) & (\ms) & (\kms) & & \\
\hline
\noalign{\smallskip}
\endhead
\hline
\endfoot
\hline
\hline
\endlastfoot
52942.8215 &	31.3372 &	0.0009 &	8.98 &	5.9072 &	-4.9460 &	0.0029 \\
52999.7639 &	31.3340 &	0.0011 &	8.84 &	5.9011 &	-4.9971 &	0.0045 \\
53000.7606 &	31.3375 &	0.0012 &	11.18 &	5.9002 &	-4.9849 &	0.0054 \\
53001.6684 &	31.3397 &	0.0012 &	8.41 &	5.8989 &	-4.9859 &	0.0054 \\
53002.6686 &	31.3393 &	0.0011 &	8.40 &	5.8993 &	-4.9689 &	0.0034 \\
53054.5912 &	31.3345 &	0.0011 &	9.32 &	5.9008 &	-4.9780 &	0.0049 \\
53692.7333 &	31.3355 &	0.0002 &	7.93 &	5.8949 &	-5.0151 &	0.0010 \\
53693.6377 &	31.3357 &	0.0002 &	8.45 &	5.8973 &	-5.0225 &	0.0013 \\
53694.7298 &	31.3370 &	0.0003 &	7.28 &	5.8981 &	-5.0151 &	0.0013 \\
53695.6997 &	31.3383 &	0.0002 &	8.49 &	5.8932 &	-5.0016 &	0.0008 \\
53696.7069 &	31.3375 &	0.0002 &	7.94 &	5.8959 &	-5.0112 &	0.0009 \\
53697.7323 &	31.3330 &	0.0004 &	6.68 &	5.9002 &	-5.0295 &	0.0035 \\
53698.7371 &	31.3309 &	0.0003 &	7.01 &	5.8990 &	-5.0104 &	0.0015 \\
53699.7361 &	31.3312 &	0.0003 &	7.70 &	5.8968 &	-5.0043 &	0.0011 \\
53700.7686 &	31.3311 &	0.0002 &	6.49 &	5.8978 &	-4.9999 &	0.0010 \\
53721.7446 &	31.3339 &	0.0002 &	7.22 &	5.8947 &	-5.0047 &	0.0009 \\
53722.7439 &	31.3351 &	0.0002 &	6.81 &	5.8963 &	-5.0047 &	0.0009 \\
53724.7067 &	31.3326 &	0.0002 &	7.29 &	5.8950 &	-5.0015 &	0.0008 \\
53725.7143 &	31.3343 &	0.0002 &	7.84 &	5.8962 &	-4.9923 &	0.0008 \\
53726.7106 &	31.3336 &	0.0002 &	7.45 &	5.8962 &	-4.9866 &	0.0008 \\
53727.6892 &	31.3304 &	0.0002 &	7.55 &	5.8961 &	-4.9859 &	0.0008 \\
53728.7664 &	31.3278 &	0.0003 &	7.35 &	5.8935 &	-4.9869 &	0.0009 \\
53729.7554 &	31.3321 &	0.0002 &	7.62 &	5.8949 &	-4.9892 &	0.0008 \\
53757.5916 &	31.3321 &	0.0003 &	8.88 &	5.8961 &	-4.9902 &	0.0016 \\
53758.5762 &	31.3286 &	0.0003 &	7.96 &	5.8966 &	-4.9954 &	0.0015 \\
53759.5854 &	31.3305 &	0.0003 &	6.74 &	5.8946 &	-4.9949 &	0.0014 \\
53760.6669 &	31.3339 &	0.0004 &	7.26 &	5.8967 &	-4.9992 &	0.0020 \\
53761.6032 &	31.3331 &	0.0003 &	7.00 &	5.8962 &	-4.9991 &	0.0015 \\
53762.5932 &	31.3323 &	0.0004 &	7.69 &	5.8957 &	-4.9949 &	0.0020 \\
53763.5947 &	31.3304 &	0.0003 &	7.03 &	5.8955 &	-4.9970 &	0.0013 \\
53764.6167 &	31.3317 &	0.0003 &	7.01 &	5.8925 &	-4.9947 &	0.0011 \\
53765.6021 &	31.3302 &	0.0003 &	5.86 &	5.8946 &	-4.9937 &	0.0012 \\
53782.5590 &	31.3366 &	0.0003 &	8.26 &	5.8979 &	-4.9953 &	0.0015 \\
53784.5865 &	31.3283 &	0.0004 &	6.35 &	5.8975 &	-4.9871 &	0.0019 \\
53786.6115 &	31.3297 &	0.0003 &	5.99 &	5.9009 &	-4.9825 &	0.0015 \\
53788.6003 &	31.3290 &	0.0003 &	7.47 &	5.9010 &	-4.9817 &	0.0016 \\
53790.5911 &	31.3357 &	0.0003 &	8.81 &	5.9011 &	-4.9768 &	0.0014 \\
53810.5838 &	31.3335 &	0.0003 &	6.93 &	5.8990 &	-4.9811 &	0.0015 \\
53811.5846 &	31.3358 &	0.0003 &	8.10 &	5.9009 &	-4.9799 &	0.0010 \\
53812.5590 &	31.3369 &	0.0004 &	8.32 &	5.9032 &	-4.9833 &	0.0018 \\
53814.6012 &	31.3304 &	0.0003 &	9.35 &	5.9007 &	-4.9867 &	0.0016 \\
53817.5938 &	31.3368 &	0.0004 &	8.38 &	5.8995 &	-4.9865 &	0.0018 \\
53829.5208 &	31.3372 &	0.0004 &	6.58 &	5.9003 &	-4.9881 &	0.0028 \\
53831.5178 &	31.3332 &	0.0003 &	7.25 &	5.8957 &	-4.9903 &	0.0017 \\
53835.5200 &	31.3345 &	0.0004 &	7.49 &	5.9013 &	-4.9913 &	0.0024 \\
53861.4619 &	31.3350 &	0.0004 &	9.85 &	5.9035 &	-4.9795 &	0.0020 \\
53862.4676 &	31.3308 &	0.0003 &	9.09 &	5.9024 &	-4.9811 &	0.0016 \\
53863.4623 &	31.3301 &	0.0003 &	8.26 &	5.8993 &	-4.9867 &	0.0015 \\
53864.4569 &	31.3331 &	0.0004 &	7.85 &	5.9013 &	-4.9817 &	0.0020 \\
53865.4551 &	31.3303 &	0.0003 &	8.41 &	5.8962 &	-4.9860 &	0.0013 \\
53866.4564 &	31.3306 &	0.0003 &	6.29 &	5.8964 &	-4.9851 &	0.0014 \\
53867.4745 &	31.3326 &	0.0003 &	7.80 &	5.8990 &	-4.9909 &	0.0015 \\
53868.4550 &	31.3358 &	0.0004 &	7.68 &	5.9006 &	-4.9882 &	0.0017 \\
53869.4563 &	31.3343 &	0.0003 &	7.54 &	5.8964 &	-4.9886 &	0.0013 \\
53870.4549 &	31.3316 &	0.0004 &	6.00 &	5.9018 &	-4.9981 &	0.0021 \\
53871.4559 &	31.3305 &	0.0004 &	6.86 &	5.8999 &	-4.9939 &	0.0017 \\
53882.4655 &	31.3363 &	0.0005 &	4.08 &	5.9093 &	-4.9600 &	0.0032 \\
53974.9226 &	31.3351 &	0.0005 &	8.82 &	5.9086 &	-4.9893 &	0.0027 \\
53980.9124 &	31.3376 &	0.0004 &	7.96 &	5.9076 &	-4.9772 &	0.0020 \\
53981.9228 &	31.3364 &	0.0004 &	6.55 &	5.9091 &	-4.9658 &	0.0025 \\
54047.8047 &	31.3294 &	0.0004 &	7.85 &	5.9048 &	-4.9978 &	0.0020 \\
54049.8006 &	31.3348 &	0.0004 &	8.67 &	5.9021 &	-4.9954 &	0.0018 \\
54051.8153 &	31.3327 &	0.0005 &	5.65 &	5.9012 &	-5.0094 &	0.0026 \\
54053.8252 &	31.3324 &	0.0004 &	6.84 &	5.8986 &	-4.9912 &	0.0016 \\
54055.8301 &	31.3292 &	0.0004 &	7.59 &	5.8995 &	-5.0001 &	0.0017 \\
54077.7374 &	31.3310 &	0.0003 &	8.58 &	5.9004 &	-4.9863 &	0.0015 \\
54079.7067 &	31.3404 &	0.0004 &	7.17 &	5.9014 &	-4.9953 &	0.0021 \\
54081.7148 &	31.3357 &	0.0005 &	6.09 &	5.9024 &	-5.0027 &	0.0030 \\
54083.7635 &	31.3377 &	0.0005 &	7.02 &	5.9034 &	-5.0039 &	0.0033 \\
54115.6153 &	31.3318 &	0.0007 &	9.70 &	5.9094 &	-4.9927 &	0.0063 \\
54120.6782 &	31.3350 &	0.0004 &	8.87 &	5.9084 &	-4.9859 &	0.0018 \\
54121.6436 &	31.3344 &	0.0004 &	7.47 &	5.9071 &	-4.9846 &	0.0020 \\
54136.6111 &	31.3354 &	0.0003 &	6.80 &	5.9032 &	-4.9942 &	0.0014 \\
54137.6026 &	31.3339 &	0.0003 &	7.15 &	5.9034 &	-4.9900 &	0.0014 \\
54141.6090 &	31.3330 &	0.0004 &	6.33 &	5.9096 &	-4.9738 &	0.0024 \\
54143.5605 &	31.3356 &	0.0004 &	7.73 &	5.9062 &	-4.9769 &	0.0019 \\
54167.5436 &	31.3352 &	0.0003 &	8.47 &	5.9015 &	-4.9842 &	0.0014 \\
54169.5179 &	31.3316 &	0.0003 &	8.33 &	5.8994 &	-4.9842 &	0.0012 \\
54171.5389 &	31.3287 &	0.0003 &	6.51 &	5.9024 &	-4.9892 &	0.0013 \\
54173.5437 &	31.3306 &	0.0003 &	8.01 &	5.9051 &	-4.9825 &	0.0013 \\
54194.5009 &	31.3315 &	0.0003 &	8.47 &	5.9035 &	-4.9793 &	0.0018 \\
54196.5009 &	31.3381 &	0.0004 &	8.25 &	5.9047 &	-4.9724 &	0.0022 \\
54197.4970 &	31.3336 &	0.0004 &	7.31 &	5.9032 &	-4.9760 &	0.0020 \\
54198.5098 &	31.3297 &	0.0003 &	8.48 &	5.9049 &	-4.9798 &	0.0017 \\
54199.4972 &	31.3314 &	0.0003 &	6.88 &	5.9044 &	-4.9828 &	0.0016 \\
54200.4850 &	31.3343 &	0.0003 &	7.41 &	5.9044 &	-4.9811 &	0.0017 \\
54202.4962 &	31.3343 &	0.0004 &	8.19 &	5.9073 &	-4.9814 &	0.0024 \\
54225.4888 &	31.3403 &	0.0006 &	7.82 &	5.9098 &	-4.9614 &	0.0048 \\
54228.4821 &	31.3334 &	0.0005 &	9.10 &	5.9107 &	-4.9763 &	0.0039 \\
54229.4847 &	31.3329 &	0.0005 &	9.26 &	5.9140 &	-4.9746 &	0.0036 \\
54231.4644 &	31.3348 &	0.0004 &	7.66 &	5.9077 &	-4.9659 &	0.0025 \\
54232.4735 &	31.3330 &	0.0004 &	8.73 &	5.9063 &	-4.9708 &	0.0029 \\
54233.4609 &	31.3354 &	0.0007 &	6.25 &	5.9113 &	-4.9676 &	0.0061 \\
54234.4906 &	31.3359 &	0.0005 &	8.53 &	5.9094 &	-4.9732 &	0.0032 \\
54315.8883 &	31.3282 &	0.0010 &	7.07 &	5.9209 &	-4.9578 &	0.0092 \\
54319.9088 &	31.3320 &	0.0005 &	7.82 &	5.9142 &	-4.9902 &	0.0025 \\
54342.8743 &	31.3346 &	0.0004 &	6.85 &	5.9110 &	-5.0019 &	0.0016 \\
54346.9095 &	31.3385 &	0.0004 &	6.32 &	5.9084 &	-4.9896 &	0.0031 \\
54347.8897 &	31.3391 &	0.0004 &	6.70 &	5.9040 &	-4.9852 &	0.0028 \\
54349.8774 &	31.3366 &	0.0004 &	9.25 &	5.9071 &	-4.9864 &	0.0016 \\
54385.7935 &	31.3362 &	0.0004 &	6.53 &	5.9165 &	-4.9940 &	0.0021 \\
54386.7498 &	31.3372 &	0.0004 &	7.29 &	5.9141 &	-4.9873 &	0.0015 \\
54387.7896 &	31.3369 &	0.0004 &	8.82 &	5.9091 &	-4.9857 &	0.0013 \\
54390.7976 &	31.3369 &	0.0004 &	7.78 &	5.9046 &	-4.9866 &	0.0016 \\
54392.7352 &	31.3312 &	0.0004 &	6.17 &	5.9088 &	-4.9825 &	0.0021 \\
54393.7681 &	31.3332 &	0.0004 &	7.46 &	5.9099 &	-4.9815 &	0.0013 \\
54394.7600 &	31.3359 &	0.0004 &	6.82 &	5.9086 &	-4.9828 &	0.0016 \\
54419.7942 &	31.3301 &	0.0004 &	8.10 &	5.9076 &	-4.9755 &	0.0016 \\
54420.7671 &	31.3297 &	0.0004 &	8.16 &	5.9039 &	-4.9753 &	0.0012 \\
54421.7266 &	31.3282 &	0.0003 &	8.33 &	5.9033 &	-4.9763 &	0.0011 \\
54422.7412 &	31.3278 &	0.0004 &	8.22 &	5.9076 &	-4.9829 &	0.0015 \\
54423.7683 &	31.3317 &	0.0004 &	6.74 &	5.9027 &	-4.9854 &	0.0016 \\
54424.7418 &	31.3366 &	0.0004 &	7.43 &	5.9036 &	-4.9861 &	0.0016 \\
54425.7368 &	31.3363 &	0.0004 &	6.84 &	5.9029 &	-4.9873 &	0.0018 \\
54426.7164 &	31.3343 &	0.0003 &	7.64 &	5.9011 &	-4.9848 &	0.0011 \\
54427.7460 &	31.3367 &	0.0004 &	7.52 &	5.9032 &	-4.9855 &	0.0013 \\
54428.7555 &	31.3386 &	0.0004 &	7.91 &	5.9030 &	-4.9824 &	0.0013 \\
54429.7291 &	31.3382 &	0.0005 &	6.26 &	5.9052 &	-4.9858 &	0.0016 \\
54445.7340 &	31.3388 &	0.0004 &	7.91 &	5.9044 &	-4.9745 &	0.0014 \\
54451.7602 &	31.3361 &	0.0004 &	7.54 &	5.9084 &	-4.9684 &	0.0020 \\
54454.7705 &	31.3379 &	0.0004 &	6.67 &	5.9112 &	-4.9591 &	0.0015 \\
54478.6654 &	31.3279 &	0.0004 &	7.08 &	5.9115 &	-4.9609 &	0.0013 \\
54479.6221 &	31.3278 &	0.0004 &	7.47 &	5.9126 &	-4.9605 &	0.0013 \\
54480.5850 &	31.3311 &	0.0004 &	8.64 &	5.9116 &	-4.9628 &	0.0016 \\
54481.6881 &	31.3309 &	0.0004 &	8.18 &	5.9115 &	-4.9595 &	0.0020 \\
54483.5935 &	31.3348 &	0.0004 &	8.49 &	5.9108 &	-4.9615 &	0.0014 \\
54484.6271 &	31.3365 &	0.0004 &	8.24 &	5.9128 &	-4.9535 &	0.0011 \\
54486.5871 &	31.3336 &	0.0004 &	9.10 &	5.9127 &	-4.9558 &	0.0016 \\
54529.5416 &	31.3361 &	0.0004 &	10.47 &	5.9167 &	-4.9496 &	0.0016 \\
54555.4916 &	31.3339 &	0.0004 &	8.38 &	5.9059 &	-4.9813 &	0.0014 \\
54556.4833 &	31.3309 &	0.0005 &	8.84 &	5.9029 &	-4.9806 &	0.0014 \\
54557.4845 &	31.3346 &	0.0004 &	8.94 &	5.9054 &	-4.9860 &	0.0020 \\
54562.4792 &	31.3340 &	0.0003 &	6.69 &	5.9017 &	-4.9824 &	0.0012 \\
54566.4761 &	31.3323 &	0.0004 &	7.07 &	5.9074 &	-4.9691 &	0.0015 \\
54570.4748 &	31.3388 &	0.0004 &	7.08 &	5.9156 &	-4.9619 &	0.0019 \\
54736.8753 &	31.3343 &	0.0005 &	7.86 &	5.9209 &	-4.9531 &	0.0023 \\
54852.6937 &	31.3335 &	0.0004 &	10.00 &	5.9260 &	-4.9036 &	0.0012 \\
54854.6131 &	31.3325 &	0.0003 &	10.36 &	5.9269 &	-4.9007 &	0.0012 \\
54932.5043 &	31.3327 &	0.0003 &	11.38 &	5.9296 &	-4.8986 &	0.0011 \\
54934.4750 &	31.3372 &	0.0004 &	13.14 &	5.9346 &	-4.9021 &	0.0016 \\
54935.4878 &	31.3356 &	0.0004 &	12.64 &	5.9297 &	-4.9011 &	0.0019 \\
54937.4739 &	31.3377 &	0.0004 &	11.02 &	5.9284 &	-4.9082 &	0.0015 \\
54939.4838 &	31.3332 &	0.0004 &	8.71 &	5.9242 &	-4.9101 &	0.0014 \\
54941.5025 &	31.3339 &	0.0003 &	9.81 &	5.9219 &	-4.9161 &	0.0011 \\
54947.4620 &	31.3346 &	0.0003 &	9.84 &	5.9257 &	-4.8960 &	0.0011 \\
54950.4825 &	31.3331 &	0.0003 &	10.23 &	5.9311 &	-4.8919 &	0.0009 \\
54951.4656 &	31.3333 &	0.0004 &	8.22 &	5.9340 &	-4.8918 &	0.0017 \\
54952.4697 &	31.3322 &	0.0004 &	9.51 &	5.9334 &	-4.8847 &	0.0015 \\
54953.4638 &	31.3342 &	0.0003 &	9.79 &	5.9302 &	-4.8851 &	0.0012 \\
54954.4592 &	31.3350 &	0.0004 &	10.51 &	5.9310 &	-4.8891 &	0.0015 \\
54955.4582 &	31.3406 &	0.0003 &	11.73 &	5.9299 &	-4.8829 &	0.0011 \\
54959.5019 &	31.3385 &	0.0005 &	12.80 &	5.9358 &	-4.8900 &	0.0028 \\
55096.8913 &	31.3348 &	0.0005 &	9.39 &	5.9338 &	-4.9093 &	0.0022 \\
55098.8873 &	31.3345 &	0.0004 &	8.85 &	5.9371 &	-4.8992 &	0.0016 \\
55100.8406 &	31.3386 &	0.0004 &	7.87 &	5.9382 &	-4.8958 &	0.0016 \\
55103.8566 &	31.3343 &	0.0006 &	9.75 &	5.9480 &	-4.8992 &	0.0034 \\
55106.8475 &	31.3379 &	0.0005 &	11.68 &	5.9439 &	-4.8758 &	0.0017 \\
55112.8464 &	31.3311 &	0.0004 &	11.62 &	5.9468 &	-4.8458 &	0.0012 \\
55134.8640 &	31.3321 &	0.0004 &	10.68 &	5.9310 &	-4.8926 &	0.0013 \\
55138.7954 &	31.3367 &	0.0005 &	10.07 &	5.9345 &	-4.8840 &	0.0013 \\
55141.8085 &	31.3366 &	0.0004 &	9.35 &	5.9467 &	-4.8616 &	0.0013 \\
55166.6311 &	31.3415 &	0.0004 &	13.81 &	5.9414 &	-4.8591 &	0.0013 \\
55168.6519 &	31.3343 &	0.0003 &	12.54 &	5.9353 &	-4.8688 &	0.0012 \\
55169.6449 &	31.3364 &	0.0004 &	13.33 &	5.9368 &	-4.8741 &	0.0014 \\
55272.5594 &	31.3403 &	0.0004 &	9.38 &	5.9496 &	-4.8411 &	0.0016 \\
55279.5027 &	31.3394 &	0.0005 &	14.26 &	5.9506 &	-4.8365 &	0.0019 \\
55284.5392 &	31.3376 &	0.0003 &	14.06 &	5.9470 &	-4.8377 &	0.0010 \\
55424.8913 &	31.3379 &	0.0007 &	14.18 &	5.9630 &	-4.8315 &	0.0040 \\
55456.8925 &	31.3382 &	0.0004 &	10.24 &	5.9483 &	-4.8483 &	0.0014 \\
55484.8377 &	31.3398 &	0.0004 &	10.28 &	5.9506 &	-4.8423 &	0.0018 \\
55487.8252 &	31.3372 &	0.0004 &	11.17 &	5.9495 &	-4.8396 &	0.0016 \\
55519.6932 &	31.3340 &	0.0004 &	13.37 &	5.9433 &	-4.8700 &	0.0021 \\
55523.7622 &	31.3383 &	0.0004 &	10.47 &	5.9420 &	-4.8612 &	0.0017 \\
55538.6289 &	31.3331 &	0.0004 &	13.92 &	5.9499 &	-4.8462 &	0.0018 \\
55544.6485 &	31.3353 &	0.0004 &	12.59 &	5.9471 &	-4.8551 &	0.0014 \\
55549.7331 &	31.3382 &	0.0003 &	13.47 &	5.9455 &	-4.8503 &	0.0013 \\
55579.7549 &	31.3365 &	0.0004 &	13.77 &	5.9402 &	-4.8590 &	0.0015 \\
55581.7226 &	31.3359 &	0.0003 &	13.36 &	5.9394 &	-4.8697 &	0.0014 \\
55585.7401 &	31.3334 &	0.0004 &	10.59 &	5.9402 &	-4.8749 &	0.0020 \\
55612.5951 &	31.3349 &	0.0004 &	11.44 &	5.9449 &	-4.8722 &	0.0016 \\
55635.5731 &	31.3417 &	0.0004 &	10.88 &	5.9636 &	-4.8333 &	0.0017 \\
55637.5622 &	31.3433 &	0.0008 &	12.91 &	5.9637 &	-4.8399 &	0.0054 \\
55640.5691 &	31.3418 &	0.0006 &	13.30 &	5.9610 &	-4.8288 &	0.0025 \\
55653.5479 &	31.3355 &	0.0004 &	11.89 &	5.9388 &	-4.8779 &	0.0016 \\
55662.5239 &	31.3366 &	0.0004 &	10.17 &	5.9483 &	-4.8648 &	0.0021 \\
55835.8160 &	31.3356 &	0.0004 &	12.98 &	5.9450 &	-4.8817 &	0.0019 \\
55844.8008 &	31.3337 &	0.0005 &	6.80 &	5.9642 &	-4.8614 &	0.0024 \\
55871.7670 &	31.3379 &	0.0004 &	11.46 &	5.9603 &	-4.8608 &	0.0020 \\
55879.8273 &	31.3397 &	0.0006 &	11.94 &	5.9560 &	-4.8721 &	0.0031 \\
55894.7106 &	31.3387 &	0.0003 &	16.52 &	5.9611 &	-4.8518 &	0.0013 \\
55895.7287 &	31.3404 &	0.0003 &	15.14 &	5.9609 &	-4.8455 &	0.0011 \\
55896.7342 &	31.3382 &	0.0003 &	14.44 &	5.9606 &	-4.8447 &	0.0016 \\
55897.7444 &	31.3398 &	0.0003 &	14.17 &	5.9586 &	-4.8502 &	0.0012 \\
55902.6204 &	31.3388 &	0.0004 &	11.47 &	5.9570 &	-4.8619 &	0.0017 \\
55923.6724 &	31.3336 &	0.0004 &	9.66 &	5.9553 &	-4.8494 &	0.0015 \\
55926.6657 &	31.3364 &	0.0003 &	9.85 &	5.9574 &	-4.8454 &	0.0013 \\
55933.7134 &	31.3370 &	0.0003 &	12.31 &	5.9588 &	-4.8315 &	0.0011 \\
56026.5410 &	31.3326 &	0.0004 &	11.61 &	5.9413 &	-4.8778 &	0.0021 \\
56030.5420 &	31.3319 &	0.0008 &	11.47 &	5.9474 &	-4.8635 &	0.0059 \\
56173.8658 &	31.3336 &	0.0006 &	13.87 &	5.9454 &	-4.8772 &	0.0025 \\
56198.8785 &	31.3404 &	0.0007 &	14.42 &	5.9792 &	-4.8108 &	0.0033 \\
56224.8548 &	31.3377 &	0.0006 &	10.48 &	5.9565 &	-4.8800 &	0.0020 \\
56226.8659 &	31.3409 &	0.0006 &	12.44 &	5.9566 &	-4.8732 &	0.0028 \\
56229.8033 &	31.3355 &	0.0006 &	9.51 &	5.9576 &	-4.8636 &	0.0019 \\
56258.7458 &	31.3305 &	0.0006 &	8.99 &	5.9408 &	-4.8932 &	0.0024 \\
56271.5920 &	31.3408 &	0.0006 &	11.72 &	5.9636 &	-4.8585 &	0.0016 \\
56285.7219 &	31.3347 &	0.0005 &	11.59 &	5.9436 &	-4.8830 &	0.0015 \\
56292.7043 &	31.3380 &	0.0005 &	10.57 &	5.9505 &	-4.8821 &	0.0016 \\
56320.6434 &	31.3357 &	0.0005 &	12.04 &	5.9477 &	-4.8828 &	0.0017 \\
56322.6348 &	31.3405 &	0.0006 &	11.26 &	5.9543 &	-4.8813 &	0.0022 \\
56370.5372 &	31.3407 &	0.0007 &	12.70 &	5.9484 &	-4.8691 &	0.0037 \\
56371.5657 &	31.3358 &	0.0005 &	12.44 &	5.9489 &	-4.8820 &	0.0020 \\
56373.5140 &	31.3368 &	0.0005 &	12.61 &	5.9437 &	-4.8917 &	0.0020 \\
56399.5113 &	31.3356 &	0.0005 &	10.51 &	5.9493 &	-4.8855 &	0.0019 \\
56402.4894 &	31.3321 &	0.0005 &	10.70 &	5.9458 &	-4.8817 &	0.0019 \\
56403.4869 &	31.3333 &	0.0006 &	12.65 &	5.9493 &	-4.8810 &	0.0021 \\
56596.8245 &	31.3350 &	0.0005 &	7.58 &	5.9411 &	-4.9306 &	0.0026 \\
56602.8710 &	31.3374 &	0.0004 &	10.69 &	5.9487 &	-4.8977 &	0.0017 \\
56613.7968 &	31.3384 &	0.0006 &	11.98 &	5.9506 &	-4.8950 &	0.0020 \\
56617.7439 &	31.3349 &	0.0006 &	13.12 &	5.9516 &	-4.9089 &	0.0023 \\
56677.5277 &	31.3387 &	0.0005 &	10.01 &	5.9459 &	-4.9118 &	0.0025 \\
56698.5976 &	31.3371 &	0.0004 &	10.94 &	5.9432 &	-4.9224 &	0.0018 \\
56702.5240 &	31.3306 &	0.0006 &	5.71 &	5.9471 &	-4.9145 &	0.0034 \\
56713.6045 &	31.3402 &	0.0005 &	11.81 &	5.9542 &	-4.8828 &	0.0025 \\
56748.5000 &	31.3379 &	0.0006 &	8.02 &	5.9555 &	-4.8921 &	0.0023 \\
56753.5467 &	31.3427 &	0.0007 &	12.16 &	5.9683 &	-4.8696 &	0.0034 \\
\end{longtable}
}

\begin{figure}
\input{HD40307_activity_timeseries_withRV.pgf}
\input{HD40307_activityGLS_short.pgf}
\caption{Upper panel: HARPS time series of HD40307. For \logR, the empty circles are the data included in \M, and have $<$\logR$> = -4.99$. The filled circles are the new data presented here, with $<$\logR$>=-4.87$. { The vertical dashed line separates the active (BJD < 2'454'800) from the inactive data set.} Lower panel: corresponding GLS periodograms for periods longer than 400 days. The vertical dotted lines represent the time span of observations and twice this value. \label{fig.40307dataGLS}}
\end{figure}

The radial velocity data set is plotted in Fig.~\ref{fig.40307dataGLS} together with the time series of the \logR, the bisector velocity span, and the FWHM of the CCF. A low-frequency signal is clearly visible in all the observables. As for HD1461, it seems reasonable to assume that this long-term trend is linked to a stellar magnetic cycle. This signal is currently stronger than the reflex motion produced by the known companion at a four-day period, which illustrates the hindering effect magnetic cycles have on the detection of low-mass planets. The period of the signal is largely unconstrained, and we therefore decided to model it with a third-degree polynomial instead of a periodic function.

It is interesting to compare the evolution of the time series as the level of activity changes. Throughout this section we consider the inactive and active data sets, corresponding to the observations obtained before and after $BJD \sim 2'454'800$, respectively {(Fig.~\ref{fig.40307dataGLS})}. The inactive data set includes only one additional observing night with respect to the data presented by \M, and two less than the data analysed by \T. The inactive period has $<$\logR$> = -4.99$, and a typical dispersion of 0.014. In the active data set we find $<$\logR$> = -4.87$ and rms = 0.024 dex over 90 observations, that is, an increased level of activity and significantly larger dispersion. We expect these differences to reflect on the radial velocities. As illustrated in Fig.~\ref{fig.HD40307_jitter}, the additional noise increases from 1.0 \ms\ to 1.7 \ms\ between the inactive and active data sets and justifies the use of the varying-jitter model.

\begin{figure}
\includegraphics[width=\columnwidth]{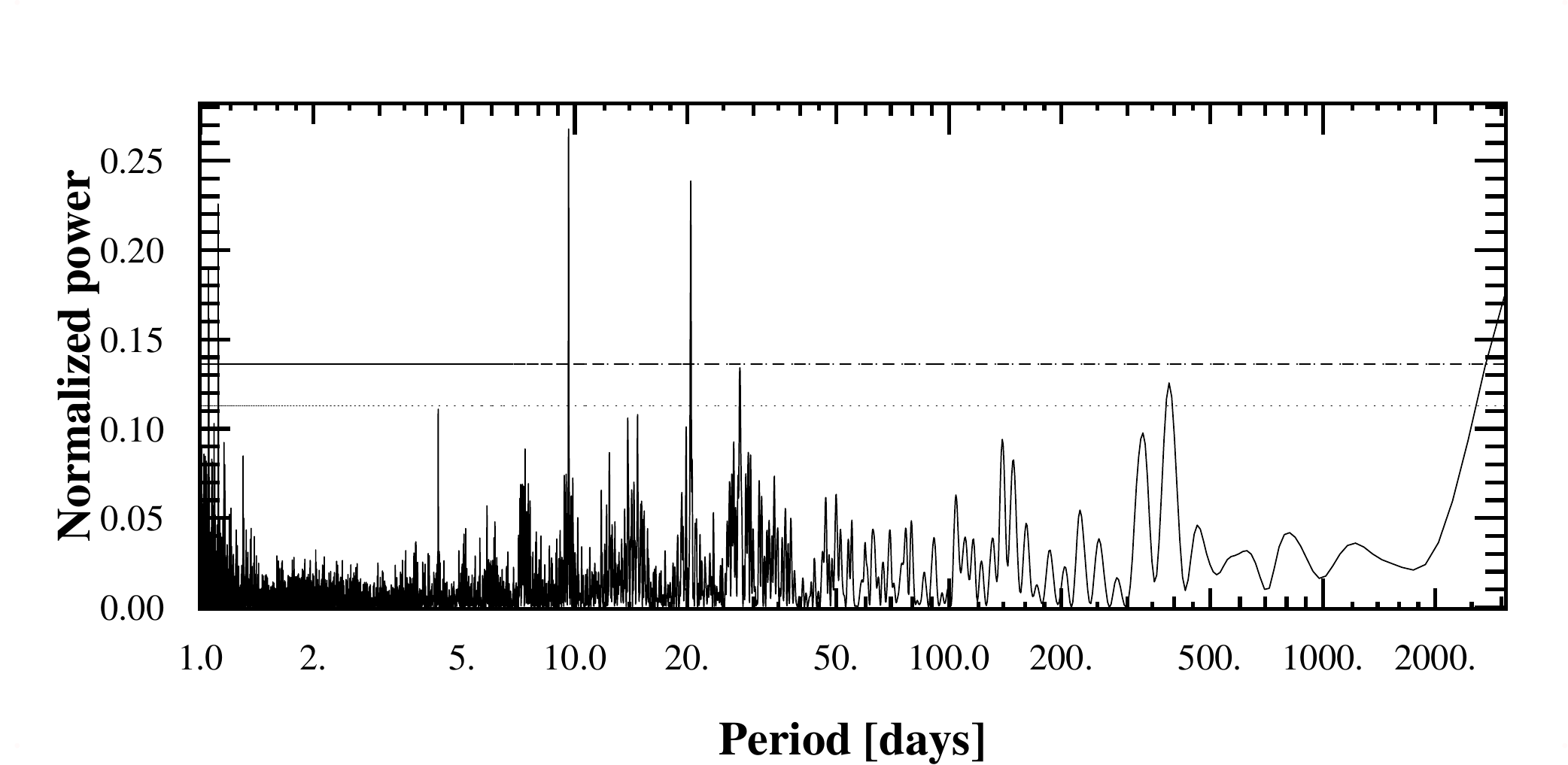}
\includegraphics[width=\columnwidth]{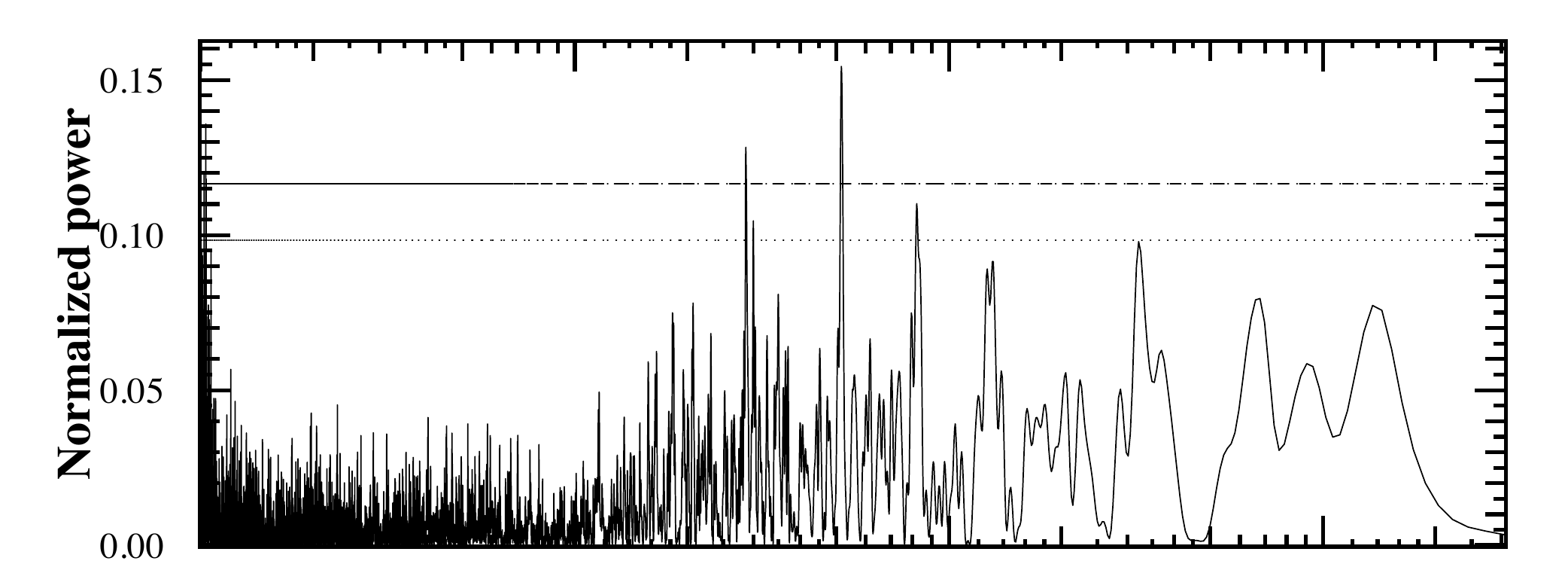}
\includegraphics[width=\columnwidth]{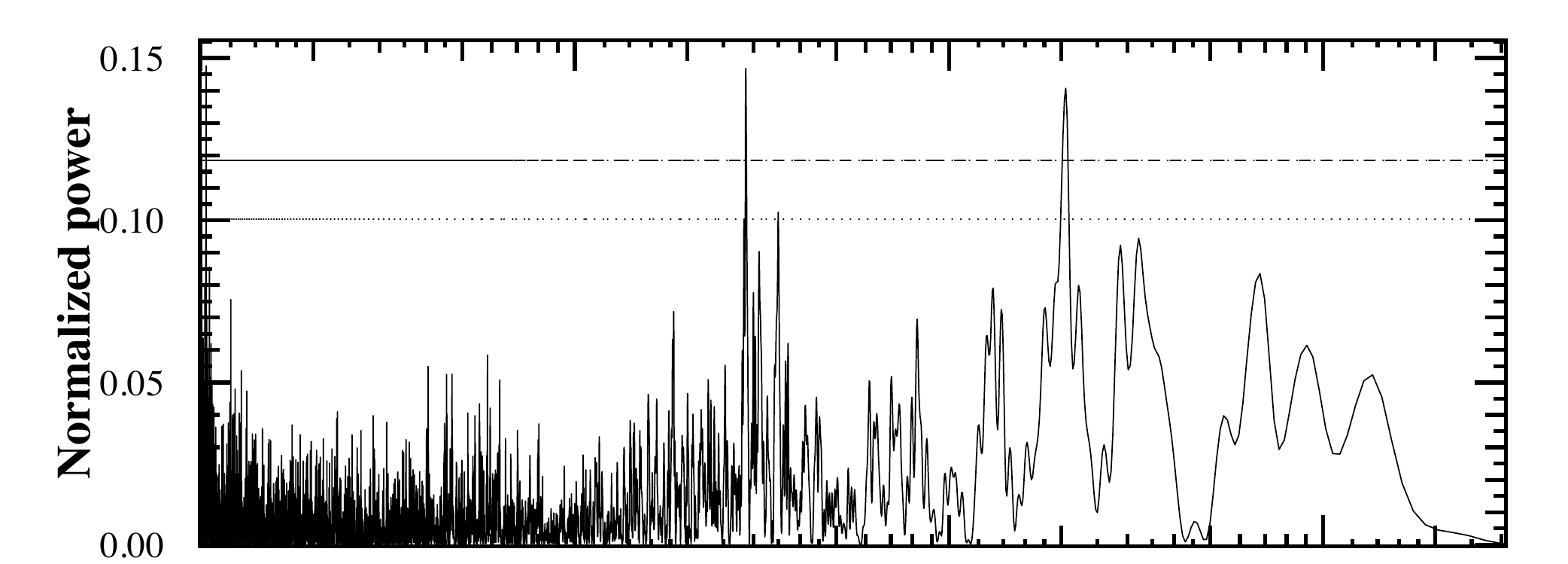}
\includegraphics[width=\columnwidth]{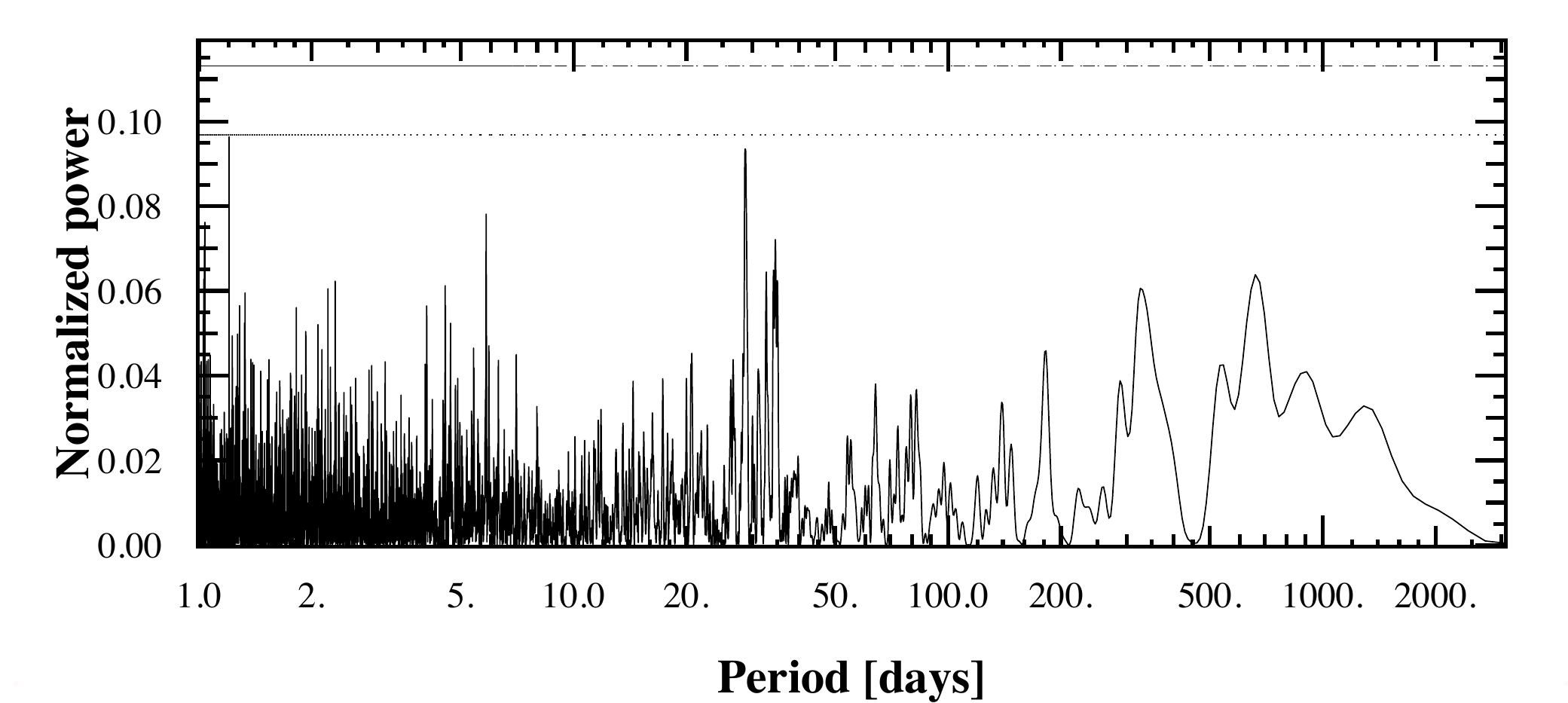}
\caption{Periodogram of the RV data of HD40307 (top panel), and of the residuals around models with three (second from top), four (second from bottom), and five (bottom) Keplerian signals in addition to a cubic function to take into account the long-term trend produced by the magnetic cycle of the star. The horizontal dotted and dashed lines represent the 10\% and 1\% false alarm probability levels, respectively.\label{fig.40307ocGLS}}
\end{figure}

In the top panel of  Fig.~\ref{fig.40307ocGLS} we present the periodogram of the HARPS RV data. The periods of the three planets reported by \M\ (4.3, 9.6, and 20.5 days) are seen as narrow spikes in the periodogram. The long-term trend is also present. The remaning panels in Fig.~\ref{fig.40307ocGLS} present the GLS of the residuals to fits with three, four, and five Keplerian signals, plus an additional third-degree polynomial to account for the activity cycle. The corresponding model evidence { estimates are listed} in Table~\ref{table.HD40307evidences} { and plotted in Fig.~\ref{fig.40307bayes}}. The methods of \CJ\ and \Perr\  agree remarkably well for models with up to four planets, with differences smaller than 1.4 in $\log \prob{D}{M_n, I}$. As expected, the TPM estimator of \citet{tuomijones2012} largely overestimates the evidences.

\begin{figure}
\includegraphics[width=\columnwidth]{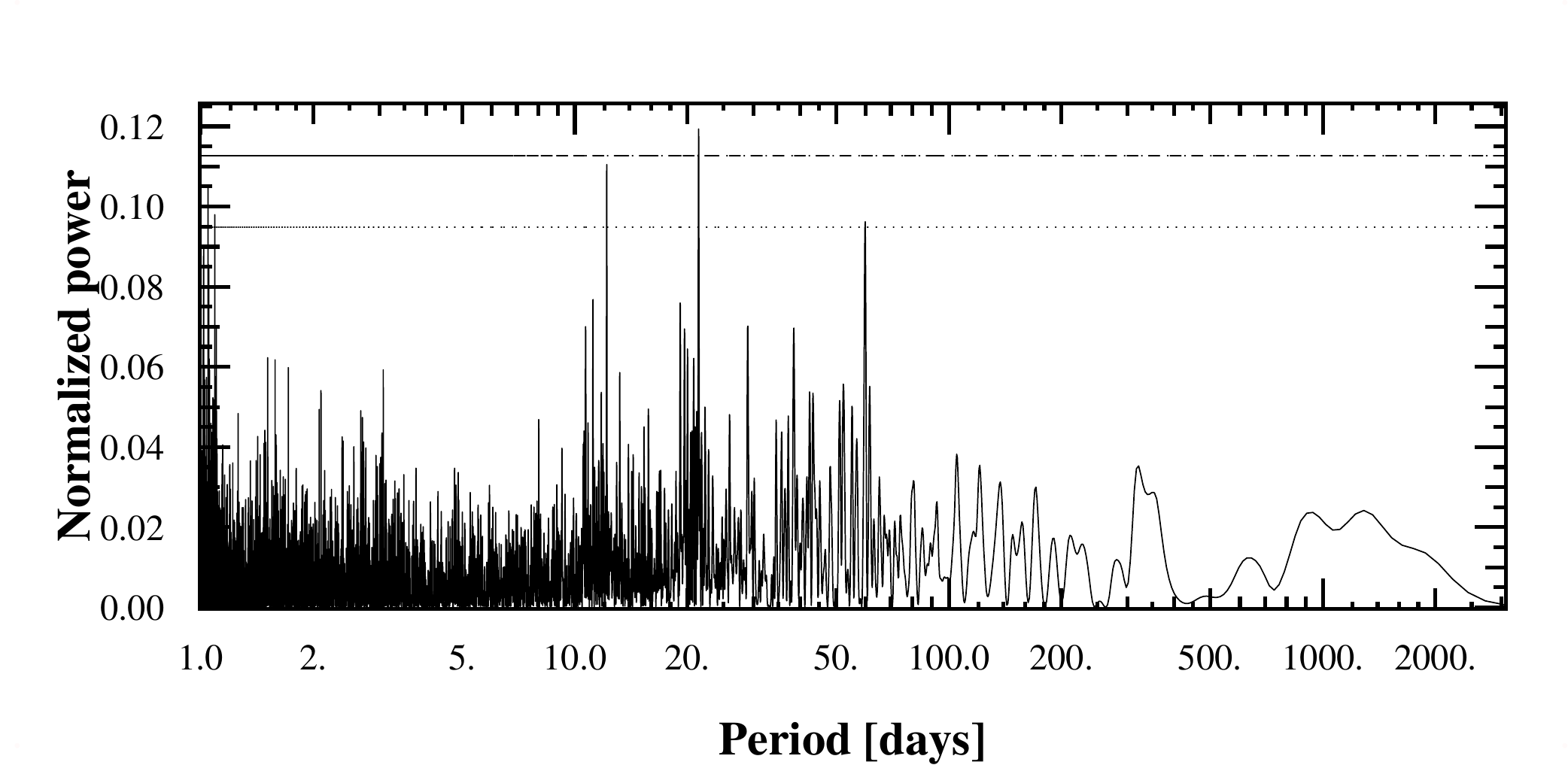}
\caption{GLS periodogram of the bisector velocity span of HD40307. A peak at the period of planet d is clearly detected.
\label{fig.HD40307_GLSbis}
}
\end{figure}

\begin{figure}
\center
\input{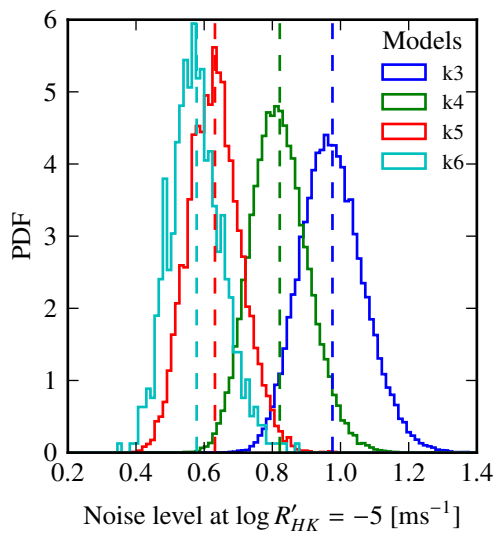}
\caption{HD40307. Posterior distribution of the amplitude the additional white noise for an inactive star \logR = -5.0. The dashed vertical lines represent the mean of each distribution.\label{fig.HD40307minjitter}}
\end{figure}

In all models with at least three planets, the system announced by \M\ is recovered, albeit with a slightly shorter period for planet d ($P\sim20.42$ d instead of $P\sim20.46$ d). This is probably due to the effect of activity at a similar period. Indeed, a significant peak appears at $P=21.4$ days in the bisector time series (Fig.~\ref{fig.HD40307_GLSbis}) when the long-term trend is corrected. However, when a least-square fit is performed on each observing season individually, the amplitude of the bisector signal is seen to anti-correlate with the one in the RVs. The bisector amplitude varies from below 50 \cms\ during the first three seasons to around 2.5 \ms\ when the activity increases. If the signal in the RV data were produced by magnetic activity, we would expect a correlation to exist between its amplitude and that of the bisector signal. The fact that an anti-correlation is seen indicates that the activity signal is scrambling the signal seen in the RV, but does not cast doubt on its interpretation as a planetary companion. Otherwise, the amplitude and eccentricity distributions of the three companions are compatible in all models. The base-level additional noise is below 1 \ms\ for all models with at least three signals (Fig.~\ref{fig.HD40307minjitter}), illustrating the high precision of HARPS. As the level of complexity of the model increases, the needed additional noise level decreases. For the five- and six-signal models, the noise level is around 60 \cms. On the other hand, the models with a weaker base-level jitter have a higher sensitivity to \logR, that is, a larger slope parameter.

\subsection{Four-Keplerian model. A super-Earth companion on a 51.6-day period orbit.}
\begin{figure}
\centering
\input{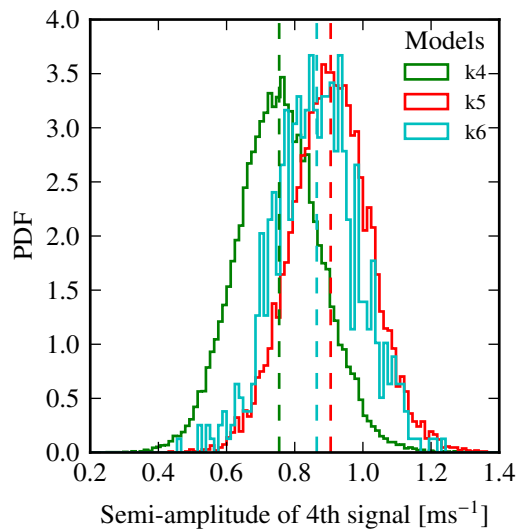}
\caption{HD40307. Posterior distribution of the amplitude of the fourth signal. The dashed vertical lines represent the mean of each distribution. The model with four Keplerians produces a slightly weaker signal than the more complex models.\label{fig.histK51days}}
\end{figure}

\begin{figure}[t]
\input{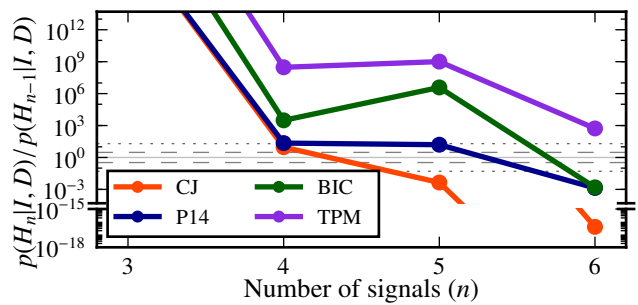}
\caption{HD40307. Odds ratio for models with $n$ Keplerian curves with respect to models with $n-1$ Keplerian curves as a function of model complexity $n$, assuming equal unity prior odds in all cases. The estimates using different techniques are shown and the customary limits for positive ($O_{n+1, n} = 3$) and strong ($O_{n+1, n} = 20$) and their inverses are shown as dashed and dotted lines, respectively. The odds ratio in favour of the model with three Keplerians is out of scale.}
\label{fig.40307bayes}
\end{figure}

A 51.6-day period signal appears as significant when the model with three Keplerian signals and a degree-three polynomial is subtracted (Fig.~\ref{fig.40307ocGLS}). Given that a long-term signal was subtracted, particular care should be given to the spectral window: as seen for HD1461, if the signal is not correctly corrected for and some power remains at very long periods, peaks will appear at the frequencies present in the spectral window function. In this case, no peak is present at frequencies corresponding to $\sim 51$ days in the window function of the HD40307 data. The 51.6-day signal has an amplitude of 75 \cms. All models with at least four Keplerians converge to a period of $P\sim51.6$ days, with an eccentricity distribution that in all cases is compatible with a circular orbit. Its amplitude, however, depends mildly on the model (Fig.~\ref{fig.histK51days}). In the model with four signals, the amplitude is around 75 \cms, while in more complex models the amplitude is closer to 85 - 90 \cms.

When this fourth Keplerian is included, the model probability increases by a factor of $9.3\pm1.7$ or $22.6\pm2.8$ using the \CJ\ and \Perr\ estimates, respectively. This corresponds to positive and strong evidence in favour of the fourth signal, according to the scale presented by \citet{kassraftery1995}. The evidence estimates based on these two techniques agree within 30\%, which is remarkable given the difficulties associated with estimating the evidence in high-dimensional spaces \citep{gregory2007}.

We note that this signal is not far from the rotational period estimated based on the \logR\ level (Table~\ref{table.stellarparams}). Indeed, the active period of the bisector velocity span exhibits a significant peak ($p$-value < 0.01) at $P=51.5$ days. However, the peak power is reduced to below the { level of the $p$-value = 0.1} when the data are detrended to account for the long-term evolution. Additionally, no equivalent peak is seen in the inactive period. Although this may cause concern at first sight, the 51.6-day signal is present in the RV data set even after the long-term effect has been removed. The fourth Keplerian signal is therefore
probably not attributable to stellar activity. The parameters are presented in Table~\ref{table.params40307}.

\subsection{Five-Keplerian model. A doubtful 205-day period signal.}

The periodogram of the residuals around the four-Keplerian model exhibits peaks at $\sim$ 200 days and  28.6 days with power above the { $p$-value=0.01 level} (Fig.~\ref{fig.40307ocGLS}). The signal at 28.6 days is most probably an alias introduced by an incomplete correction of the long-term variability. The most prominent peak of the window function after the one-year peak is at 27.8 days, probably linked to the Moon orbital period\footnote{Observations are usually scarcer during full Moon.}. Including a fifth Keplerian in the MCMC model leads to a period of $P\sim205$ d for the fifth signal, with an amplitude of 80 \cms and an eccentricity posterior distribution peaked at 0.32 but compatible with zero at the 90\%-level.

Up to this point, the estimates of the model evidence provided by the \CJ\ and \Perr\ techniques have agreed, at least qualitatively. However, they disagree strongly concerning the significance of the fifth Keplerian signal. While based on the \Perr\ estimator the five-Keplerian model is around 15.8 times more probable than the model with only four signals, the \CJ\ estimate indicates that the the four-Keplerian model is 2.6 times more probable than the five-Keplerian model.

The difference is certainly due to the approximations and limitations of each estimator and highlights the difficulty of obtaining a robust estimate of the integral in Eq.~\ref{eq.evidence} for highly dimensional models. This problem was also reported by \citet{gregory2007} in his study of the radial velocity data of HD11964. The author compared three different techniques for computing the evidence and found large differences (factors of around 20) for models with more than two planets. In our case, the estimates for models with $n = 2, 3$ differ by a factor of a few (3 - 4), and even less for the model with four planets. On the other hand, a difference of three orders of magnitudes exists for the model with five Keplerians.

As final value for the evidence of the four-Keplerian model, we adopted the geometric mean between both estimates, with an uncertainty that contains both estimates (Table~\ref{table.bayesfactors}). We did this to account for the systematic errors associated with the estimation technique. The values in Table~\ref{table.bayesfactors} indicate that it is not possible to decide between models with four and five signals with the current data. However, it can still be claimed that the model with four signals is more probable than the model with three planets (between 7.4 and 28 times).

On the other hand, the spectral window function of the analysed data exhibit peaks near this period (Fig.~\ref{fig.40307windowzoom})\footnote{There is significant power at 180 days and 240 days, the latter probably related to the duration of the observing season.}. Even if present in the RV data, the observed signal at 205 days could also be an alias introduced by an incomplete correction of the activity cycle. Moreover, with a period of around 200 days, the putative planet would be well inside the habitable zone of HD40307 (\T). For such planet it is particularly important to have strong and robust evidence for this signal and its nature. We therefore decided to remain cautious and to conservatively retain the model with four signals over the one included the 205-day variability.

\subsection{Six-Keplerian model.}

The residuals of the model with five Keplerians still show power at 28.6 days, but with a much smaller amplitude (Fig.~\ref{fig.40307ocGLS}, lower panel). No additional signal is significant according to the periodogram analysis. However, as discussed above, the periodogram can produce an incorrect estimate of the significance of a signal, and we therefore estimated the Bayesian evidence for a model with six signals.

When adopting this model for the MCMC algorithm the fifth Keplerian does not converge to a clear maximum. There is posterior mass at periods of $\sim$200 days, but also at around 28 days and 330 days (see Fig.~\ref{fig.40307ocGLS}). The sixth Keplerian also exhibits a multimodal behaviour, with power at 200 days, 34.6 days, and 20.5 days. This shows that the fifth and sixth Keplerians, if present in the data, do not have a clearly defined period in this model. We note that the signal at 34.6 days and the one at around 200 days have been announced by \T\ as planets e and g, respectively. We discuss this further in Sect.~\ref{sect.hd40307system}.

The Bayesian model comparison rules out the six-signal model as less probable than models with three or four Keplerians (Table~\ref{table.HD40307evidences}). The discrepancy between the two estimates becomes much stronger for this model than for the four-Keplerian one, but the conclusion holds even if the systematic error between the estimators is taken into account (Table~\ref{table.bayesfactors}).

\begin{table*}
{\tiny
\caption{Model probabilities for HD40307. Estimate of the evidence (marginal likelihood) for models with $n = 2, ...,  5$ Keplerians and an additional long-term drift, common to all of them. Columns are the same as in Table~\ref{table.HD1461evidences}.
\label{table.HD40307evidences}}
\begin{tabular}{c p{1.3cm} | c c c c | c c  | c c }
\hline
\hline
                        &                       &\multicolumn{4}{c|}{$\log \prob{D}{M_n, I} - 1000$}                                                       &\multicolumn{2}{c|}{$\prob{M_n}{D, I}/\prob{M_3}{D, I}$}           &$\sigma_{J_n}|_{\log R^\prime_{HK} = -5}$      &$\sigma_{O-C}^\mathrm{(inactive)}$\\
$n$                     &Periods[d]     & \CJ                   &\Perr                  & TPM                     &BIC            & \CJ                                           &\Perr                          &[ms$^{-1}$]                                                    &[ms$^{-1}$]\\
\hline
2                       &\{9.6, 20.4\}  &$1.95\pm0.09$  &$0.57\pm0.04$  &52.41                  &16.60  &$(1.93\pm0.23)\times10^{-21}$  &$(1.65\pm0.12)\times10^{-21}$&$1.70 \pm 0.15$                                       &$1.81 \pm0.04$\\
\hline
3                       &\{4.3, 9.6, 20.4\}&$49.65\pm0.10$      &$48.43\pm0.07$ &123.90                 &76.73  &$1.0$                          &$1.0$                          &$0.972 \pm 0.099$                                                              &$1.19 \pm 0.03$\\
\hline
4                       &\{4.3, 9.6, 20.4, 51.6\}&$51.88\pm0.16$        &$51.54\pm0.12$ &143.41                 &84.76  &$9.3\pm1.7$                    &$22.6\pm2.8$                   &$0.818 \pm 0.090$                                                              &$1.05 \pm 0.03$\\
\hline
5                       &\{4.3, 9.6, 20.4, 51.6, 204.9\}&$46.46^{+0.42}_{-0.72}$&$54.30^{+0.19}_{-0.28}$        &164.15 &99.91  &$0.041\pm0.026$                &$356\pm11$             &$0.630 \pm 0.079$                                                                      &$0.92 \pm 0.03$\\
\hline
6                       &\{4.3, 9.6, 20.4, 51.6, ...$^\dagger$\}&$8.7\pm7.0$            &$47.7\pm2.3$           & 170.43                  &93.44  &$1.64\times10^{-18}$   &$0.46^{+4.2}_{-0.41}$          &$0.574 \pm 0.084$                                                      &$0.85 \pm 0.04$\\
\hline
\end{tabular}
}
\tablefoot{
$\dagger$: the orbital periods of the fifth and sixth Keplerian curves are not well constrained in this model. See text for details.
}
\end{table*}

\begin{table}
\caption{Summary of the Bayesian evidence computation for each model. The geometric mean between the estimates by \CJ\ and \Perr\ is computed. The uncertainties are the quadratic mean between the statistical errors reported in Table~\ref{table.HD40307evidences} and the difference between the two estimates.
\label{table.bayesfactors}}
\begin{tabular}{c | c | c }
\hline
\hline
$n$     &$\log \prob{D}{M_n, I} - 1000$& $\log_{10}\prob{M_n}{D, I} / \prob{M_3}{D, I}$\\
\hline
2       &$1.26\pm0.70$  &$-20.75\pm0.61$\\
3       &$49.04\pm0.62$ &$0.0$\\
4       &$51.71\pm0.23$ &$1.16\pm0.29$\\
5       &$50.38\pm3.98$ &$0.6\pm1.8$\\
6       &$28.2\pm20.7$  &$-9.0\pm9.0$\\
\end{tabular}
\end{table}

\subsection{Signals in other observables. The stellar rotation period.}
The remaining observables obtained routinely from the HARPS spectra are used to validate the nature of the planetary signals. They often exhibit a high stability. In the case of HD40307, for example, the dispersion of the bisector measurements is 1.25 \ms\ over more than ten years and only 0.86 \ms\ for the inactive data (defined in Sect.~\ref{sect.HD40307}; around three years of data). This shows the exquisite precision and long-term stability of the HARPS spectrograph.

\begin{figure}
\includegraphics[width=\columnwidth]{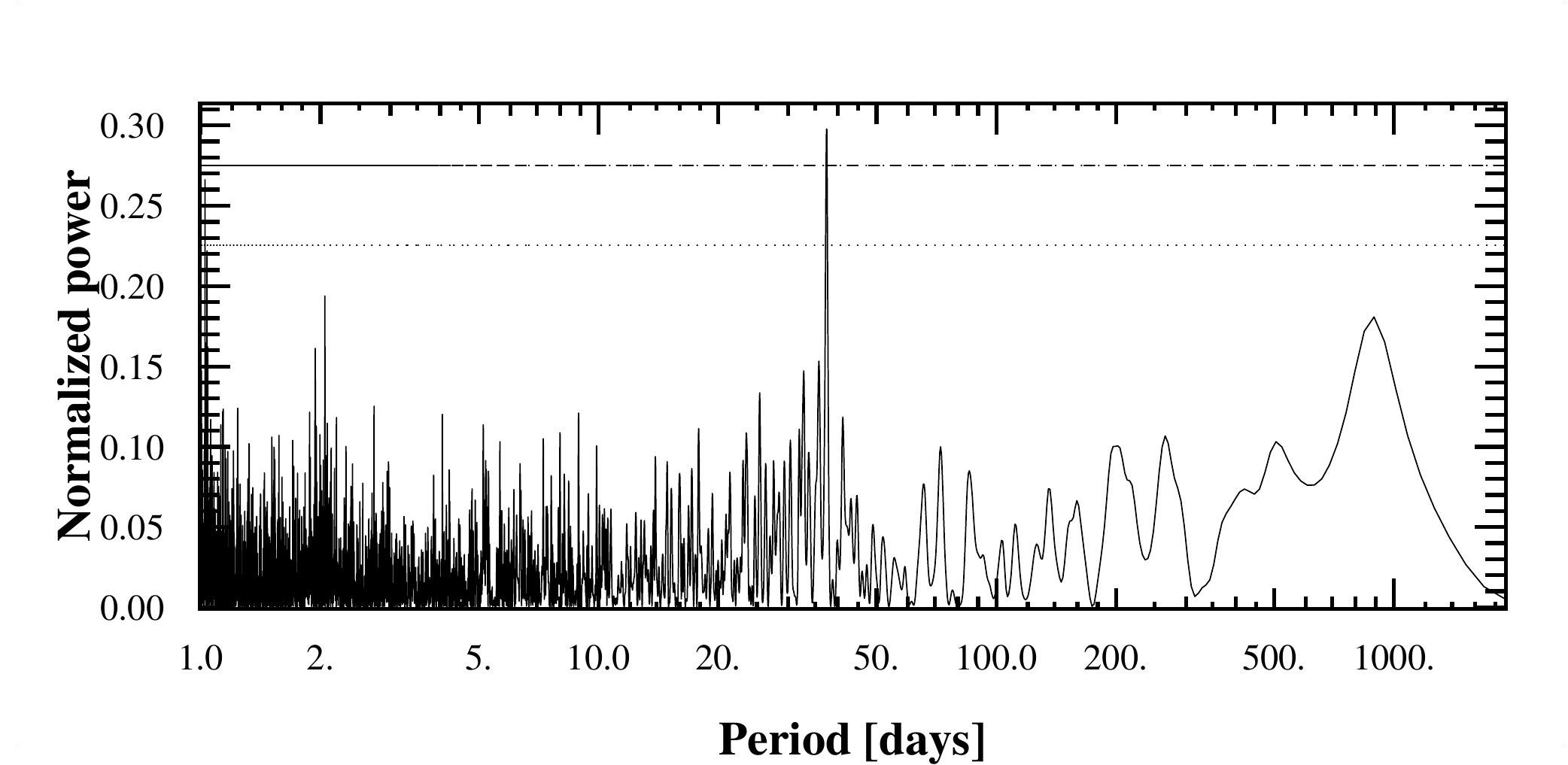}
\caption{Periodogram of the active FWHM time series of HD40307 after detrending using a third-degree polynomial. The single peak reaching the { $p$-value=0.01 level} (indicated by the dot-dashed lines) has $P=37.4$ days. We interpret it as a signature of the stellar rotation period.
\label{fig.40307fwhm_actGLS}
}
\end{figure}

A significant period of around 37.4 days is found in the active time series of the FWHM, after correcting for the long-term evolution using a second-degree polynomial fit (Fig.~\ref{fig.40307fwhm_actGLS}). A peak at the same period appears in the GLS periodogram of the active \logR\ time series, albeit with { $p$-value > 0.01}\citep[see, however,][]{tuomi2012, lovis2011}, and is also seen in the BIS time series at a slightly different period (39.1 days). We interpret this as the signature of the rotational period of the star, which is estimated to be around $P = 47.9 \pm 6.4$ days (Table~\ref{table.stellarparams}) taking the entire data set, and $P = 41.6 \pm 1.7$ days \citep{mamajekhillenbrand2008} if only the active part of the \logR\ is considered.

Other signals are present in the \logR\ time series. Considering the entire data set, a significant periodicity at 27.2 days appears, which is probably related to the first harmonic of the rotational period. The inactive time series exhibits power at 340 days and 41.2 days. Both periods have also been reported in the analysis of the S index by \T. The former is probably due to the long-term evolution of the stellar activity, while the latter is related to the rotational period as well. It is tempting to hypothesise that the difference with the period found in the inactive data set ($P=39.4$ days) is due to the migration of active regions towards lower (faster) latitudes as the cycle progresses.

Signals at periods of between 1000 - 2000 days are seen in the \logR\ and FWHM time series as well. Most, if not all of them, are probably caused by an imperfect detrending of the long-term evolution, which introduces periodicities at frequencies associated with the spectral window function.

\subsection{{ The} planetary system around HD40307 \label{sect.hd40307system}}
\begin{figure*}
\input{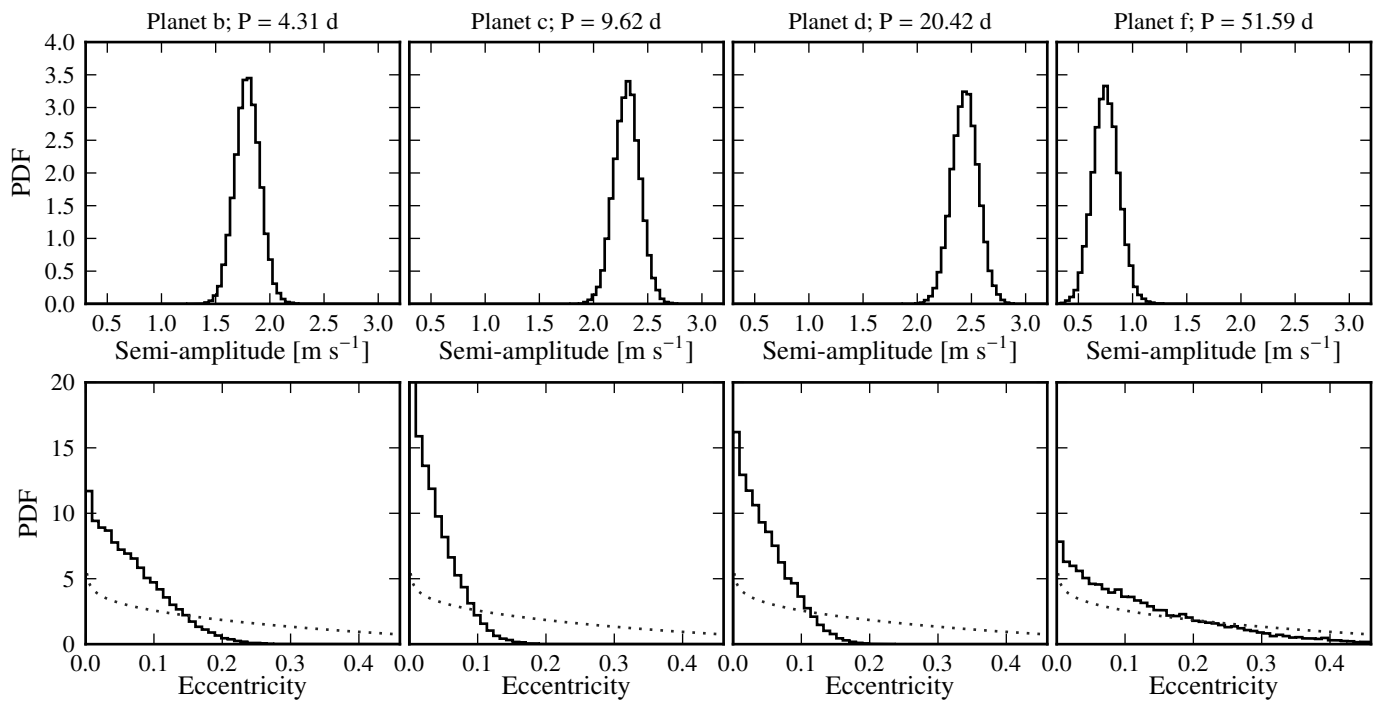}
\caption{Posterior distributions of the amplitude (top row) and eccentricity (bottom row) of the four Keplerian curves used to model the HARPS radial velocities of HD40307. The grey dotted curves represent the eccentricity prior. To facilitate comparison, the axis scales are the same for all signals. \label{fig.HD40307_posteriors}}
\end{figure*}

\begin{figure*}
\input{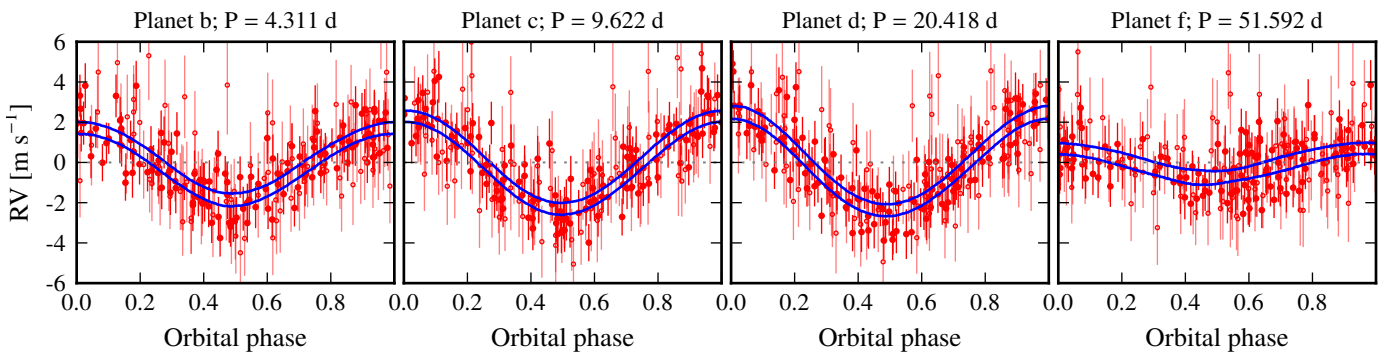}
\caption{Radial velocity data phase-folded to the best-fit period of each of the four Keplerian curves used in the final modelling of HD40307 after subtracting the effect of the remaining signals and the long-term drift. The error bars include the additional noise term (see text). The inactive data set is plotted using filled red circles, while for the active data set we chose lighter empty circles. This promptly shows that the dispersion around the model is largely caused by the active data set. The blue lines represent the 95-\% highest density interval (HDI)\label{fig.HD40307_orbits}}
\end{figure*}

HD40307 hosts four planetary candidates. We confirm the candidate announced by \T\ with a period of 51.6 days, although the amplitude is slightly smaller in our analysis, implying a decrease of the minimum mass from 5.2 \Me\ to 3.6 \Me. The posterior distributions of the model parameters are presented in Table~\ref{table.params40307}. The histogram of the MCMC sample for the signal amplitude and eccentricity are plotted in Fig.~\ref{fig.HD40307_posteriors}. All planets are on orbits that are compatible with circular. Indeed, the circular model explains the data equally well, and
because it has fewer parameters, it is about 40 times more probable than the full Keplerian solution. We have, however, retained the eccentric model to provide upper limits to the eccentricity. In Fig.~\ref{fig.HD40307_posteriors} the posterior distributions are more concentrated than the corresponding prior distributions, indicating that that the current data constrain the planet eccentricities beyond the prior level, albeit mildly for companion f. All the planets in the system have minimum masses below 10 \Me\ and can therefore be classified as super-Earths. The least massive planet of the system is planet $f$, with a mass similar to that of the innermost planet (around 3.7 \Me). In Fig.~\ref{fig.HD40307_orbits} we plot the phase-folded velocity variation for each planet candidate after subtracting the effect of the remaining ones and the long-term drift. We chose different symbols for the active and inactive data set to show the increased dispersion during the active period (see also Fig.~\ref{fig.HD40307_jitter}).

{ Dynamical integrations of the system were preformed over half a million years similar as for HD1461. The mutual inclination of the planetary orbits were randomly drawn from a uniform distribution extending between -5 and +5 degrees, which is a conservative assumption in the light of the observed distribution of mutual inclinations \citep{figueira2010}. The true masses were set to twice the minimum mass. The system is stable over the explored timescale, the eccentricities remain lower than 0.34 --for the outer planet--, and the semi-major axes do not evolve significantly: the fractional changes are around $10^{-6}$, except for the inner planet, which exhibits a fractional change of $\sim 10^{-4}$, related with the precision in the energy conservation of the system.}

\begin{figure}
\center
\input{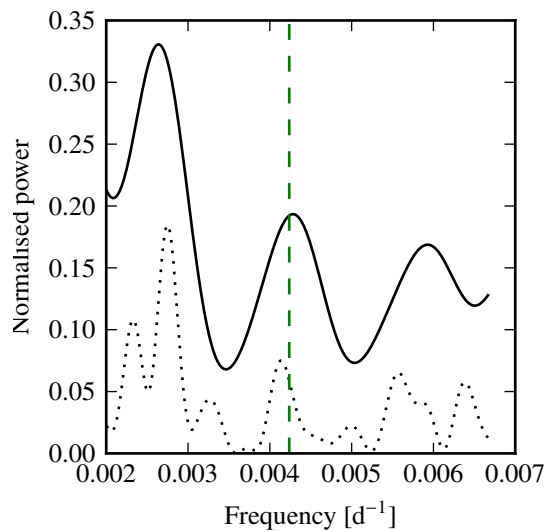}
\caption{HD40307. Window function around the one-year alias (0.00273 cycles/day) for the data set used by \T\ (solid curve), consisting of the inactive data set plus a few points taken during commissioning. The corresponding spectral window function for the data set analysed here is shown as a dotted curve. The frequency corresponding to 236 days is indicated by the vertical dashed line.\label{fig.40307windowzoom}}
\end{figure}

Two other signals were reported in this system by \T. Their companion g at $P=197.8$ days could correspond roughly to the signal detected at 205 days --although { the upper limit of their 99\% HDI is 203.5 days} -- but as discussed above, its significance and interpretation are doubtful. The periodicity of candidate e is found in the six-Keplerian model, but this model is clearly disfavoured by the data. Additionally, signals with periods $P\sim 41 - 42$ days and $P\sim37 - 39$ days were found in the \logR, bisector, and FWHM time series, indicating that these frequencies are associated with activity phenomena; they are close to the period of candidate e. More importantly, in the spectral window function of the data set employed by \T\ the $\sim 236$-day peak is much more prominent than in our data set, probably because of the longer observation time-span (Fig.~\ref{fig.40307windowzoom}). The 236-day alias of $P=41.9$ days (the rotational period present in the \emph{\textup{inactive}} \logR\ time series) is $P=35.6$ days, not far from the frequency of candidate e. Depending on the detailed structure of the sampling window, the aliased period may have more power than the real signal \citep{dawsonfabrycky2010}.

Concerning the long-term drift attributed to stellar activity, the posterior distribution of the polynomial coefficients resemble closely the priors imposed from the analysis of the \logR\ time series (Table~\ref{table.priors1461}). The conversion constant between \logR\ and RV is found to be $(22.2\pm2.2)$ \ms/dex, which is higher than the value expected for this type of star according to \citet{lovis2011b}, which is around 12.1 \ms\ per unit of \logR, but where the error in the calibration coefficients has not been considered. This could indicate the presence of long-period planets in the system. However, the dispersion around the fit by \citet{lovis2011b} is large, and it would be premature
to conclude based on this discrepancy.

\begin{table*}[t]
{\small\center
\caption{Parameter posteriors for the HD40307 system. The epoch is BJD=2,454,521.6791.\label{table.params40307}}            
\begin{tabular}{l l c c }        
\hline\hline                 
\noalign{\smallskip}

\multicolumn{2}{c}{ Orbital parameters } 		&Planet b	&Planet c\\
\hline
\noalign{\smallskip}
Orbital period, $P^{\bullet}$ 	&[days]	 	&$4.31150\pm0.00027$ &$9.62070\pm0.0012$\\
RV amplitude, $K^{\bullet}$ 	&[\ms]		&$1.79\pm0.13$ &$2.31\pm0.13$\\
Eccentricity, $e$$^{\bullet}$    	& 			&$<0.161; <0.168^\dagger$ &$<0.098; <0.103^\dagger$\\
%
Mean longitude at epoch, $L_0^{\bullet}$ &[deg] &$163.5\pm4.0$ &$321.5\pm3.2$\\
Semi-major axis of relative orbit, $a$		&[AU] &$0.0475\pm0.0011$ &$0.0812\pm0.0018$\\
Minimum mass, $M \sin i$ 			&[\Me] &$3.81\pm0.30$ &$6.43\pm0.44$\\
\noalign{\smallskip}
\multicolumn{2}{c}{ Orbital parameters } 		&Planet d &Planet f\\
\hline
\noalign{\smallskip}
Orbital period, $P^{\bullet}$ 	&[days]	 	&$20.4185\pm0.0052$ &$51.56\pm0.14$\\
RV amplitude, $K^{\bullet}$ 	&[\ms]		&$2.44\pm0.13$ &$0.75\pm0.13$\\
Eccentricity, $e$$^{\bullet}$    	& 			&$<0.117; <0.122^\dagger$ &$<0.335; <0.352^\dagger$\\
%
Mean longitude at epoch, $L_0^{\bullet}$ &[deg] &$185.4\pm3.1$ &$142\pm11$\\
Semi-major axis of relative orbit, $a$		&[AU] &$0.1340\pm0.0029$ &$0.2485\pm0.0054$\\
Minimum mass, $M \sin i$ 			&[\Me] &$8.74\pm0.58$ &$3.63\pm0.60$\\
\noalign{\smallskip}
\multicolumn{2}{c}{Velocity drift$\star$} 		\\
\hline
\noalign{\smallskip}
Systemic velocity, $V_0^{\bullet}$		&[\kms]		&\multicolumn{2}{c}{$31.334376 \pm 9.9\times10^{-5}$}\\
Scaling constant, $\alpha^{\bullet}$	&[\ms/dex]	&\multicolumn{2}{c}{$22.1\pm2.1$}\\
Linear$^{\bullet}$					&[dex yr$^{-1}$]	&\multicolumn{2}{c}{$0.03761 \pm 8.9\times10^{-4}$}\\
Quadratic$^{\bullet}$					&[dex yr$^{-2}$]	&\multicolumn{2}{c}{$0.00488\pm4.2\times10^{-4}$}\\
Cubic$^{\bullet}$					&[dex yr$^{-3}$]	&\multicolumn{2}{c}{$-0.001617\pm8.5\times10^{-4}$}\\
\noalign{\smallskip}
\multicolumn{2}{c}{Noise model$\ddag$} 		\\
\hline
\noalign{\smallskip}
Additional noise at \logR=-5, $\sigma_J|_{-5.0}^{\bullet}$	&[\cms]		&\multicolumn{2}{c}{$81.8 \pm 8.5$}\\
Slope, $\alpha_J^{\bullet}$						&[\ms/dex]	&\multicolumn{2}{c}{$8.1\pm1.6$}\\
$\mathrm{rms}(\mathrm{O-C})$				&[\ms]		&\multicolumn{2}{c}{$1.733\pm0.022$}\\
\hline
\hline
\end{tabular}
\tablefoot{

The argument of periastron $\omega$ is unconstrained for these nearly-circular orbits.

$\bullet$: MCMC jump parameter.

$\dagger$: eccentricity does not differ significantly from zero; the 95\% and 99\% upper limits are reported.

$\star$: see Sect.~\ref{sect.priors40307}.

$\ddag$: the additional (stellar) noise for measurement $i$ is $\sigma_{Ji} = \sigma_J|_{-5.0} + \alpha_J \cdot (\log{(R'_{\rm HK})}_i + 5.0)$.
}
}
\end{table*}

\section{HD204313} 
\citet{segransan2010} discovered a companion to HD204313 with $m\sin i = 4$ \MJ on a 5-yr period orbit  based on CORALIE measurements. HARPS data were obtained on this system starting in 2006, and
they permitted \citet{mayor2011} to announce the presence of an additional Neptune-mass planet on an inner orbit of 34.9 days. More recently, \citet{robertson2012} combined the \citet{segransan2010} data set with 36 RV measurements obtained at McDonald observatory that span almost eight years. They reported an additional 1.7-\MJ companion on an outer orbit with $P \sim 2800$ days. The authors stated that the combined baseline of both data sets is responsible for the detection of this outer planet in 3:2 resonance with the first detected companion. Here we revisited this system using 28 new CORALIE data points taken between August 2009 and October 2014 and 95 HARPS data points obtained between May 2006 and October 2014, together with the already published data from CORALIE and McDonald data. { Of the 104 CORALIE data points, 56 were obtained after the instrument upgrade in 2007, which mainly increased the efficiency of the spectrograph. This led to a reduced photon noise uncertainty on a given target and exposure time, but did not improve the overall precision of around 5\ms. Additionally, the hardware changes (see \citet{segransan2010} for details) introduced a zero-point offset of around 20 \ms. For this reason, the data taken before and after the upgrade were modelled as if they had been taken by different instruments\footnote{The instrument is referred to as CORALIE98 before the upgrade and CORALIE07 after the upgrade.}.} The data span 13 years and contain 215 points. The weighted mean error of the measurements is 63 \cms. Two HARPS data points (BJD = 2,453,951 and BJD = 2,456,468) were discarded because they had an unusually low signal-to-noise ratio or exhibited an anomalous CCF (high contrast), probably linked to an incorrect background correction. The data are presented in Table~\ref{table.rvHD204313}.

\onllongtab{
\begin{longtable}{r r r r r r r l}
\caption{HARPS and CORALIE measurements of HD204313. \label{table.rvHD204313}}\\
\hline\hline
\multicolumn{1}{c}{BJD} &	\multicolumn{1}{c}{RV} &	\multicolumn{1}{c}{$\sigma_\text{RV}$} &	\multicolumn{1}{c}{BIS} &	\multicolumn{1}{c}{FWHM} &	 \multicolumn{1}{c}{\logR} &	\multicolumn{1}{c}{$\sigma_\text{\logR}$} &	Instrument\\
-2 450 000  &(\kms) & (\kms) & (\ms) & (\kms) & & \\
\hline
\noalign{\smallskip}
\endfirsthead
\caption{Continued.}\\
\hline
\multicolumn{1}{c}{BJD} &	\multicolumn{1}{c}{RV} &	\multicolumn{1}{c}{$\sigma_\text{RV}$} &	\multicolumn{1}{c}{BIS} &	\multicolumn{1}{c}{FWHM} &	 \multicolumn{1}{c}{\logR} &	\multicolumn{1}{c}{$\sigma_\text{\logR}$} &	Instrument\\
-2 450 000  &(\kms) & (\kms) & (\ms) & (\kms) & & \\
\hline
\noalign{\smallskip}
\endhead
\hline
\endfoot
\hline
\hline
\endlastfoot
53861.9016 &	-9.7468 &	0.0004 &	-26.22 &	7.0415 &	-5.0043 &	0.0026 &	 HARPS \\
53862.8961 &	-9.7465 &	0.0004 &	-28.17 &	7.0409 &	-5.0026 &	0.0028 &	 HARPS \\
53863.9005 &	-9.7430 &	0.0004 &	-29.17 &	7.0377 &	-5.0020 &	0.0028 &	 HARPS \\
53864.8695 &	-9.7434 &	0.0004 &	-27.87 &	7.0383 &	-4.9913 &	0.0029 &	 HARPS \\
53865.9012 &	-9.7446 &	0.0004 &	-27.85 &	7.0390 &	-5.0062 &	0.0029 &	 HARPS \\
53866.8224 &	-9.7435 &	0.0004 &	-27.75 &	7.0379 &	-5.0094 &	0.0028 &	 HARPS \\
53867.9122 &	-9.7415 &	0.0004 &	-28.17 &	7.0381 &	-5.0037 &	0.0024 &	 HARPS \\
53868.8966 &	-9.7418 &	0.0005 &	-25.98 &	7.0398 &	-5.0071 &	0.0032 &	 HARPS \\
53869.8624 &	-9.7402 &	0.0004 &	-30.78 &	7.0381 &	-5.0143 &	0.0029 &	 HARPS \\
53870.9019 &	-9.7419 &	0.0005 &	-31.23 &	7.0395 &	-5.0070 &	0.0036 &	 HARPS \\
53871.9113 &	-9.7416 &	0.0004 &	-30.63 &	7.0377 &	-5.0115 &	0.0030 &	 HARPS \\
53882.9089 &	-9.7406 &	0.0006 &	-27.48 &	7.0414 &	-4.9909 &	0.0028 &	 HARPS \\
53883.8897 &	-9.7398 &	0.0006 &	-27.37 &	7.0422 &	-4.9839 &	0.0029 &	 HARPS \\
53887.8895 &	-9.7432 &	0.0005 &	-28.11 &	7.0416 &	-4.9878 &	0.0033 &	 HARPS \\
53889.9290 &	-9.7416 &	0.0006 &	-25.91 &	7.0486 &	-4.9887 &	0.0048 &	 HARPS \\
53920.8722 &	-9.7299 &	0.0005 &	-26.10 &	7.0402 &	-4.9937 &	0.0029 &	 HARPS \\
53945.7815 &	-9.7228 &	0.0005 &	-27.89 &	7.0447 &	-5.0153 &	0.0026 &	 HARPS \\
53975.6693 &	-9.7130 &	0.0005 &	-29.04 &	7.0397 &	-5.0042 &	0.0034 &	 HARPS \\
53979.7387 &	-9.7114 &	0.0007 &	-28.70 &	7.0421 &	-5.0168 &	0.0061 &	 HARPS \\
54054.5503 &	-9.6977 &	0.0005 &	-28.54 &	7.0395 &	-5.0305 &	0.0030 &	 HARPS \\
54732.6573 &	-9.7138 &	0.0006 &	-28.79 &	7.0445 &	-5.0178 &	0.0039 &	 HARPS \\
54738.6433 &	-9.7088 &	0.0008 &	-25.07 &	7.0559 &	-5.0090 &	0.0066 &	 HARPS \\
54743.5709 &	-9.7077 &	0.0006 &	-26.36 &	7.0542 &	-4.9863 &	0.0043 &	 HARPS \\
54749.5638 &	-9.7127 &	0.0006 &	-26.67 &	7.0500 &	-5.0117 &	0.0042 &	 HARPS \\
54753.5955 &	-9.7151 &	0.0006 &	-29.00 &	7.0461 &	-5.0222 &	0.0047 &	 HARPS \\
54774.5181 &	-9.7185 &	0.0005 &	-27.41 &	7.0497 &	-4.9892 &	0.0030 &	 HARPS \\
54776.4977 &	-9.7154 &	0.0006 &	-26.26 &	7.0475 &	-5.0104 &	0.0040 &	 HARPS \\
54777.5057 &	-9.7145 &	0.0006 &	-28.03 &	7.0434 &	-5.0028 &	0.0037 &	 HARPS \\
54778.4972 &	-9.7176 &	0.0004 &	-28.04 &	7.0445 &	-5.0107 &	0.0026 &	 HARPS \\
54780.4984 &	-9.7169 &	0.0005 &	-26.53 &	7.0450 &	-5.0210 &	0.0032 &	 HARPS \\
54952.9083 &	-9.7477 &	0.0007 &	-31.28 &	7.0471 &	-5.0290 &	0.0057 &	 HARPS \\
54953.9291 &	-9.7483 &	0.0005 &	-29.74 &	7.0472 &	-5.0305 &	0.0035 &	 HARPS \\
54954.8895 &	-9.7498 &	0.0006 &	-30.52 &	7.0456 &	-5.0195 &	0.0029 &	 HARPS \\
54955.8686 &	-9.7475 &	0.0007 &	-31.42 &	7.0496 &	-5.0243 &	0.0044 &	 HARPS \\
54956.9325 &	-9.7508 &	0.0007 &	-32.17 &	7.0512 &	-5.0241 &	0.0047 &	 HARPS \\
54989.9047 &	-9.7535 &	0.0004 &	-28.44 &	7.0397 &	-5.0122 &	0.0026 &	 HARPS \\
54992.8844 &	-9.7567 &	0.0005 &	-30.49 &	7.0425 &	-5.0130 &	0.0027 &	 HARPS \\
54995.8871 &	-9.7578 &	0.0004 &	-28.68 &	7.0480 &	-5.0201 &	0.0027 &	 HARPS \\
55001.8908 &	-9.7617 &	0.0006 &	-29.21 &	7.0449 &	-5.0182 &	0.0046 &	 HARPS \\
55020.8298 &	-9.7601 &	0.0005 &	-30.34 &	7.0441 &	-5.0207 &	0.0023 &	 HARPS \\
55024.8545 &	-9.7617 &	0.0006 &	-27.74 &	7.0396 &	-5.0251 &	0.0035 &	 HARPS \\
55040.7680 &	-9.7682 &	0.0005 &	-28.93 &	7.0518 &	-5.0042 &	0.0035 &	 HARPS \\
55047.8457 &	-9.7684 &	0.0005 &	-28.44 &	7.0449 &	-5.0185 &	0.0039 &	 HARPS \\
55071.5485 &	-9.7721 &	0.0006 &	-29.87 &	7.0458 &	-5.0168 &	0.0043 &	 HARPS \\
55074.6825 &	-9.7742 &	0.0005 &	-28.33 &	7.0482 &	-5.0126 &	0.0033 &	 HARPS \\
55076.6888 &	-9.7766 &	0.0009 &	-29.71 &	7.0517 &	-5.0307 &	0.0090 &	 HARPS \\
55097.6725 &	-9.7720 &	0.0005 &	-30.20 &	7.0475 &	-5.0132 &	0.0029 &	 HARPS \\
55100.6163 &	-9.7747 &	0.0006 &	-28.00 &	7.0474 &	-5.0229 &	0.0039 &	 HARPS \\
55105.5753 &	-9.7791 &	0.0007 &	-28.81 &	7.0503 &	-5.0277 &	0.0055 &	 HARPS \\
55106.5758 &	-9.7785 &	0.0006 &	-27.23 &	7.0446 &	-5.0245 &	0.0050 &	 HARPS \\
55110.6073 &	-9.7810 &	0.0006 &	-26.54 &	7.0417 &	-5.0216 &	0.0039 &	 HARPS \\
55112.6076 &	-9.7808 &	0.0005 &	-30.41 &	7.0472 &	-5.0230 &	0.0036 &	 HARPS \\
55117.6494 &	-9.7799 &	0.0011 &	-21.97 &	7.0514 &	-5.0667 &	0.0166 &	 HARPS \\
55123.6127 &	-9.7760 &	0.0007 &	-30.83 &	7.0500 &	-5.0300 &	0.0063 &	 HARPS \\
55137.5601 &	-9.7808 &	0.0007 &	-32.01 &	7.0432 &	-5.0422 &	0.0053 &	 HARPS \\
55138.5994 &	-9.7827 &	0.0005 &	-31.76 &	7.0454 &	-5.0287 &	0.0035 &	 HARPS \\
55140.5629 &	-9.7817 &	0.0005 &	-29.49 &	7.0465 &	-5.0311 &	0.0031 &	 HARPS \\
55152.5714 &	-9.7830 &	0.0006 &	-31.15 &	7.0481 &	-5.0252 &	0.0039 &	 HARPS \\
55160.5401 &	-9.7801 &	0.0005 &	-29.48 &	7.0469 &	-5.0327 &	0.0038 &	 HARPS \\
55161.5287 &	-9.7794 &	0.0005 &	-29.61 &	7.0500 &	-5.0178 &	0.0036 &	 HARPS \\
55372.8325 &	-9.8009 &	0.0006 &	-29.08 &	7.0477 &	-5.0327 &	0.0056 &	 HARPS \\
55374.7997 &	-9.8019 &	0.0007 &	-26.83 &	7.0485 &	-5.0438 &	0.0065 &	 HARPS \\
55375.7769 &	-9.8036 &	0.0007 &	-30.41 &	7.0507 &	-5.0500 &	0.0078 &	 HARPS \\
55397.8334 &	-9.8076 &	0.0010 &	-29.36 &	7.0568 &	-5.0395 &	0.0129 &	 HARPS \\
55403.7721 &	-9.8041 &	0.0009 &	-31.78 &	7.0495 &	-5.0424 &	0.0095 &	 HARPS \\
55409.7851 &	-9.8044 &	0.0006 &	-31.88 &	7.0559 &	-5.0215 &	0.0042 &	 HARPS \\
55779.7303 &	-9.7795 &	0.0006 &	-30.21 &	7.0539 &	-5.0382 &	0.0046 &	 HARPS \\
55803.5844 &	-9.7721 &	0.0009 &	-28.91 &	7.0582 &	-5.0467 &	0.0090 &	 HARPS \\
55809.6584 &	-9.7704 &	0.0005 &	-31.17 &	7.0556 &	-5.0415 &	0.0038 &	 HARPS \\
55816.6504 &	-9.7681 &	0.0006 &	-28.03 &	7.0466 &	-5.0480 &	0.0046 &	 HARPS \\
55834.7054 &	-9.7604 &	0.0005 &	-30.23 &	7.0531 &	-5.0444 &	0.0038 &	 HARPS \\
55841.5609 &	-9.7650 &	0.0004 &	-30.17 &	7.0544 &	-5.0427 &	0.0033 &	 HARPS \\
55873.5667 &	-9.7552 &	0.0005 &	-31.29 &	7.0531 &	-5.0655 &	0.0046 &	 HARPS \\
55888.5490 &	-9.7498 &	0.0005 &	-31.35 &	7.0509 &	-5.0610 &	0.0048 &	 HARPS \\
56057.8439 &	-9.7115 &	0.0007 &	-33.79 &	7.0527 &	-5.0445 &	0.0066 &	 HARPS \\
56079.8487 &	-9.7041 &	0.0005 &	-28.52 &	7.0524 &	-5.0382 &	0.0038 &	 HARPS \\
56093.9135 &	-9.7048 &	0.0006 &	-30.62 &	7.0537 &	-5.0221 &	0.0050 &	 HARPS \\
56118.8955 &	-9.6975 &	0.0005 &	-29.95 &	7.0604 &	-5.0344 &	0.0045 &	 HARPS \\
56151.7989 &	-9.6923 &	0.0006 &	-30.26 &	7.0571 &	-5.0414 &	0.0051 &	 HARPS \\
56167.6608 &	-9.6865 &	0.0006 &	-32.29 &	7.0589 &	-5.0342 &	0.0058 &	 HARPS \\
56235.5931 &	-9.6797 &	0.0007 &	-29.22 &	7.0543 &	-5.0479 &	0.0049 &	 HARPS \\
56437.9417 &	-9.6809 &	0.0007 &	-29.29 &	7.0541 &	-5.0314 &	0.0043 &	 HARPS \\
56455.9107 &	-9.6716 &	0.0008 &	-29.79 &	7.0544 &	-5.0403 &	0.0065 &	 HARPS \\
56477.8601 &	-9.6819 &	0.0006 &	-30.72 &	7.0609 &	-5.0454 &	0.0045 &	 HARPS \\
56501.8238 &	-9.6802 &	0.0009 &	-30.83 &	7.0603 &	-5.0655 &	0.0096 &	 HARPS \\
56618.5069 &	-9.6954 &	0.0006 &	-29.18 &	7.0575 &	-5.0387 &	0.0035 &	 HARPS \\
56858.8256 &	-9.7388 &	0.0006 &	-30.70 &	7.0612 &	-5.0368 &	0.0046 &	 HARPS \\
56863.7833 &	-9.7352 &	0.0007 &	-31.26 &	7.0617 &	-5.0412 &	0.0051 &	 HARPS \\
56871.7569 &	-9.7348 &	0.0007 &	-30.43 &	7.0596 &	-5.0308 &	0.0052 &	 HARPS \\
56929.6719 &	-9.7515 &	0.0007 &	-32.15 &	7.0600 &	-5.0452 &	0.0048 &	 HARPS \\
56930.5932 &	-9.7503 &	0.0007 &	-30.79 &	7.0611 &	-5.0358 &	0.0042 &	 HARPS \\
56944.6688 &	-9.7481 &	0.0008 &	-32.66 &	7.0643 &	-5.0303 &	0.0063 &	 HARPS \\
56948.6403 &	-9.7498 &	0.0006 &	-33.01 &	7.0622 &	-5.0324 &	0.0043 &	 HARPS \\
51790.5844 &	-9.8125 &	0.0056 &	-67.90 &	8.4548 &	-- &	-- &	 CORALIE98 \\
51793.6166 &	-9.8145 &	0.0054 &	-49.32 &	8.4612 &	-- &	-- &	 CORALIE98 \\
52076.9164 &	-9.7443 &	0.0048 &	-54.68 &	8.4534 &	-- &	-- &	 CORALIE98 \\
52078.9427 &	-9.7345 &	0.0083 &	-51.64 &	8.4521 &	-- &	-- &	 CORALIE98 \\
52079.9201 &	-9.7381 &	0.0061 &	-60.56 &	8.4373 &	-- &	-- &	 CORALIE98 \\
52131.8122 &	-9.7411 &	0.0040 &	-42.19 &	8.4530 &	-- &	-- &	 CORALIE98 \\
52135.8020 &	-9.7377 &	0.0049 &	-51.82 &	8.4632 &	-- &	-- &	 CORALIE98 \\
52136.7545 &	-9.7292 &	0.0041 &	-55.77 &	8.4515 &	-- &	-- &	 CORALIE98 \\
52144.7767 &	-9.7405 &	0.0040 &	-53.57 &	8.4473 &	-- &	-- &	 CORALIE98 \\
52167.6334 &	-9.7428 &	0.0034 &	-58.05 &	8.4445 &	-- &	-- &	 CORALIE98 \\
52177.6897 &	-9.7292 &	0.0041 &	-63.72 &	8.4308 &	-- &	-- &	 CORALIE98 \\
52183.5647 &	-9.7249 &	0.0042 &	-37.72 &	8.4372 &	-- &	-- &	 CORALIE98 \\
52197.6165 &	-9.7213 &	0.0037 &	-62.26 &	8.4381 &	-- &	-- &	 CORALIE98 \\
52438.8769 &	-9.7152 &	0.0046 &	-49.26 &	8.4307 &	-- &	-- &	 CORALIE98 \\
52443.8899 &	-9.7184 &	0.0047 &	-74.65 &	8.4197 &	-- &	-- &	 CORALIE98 \\
52444.8565 &	-9.7274 &	0.0048 &	-59.47 &	8.4241 &	-- &	-- &	 CORALIE98 \\
52446.8884 &	-9.7332 &	0.0046 &	-60.90 &	8.4300 &	-- &	-- &	 CORALIE98 \\
52460.8156 &	-9.7196 &	0.0061 &	-53.88 &	8.4161 &	-- &	-- &	 CORALIE98 \\
52465.8007 &	-9.7281 &	0.0051 &	-61.65 &	8.4292 &	-- &	-- &	 CORALIE98 \\
52543.6614 &	-9.7454 &	0.0047 &	-62.36 &	8.4653 &	-- &	-- &	 CORALIE98 \\
52547.6385 &	-9.7523 &	0.0055 &	-61.01 &	8.4428 &	-- &	-- &	 CORALIE98 \\
52562.5913 &	-9.7538 &	0.0047 &	-50.37 &	8.4441 &	-- &	-- &	 CORALIE98 \\
52565.5768 &	-9.7411 &	0.0058 &	-72.17 &	8.4618 &	-- &	-- &	 CORALIE98 \\
52802.8758 &	-9.7836 &	0.0055 &	-53.62 &	8.4547 &	-- &	-- &	 CORALIE98 \\
52853.7850 &	-9.8010 &	0.0046 &	-49.82 &	8.4868 &	-- &	-- &	 CORALIE98 \\
52898.6892 &	-9.7998 &	0.0053 &	-57.64 &	8.4607 &	-- &	-- &	 CORALIE98 \\
53166.8160 &	-9.8409 &	0.0078 &	-39.39 &	8.4227 &	-- &	-- &	 CORALIE98 \\
53167.7680 &	-9.8422 &	0.0108 &	-48.01 &	8.4656 &	-- &	-- &	 CORALIE98 \\
53215.7791 &	-9.8524 &	0.0054 &	-52.90 &	8.4544 &	-- &	-- &	 CORALIE98 \\
53237.7598 &	-9.8256 &	0.0074 &	-71.67 &	8.4273 &	-- &	-- &	 CORALIE98 \\
53262.7081 &	-9.8457 &	0.0039 &	-63.28 &	8.4368 &	-- &	-- &	 CORALIE98 \\
53287.6020 &	-9.8399 &	0.0043 &	-57.79 &	8.4312 &	-- &	-- &	 CORALIE98 \\
53291.5983 &	-9.8514 &	0.0043 &	-57.31 &	8.4292 &	-- &	-- &	 CORALIE98 \\
53587.8413 &	-9.8379 &	0.0058 &	-63.09 &	8.4598 &	-- &	-- &	 CORALIE98 \\
53592.7573 &	-9.8410 &	0.0032 &	-62.49 &	8.4525 &	-- &	-- &	 CORALIE98 \\
53596.6979 &	-9.8466 &	0.0037 &	-48.49 &	8.4525 &	-- &	-- &	 CORALIE98 \\
53599.7214 &	-9.8245 &	0.0061 &	-66.54 &	8.4380 &	-- &	-- &	 CORALIE98 \\
53615.6903 &	-9.8399 &	0.0045 &	-71.02 &	8.4292 &	-- &	-- &	 CORALIE98 \\
53621.6809 &	-9.8304 &	0.0038 &	-55.33 &	8.4552 &	-- &	-- &	 CORALIE98 \\
53630.7248 &	-9.8267 &	0.0049 &	-51.86 &	8.4575 &	-- &	-- &	 CORALIE98 \\
53644.6765 &	-9.8436 &	0.0042 &	-48.86 &	8.4520 &	-- &	-- &	 CORALIE98 \\
53666.5889 &	-9.8312 &	0.0034 &	-57.66 &	8.4578 &	-- &	-- &	 CORALIE98 \\
53703.5349 &	-9.8095 &	0.0050 &	-32.84 &	8.4440 &	-- &	-- &	 CORALIE98 \\
53862.9108 &	-9.7791 &	0.0055 &	-33.73 &	8.4476 &	-- &	-- &	 CORALIE98 \\
53904.9226 &	-9.7945 &	0.0090 &	-38.53 &	8.4419 &	-- &	-- &	 CORALIE98 \\
53915.8487 &	-9.7695 &	0.0047 &	-57.17 &	8.4684 &	-- &	-- &	 CORALIE98 \\
53993.6283 &	-9.7537 &	0.0068 &	-41.02 &	8.4789 &	-- &	-- &	 CORALIE98 \\
54007.6840 &	-9.7235 &	0.0066 &	-43.02 &	8.4808 &	-- &	-- &	 CORALIE98 \\
54279.8616 &	-9.6970 &	0.0028 &	-38.57 &	8.3104 &	-- &	-- &	 CORALIE07 \\
54284.8602 &	-9.7011 &	0.0029 &	-41.52 &	8.3050 &	-- &	-- &	 CORALIE07 \\
54289.8533 &	-9.6887 &	0.0037 &	-40.40 &	8.3004 &	-- &	-- &	 CORALIE07 \\
54290.8110 &	-9.6806 &	0.0031 &	-52.94 &	8.3160 &	-- &	-- &	 CORALIE07 \\
54291.7859 &	-9.6947 &	0.0029 &	-50.13 &	8.3126 &	-- &	-- &	 CORALIE07 \\
54292.8187 &	-9.7049 &	0.0026 &	-43.14 &	8.3106 &	-- &	-- &	 CORALIE07 \\
54295.8414 &	-9.7070 &	0.0028 &	-43.87 &	8.3108 &	-- &	-- &	 CORALIE07 \\
54296.8547 &	-9.7033 &	0.0032 &	-48.21 &	8.3028 &	-- &	-- &	 CORALIE07 \\
54300.7997 &	-9.7093 &	0.0034 &	-46.57 &	8.3274 &	-- &	-- &	 CORALIE07 \\
54301.7625 &	-9.7070 &	0.0030 &	-45.29 &	8.3035 &	-- &	-- &	 CORALIE07 \\
54325.8387 &	-9.6895 &	0.0033 &	-42.87 &	8.3079 &	-- &	-- &	 CORALIE07 \\
54348.5946 &	-9.7029 &	0.0027 &	-37.08 &	8.2967 &	-- &	-- &	 CORALIE07 \\
54351.5233 &	-9.6969 &	0.0034 &	-49.56 &	8.2948 &	-- &	-- &	 CORALIE07 \\
54389.5786 &	-9.6998 &	0.0031 &	-41.98 &	8.3038 &	-- &	-- &	 CORALIE07 \\
54398.5804 &	-9.7236 &	0.0029 &	-52.16 &	8.3255 &	-- &	-- &	 CORALIE07 \\
54412.5513 &	-9.7230 &	0.0036 &	-56.26 &	8.3100 &	-- &	-- &	 CORALIE07 \\
54593.9068 &	-9.7458 &	0.0030 &	-46.80 &	8.3514 &	-- &	-- &	 CORALIE07 \\
54615.8576 &	-9.7384 &	0.0045 &	-55.12 &	8.2960 &	-- &	-- &	 CORALIE07 \\
54662.8071 &	-9.7426 &	0.0032 &	-32.86 &	8.3025 &	-- &	-- &	 CORALIE07 \\
54687.6126 &	-9.7498 &	0.0030 &	-51.40 &	8.3058 &	-- &	-- &	 CORALIE07 \\
54688.5812 &	-9.7517 &	0.0029 &	-45.85 &	8.3120 &	-- &	-- &	 CORALIE07 \\
54704.5866 &	-9.7584 &	0.0037 &	-38.58 &	8.3092 &	-- &	-- &	 CORALIE07 \\
54729.6078 &	-9.7501 &	0.0036 &	-46.12 &	8.2689 &	-- &	-- &	 CORALIE07 \\
54733.5076 &	-9.7616 &	0.0037 &	-45.89 &	8.2958 &	-- &	-- &	 CORALIE07 \\
55074.6686 &	-9.8201 &	0.0035 &	-48.74 &	8.2983 &	-- &	-- &	 CORALIE07 \\
55100.6904 &	-9.8107 &	0.0030 &	-44.74 &	8.2953 &	-- &	-- &	 CORALIE07 \\
55128.5721 &	-9.8065 &	0.0034 &	-35.94 &	8.2816 &	-- &	-- &	 CORALIE07 \\
55706.8892 &	-9.8161 &	0.0037 &	-41.29 &	8.3049 &	-- &	-- &	 CORALIE07 \\
55762.8602 &	-9.8216 &	0.0031 &	-42.86 &	8.3132 &	-- &	-- &	 CORALIE07 \\
55807.7483 &	-9.8029 &	0.0031 &	-43.50 &	8.2919 &	-- &	-- &	 CORALIE07 \\
55843.6692 &	-9.7912 &	0.0034 &	-47.10 &	8.3080 &	-- &	-- &	 CORALIE07 \\
56119.8523 &	-9.7239 &	0.0034 &	-41.77 &	8.2839 &	-- &	-- &	 CORALIE07 \\
56126.6381 &	-9.7260 &	0.0027 &	-39.33 &	8.3140 &	-- &	-- &	 CORALIE07 \\
56254.5258 &	-9.7089 &	0.0046 &	-51.30 &	8.2924 &	-- &	-- &	 CORALIE07 \\
56489.6841 &	-9.7121 &	0.0044 &	-49.03 &	8.2822 &	-- &	-- &	 CORALIE07 \\
56522.6732 &	-9.7177 &	0.0030 &	-46.43 &	8.3216 &	-- &	-- &	 CORALIE07 \\
56523.6497 &	-9.7111 &	0.0031 &	-40.83 &	8.3031 &	-- &	-- &	 CORALIE07 \\
56528.8065 &	-9.7060 &	0.0063 &	-37.82 &	8.3206 &	-- &	-- &	 CORALIE07 \\
56529.7613 &	-9.7261 &	0.0050 &	-56.32 &	8.2991 &	-- &	-- &	 CORALIE07 \\
56581.5430 &	-9.7380 &	0.0029 &	-39.56 &	8.2920 &	-- &	-- &	 CORALIE07 \\
56582.5672 &	-9.7302 &	0.0029 &	-42.99 &	8.2891 &	-- &	-- &	 CORALIE07 \\
56596.5973 &	-9.7218 &	0.0047 &	-46.02 &	8.2844 &	-- &	-- &	 CORALIE07 \\
56772.9107 &	-9.7578 &	0.0027 &	-47.06 &	8.3429 &	-- &	-- &	 CORALIE07 \\
56782.8849 &	-9.7582 &	0.0033 &	-38.55 &	8.3070 &	-- &	-- &	 CORALIE07 \\
56799.9007 &	-9.7454 &	0.0034 &	-45.57 &	8.3298 &	-- &	-- &	 CORALIE07 \\
56832.8230 &	-9.7570 &	0.0038 &	-36.24 &	8.3170 &	-- &	-- &	 CORALIE07 \\
56854.8623 &	-9.7650 &	0.0036 &	-40.08 &	8.3138 &	-- &	-- &	 CORALIE07 \\
56886.8242 &	-9.7771 &	0.0036 &	-34.67 &	8.3113 &	-- &	-- &	 CORALIE07 \\
56916.5569 &	-9.7589 &	0.0050 &	-43.26 &	8.2864 &	-- &	-- &	 CORALIE07 \\
56931.5655 &	-9.7876 &	0.0033 &	-46.79 &	8.3111 &	-- &	-- &	 CORALIE07 \\
56936.5210 &	-9.7692 &	0.0036 &	-41.94 &	8.2972 &	-- &	-- &	 CORALIE07 \\
56953.5724 &	-9.7712 &	0.0033 &	-40.94 &	8.2820 &	-- &	-- &	 CORALIE07 \\
\end{longtable}
}

\begin{figure}
\includegraphics[width=\columnwidth]{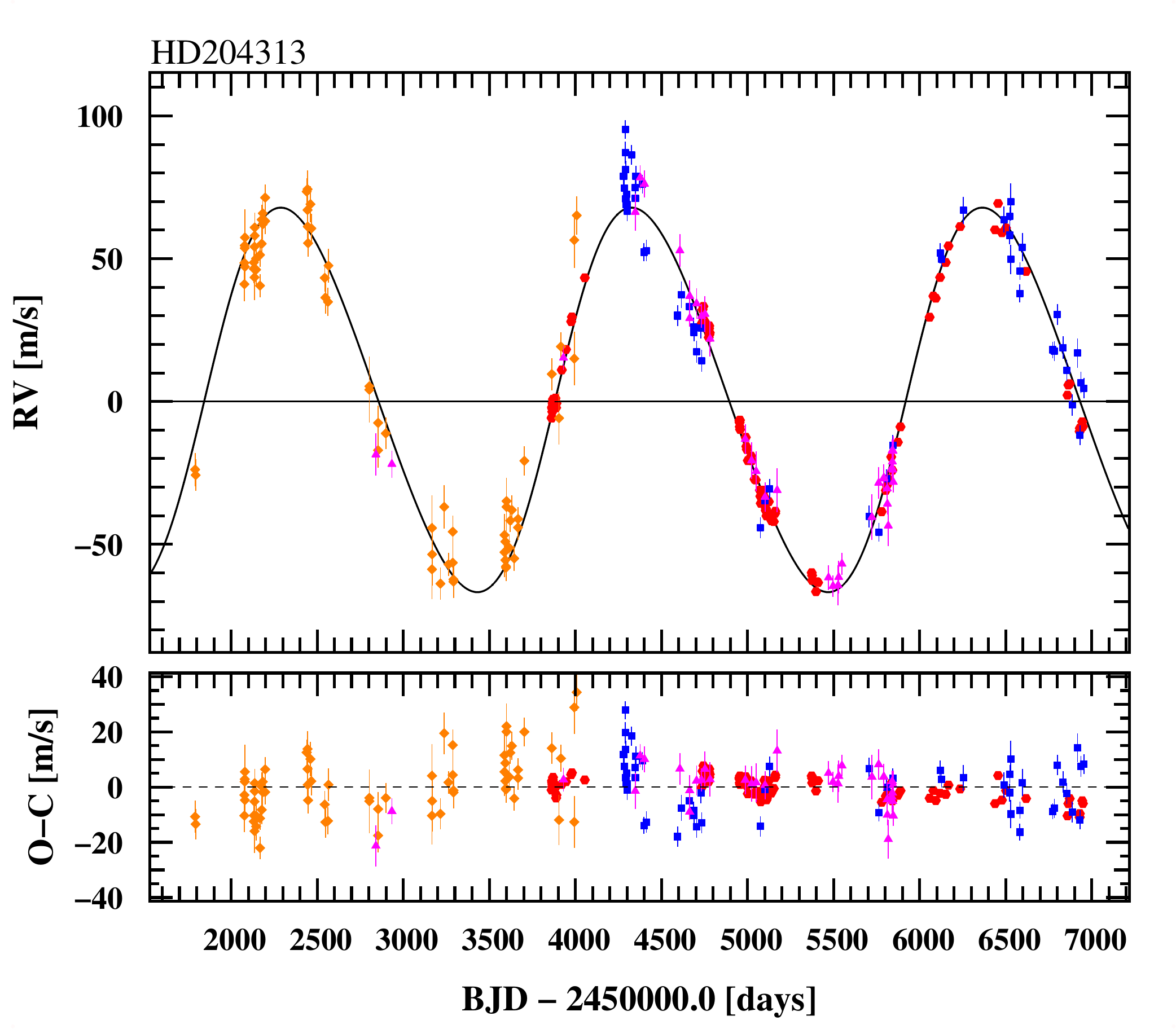}
\includegraphics[width=\columnwidth]{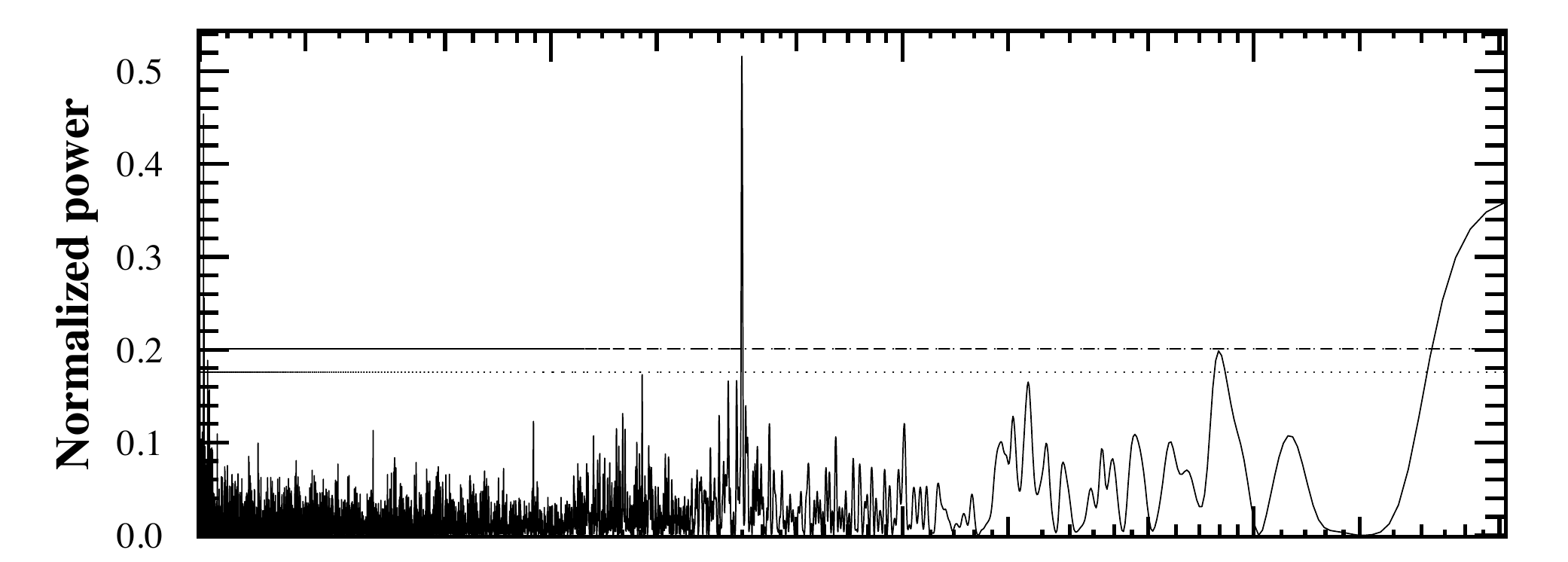}
\includegraphics[width=\columnwidth]{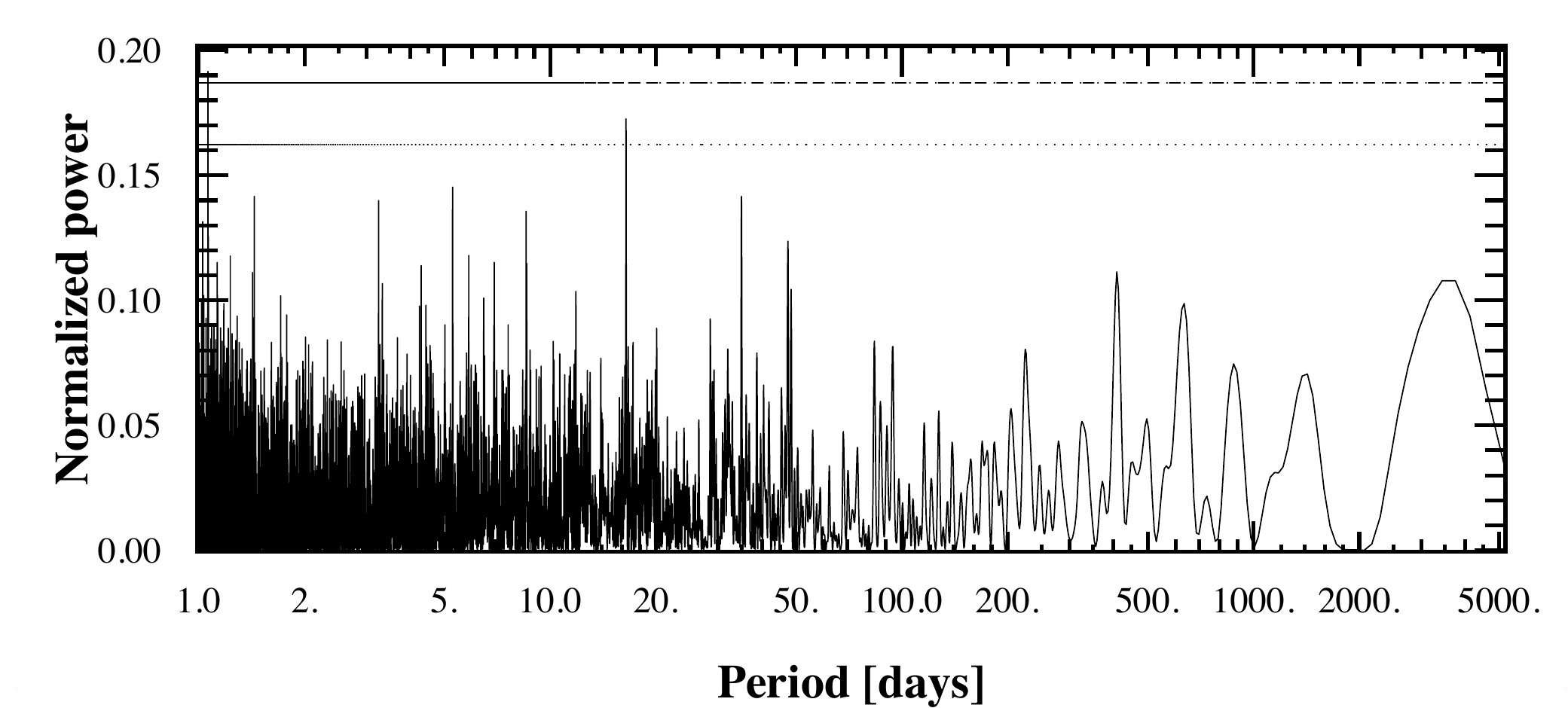}
\caption{\emph{Top:} Radial velocities of HD204313 with the corresponding one-Keplerian model. Red points are HARPS data; blue and orange points represent CORALIE07 and CORALIE98 data, respectively;  magenta points are the \citet{robertson2012} McDonald radial velocities. \emph{Middle}: Generalised Lomb-Scargle periodogram of the residuals to the model plotted in the upper panel, exhibiting excess power at $P=34.9$ days and a low-frequency trend. The horizontal dotted and dashed lines represent the 10\% and 1\% { $p$-value} levels, respectively. \emph{Bottom:} GLS periodogram of the RV residuals to the one-Keplerian model without the HARPS data. The peak at 34.9 days is no longer significant, and the long-term trend has been replaced by a definite period at $P\sim3500$ days with insignificant power. \label{fig.HD204313dataGLS}}
\end{figure}

The RV time series are plotted in Fig.~\ref{fig.HD204313dataGLS}. The variability produced by the planet first reported by \citet{segransan2010} is clearly seen by eye. When the data are blindly searched for signals using the GA, a 2050-day nearly circular orbit is found, that is, a period slightly longer than originally reported, but in agreement with the findings of \citet{robertson2012} and \citet{mayor2011}.

\begin{figure}
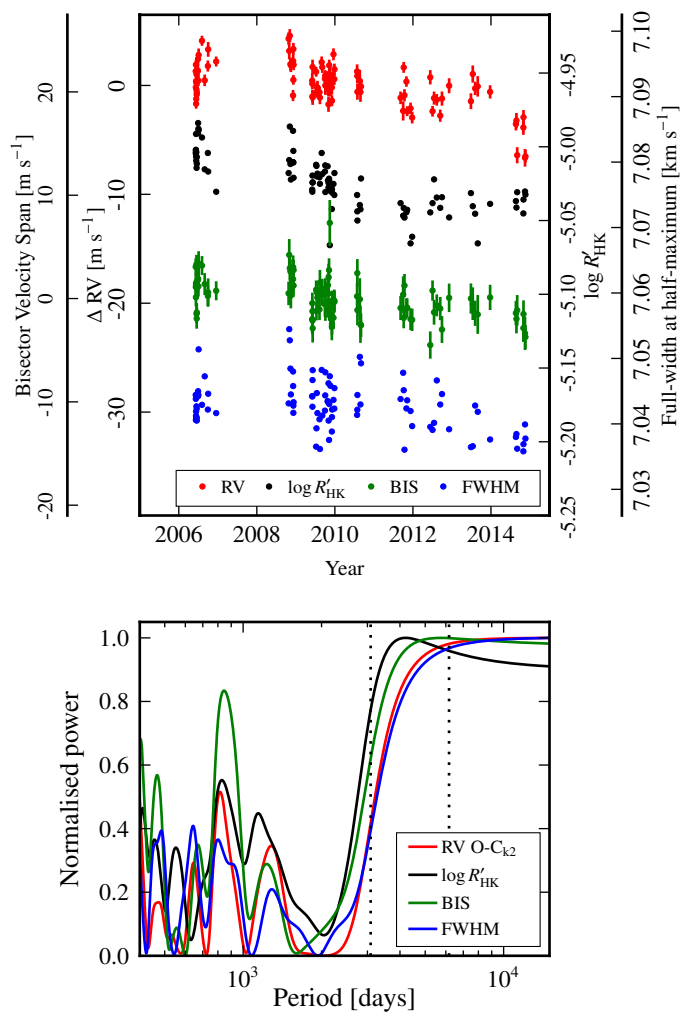

\input{HD204313_activity_timeseries_withRV.pgf}
\input{HD204313_activityGLS_short.pgf}
\caption{Same as Figs.~\ref{fig.HD1461obs} and ~\ref{fig.40307dataGLS} for HD204313, except that the RV data in the upper panel are the residuals to the two-Keplerian model and the red curve in the lower panel is the corresponding periodogram. \label{fig.HD204313activity}}
\end{figure}

\subsection{Two-Keplerian model. A Neptune-mass object on a 35-day period orbit.}
The residuals of the one-Keplerian model show a sharp significant peak at 34.9-days and an additional trend with unconstrained period (Fig.~\ref{fig.HD204313dataGLS}). The signal at 34.9 days corresponds to the Neptune-mass planet announced by \citet{mayor2011}, whose full discovery report is given here. The detection of this signal is solely due to the HARPS data. Indeed, without the 93 HARPS measurements, the GLS periodogram of the residuals to the one-Keplerian model does not exhibit any significant peak (Fig.~\ref{fig.HD204313dataGLS})\footnote{When a two-Keplerian model is fitted to the data from instruments other than HARPS, an RV amplitude significantly different from zero is found for the signal at 34.9 days. This does not mean, however, that the detection is significant. See discussion below.}. The other instruments are useful for constraining the period of the massive outer planet, but contribute only negligibly to the identification of this new planet candidate. 

\begin{figure}
\centering
\includegraphics[width=\columnwidth]{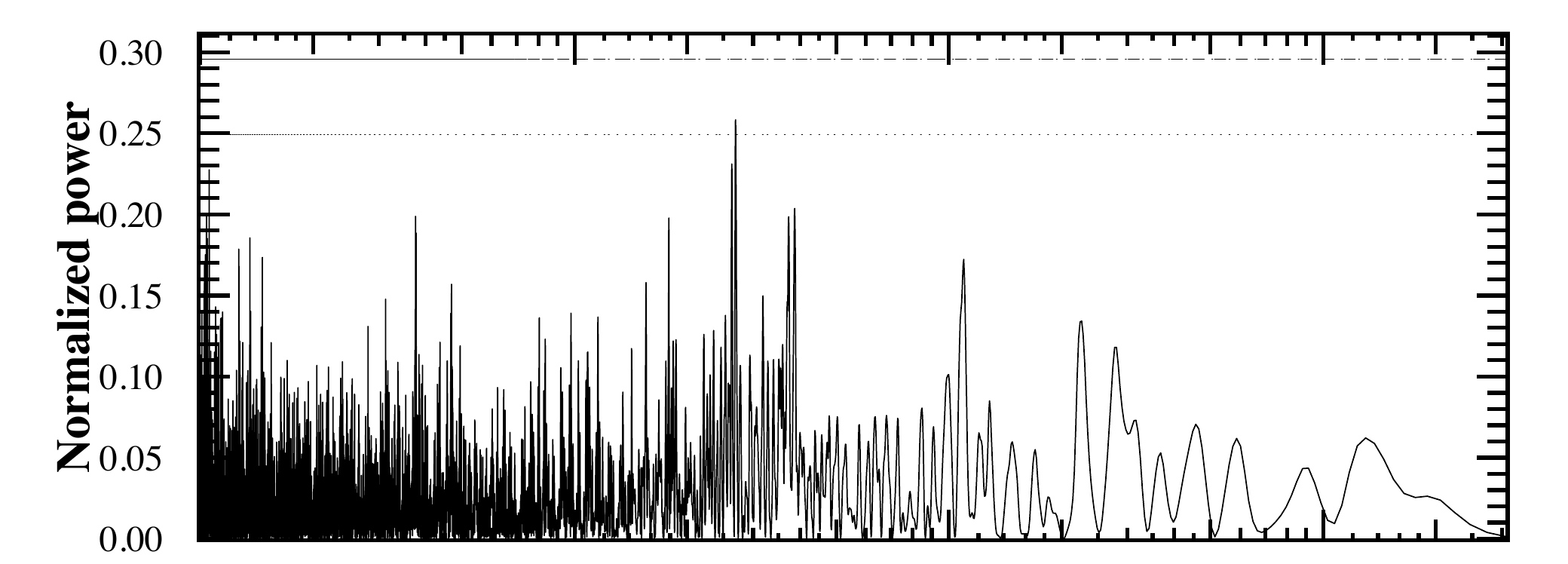}
\includegraphics[width=\columnwidth]{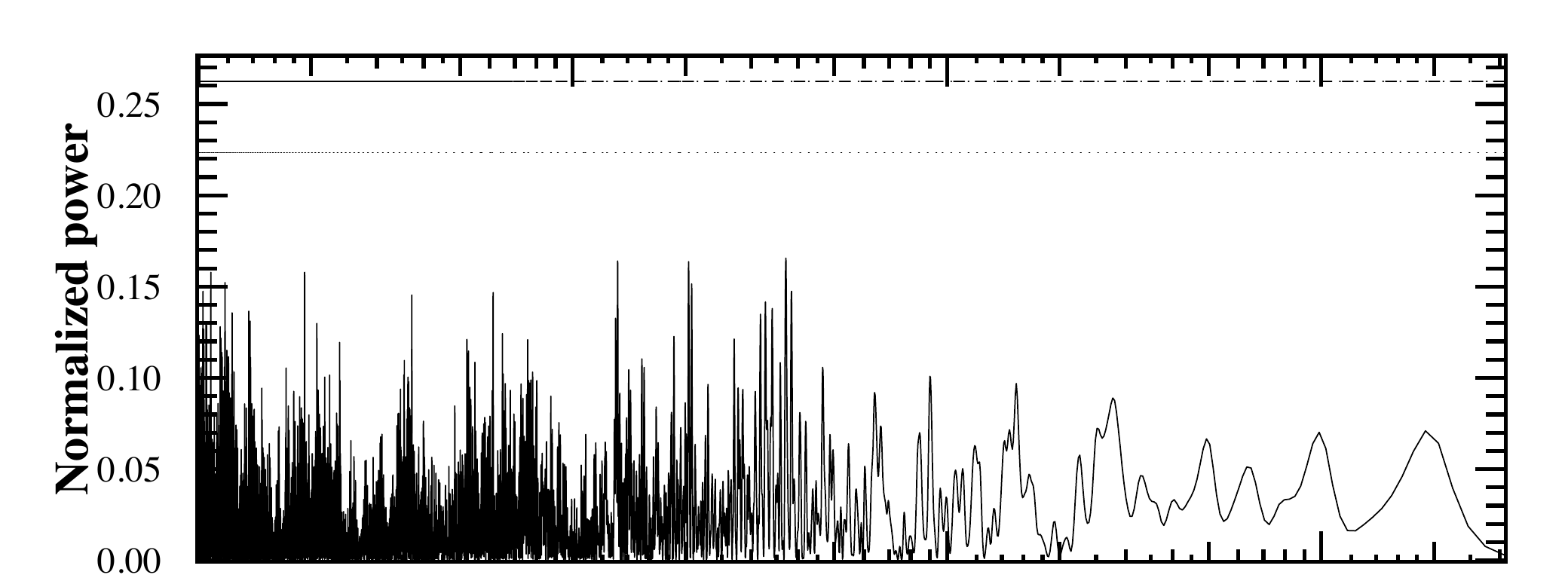}
\includegraphics[width=\columnwidth]{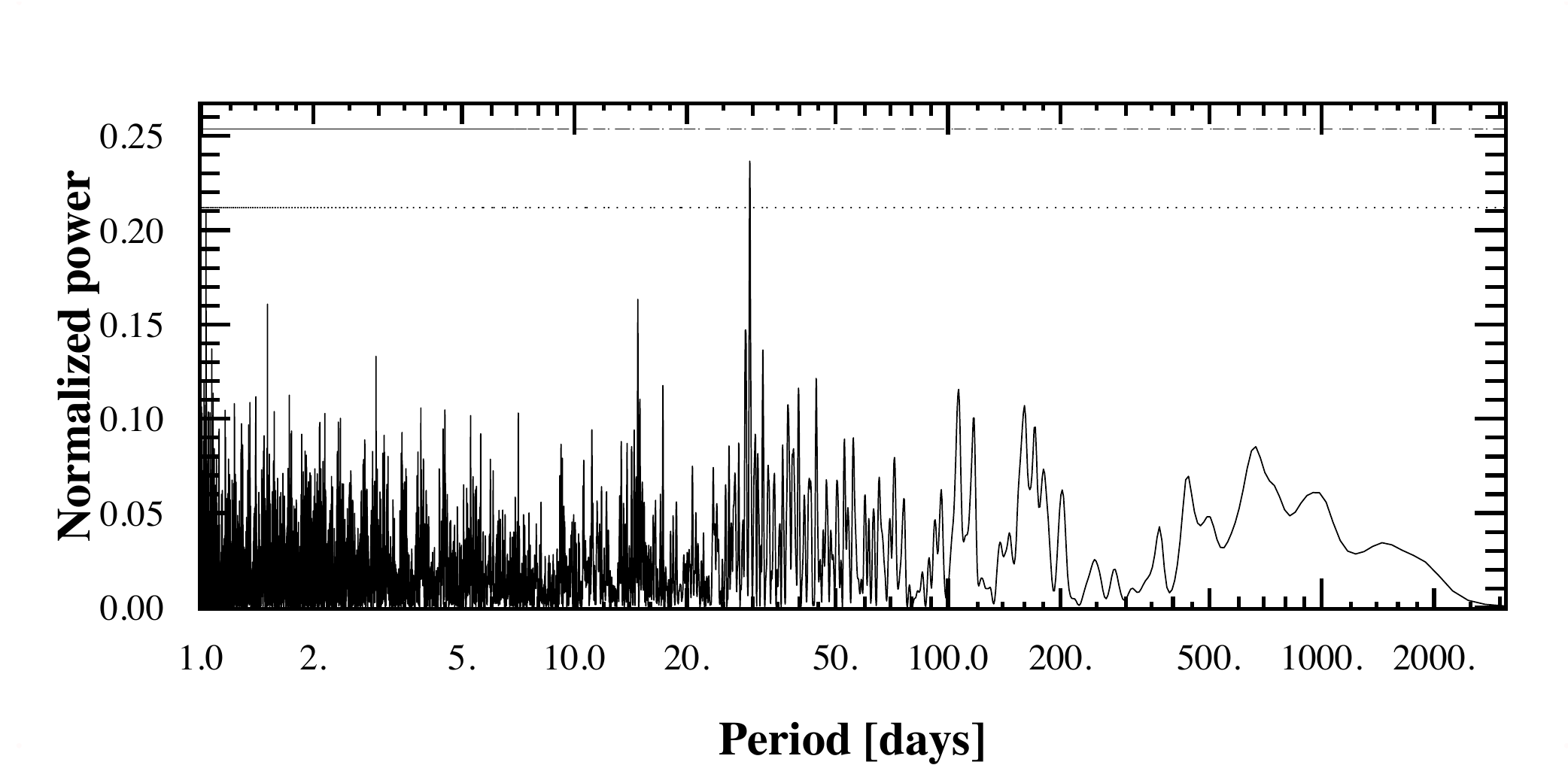}
\caption{GLS of the (from top to bottom) \logR, bisector, and FWHM time series of HD204313 obtained with HARPS after subtracting a third-degree polynomial to account for the long-term activity evolution. The horizontal lines { correspond to $p$-value = 0.1 and 0.01.} }\label{fig.HD204313_fwhmgls}
\end{figure}

The estimates of the rotational period based on \citet{noyes84} and \citet{mamajekhillenbrand2008} are around 32 days, which is close to the frequency of the signal at 34.9 days. However, the \logR\ time series does not exhibit any remaining significant peak after the long-term activity evolution is corrected for (Fig.~\ref{fig.HD204313_fwhmgls}, top panel). The same is true for the bisector velocity span (Fig.~\ref{fig.HD204313_fwhmgls}, middle panel), whose dispersion after detrending is 1.5 \ms. The GLS periodogram of the  FWHM does show a peak that could correspond to the rotational period of the star (at period $P=29.5$ days; Fig.~\ref{fig.HD204313_fwhmgls}, lower panel). However, its significance is below the 1\%-level. These facts, together with the general inactive state of the star and the relatively large amplitude of the 34.9-day signal detected in the RV lead us to conclude that it is most likely planetary in origin.
 
The long-term trend is still seen in the residuals of the two-Keplerian model. It is also detected exclusively in the HARPS data. For HD1461 and HD40307, a similar trend is also observable in the HARPS \logR, line bisector, and CCF FWHM (Fig.~\ref{fig.HD204313activity}). Once again we are led to conclude that the trend is produced by a change in the activity level of the star in the past ten years. No cyclic behaviour is detected so far, and we therefore decided, as for HD40307, to model this effect using a third-degree polynomial. We note that the RV variation in the past ten years is about 10 \ms, which is similar to the dispersion around the one-Keplerian model for the CORALIE and McDonald data. This explains why the activity trend is detected solely by HARPS. Unlike HD40307, the active and inactive periods of the target are sampled very differently, with only 28 measurements after 2011, the period that could be considered as inactive. This impedes a systematic separate analysis of the inactive and active periods.

All available data (CORALIE98, CORALIE07, McDonald, and HARPS) were employed to constrain the parameters of a two-Keplerian model with an additional third-degree polynomial, which was employed to account for the long-term effect of the magnetic activity of the star seen in the HARPS data.  As for HD40307, the priors for the coefficients of the polynomial were taken from the least-squares fit to the \logR\ time series (Table~\ref{table.priors204313}).  

As the \logR\ is not available for the other instruments, the long-term drift cannot be monitored. Instead, we decided to perform the analysis in two steps equivalent to using a model where the long-term drift is only present for HARPS data: first, we modelled data from all instruments \emph{\textup{except}} for HARPS using a two-Keplerian model, without long-term drift, and with a constant-jitter model.  We used wide, uninformative priors and sampled the parameter posterior distribution with the MCMC algorithm. The MCMC sample was used to estimate the posterior density of the system. The result from this modelling is used as the \emph{\textup{prior}} distribution for the analysis of the HARPS data, this time including the long-term third-degree polynomial and the varying-jitter model. In other words, we updated (in the Bayesian sense) the probability distributions for the parameters present in the two-Keplerian model using HARPS data, and sampled from the posterior of the new parameters associated with the long-term drift. The posterior distribution of the model without long-term trend was approximated by an uncorrelated multi-normal distribution for all parameters, except for the eccentricity of the 39.4-day candidate, which was described using a Beta distribution with parameters $a=0.93; b=5.50$. The modelled distributions are listed in Table~\ref{table.priors204313}.

The validity of the two-step procedure described above is based on two assumptions: 1) the long-term drift has a negligible effect on data from all instruments other than HARPS, and 2) the posterior distributions of the two-Keplerian model are correctly sampled and modelled. Concerning 1), the fact that no such trend is visible when the HARPS data are left out (see above) partly justifies the assumption. Concerning point 2), we rely on the excellent goodness-of-fit found and on the fact that no strong correlations are seen in the posterior distributions of the two-Keplerian model parameters.

We note that even  without the HARPS data, both the period and RV semi-amplitude of the 34.9-day signal are well constrained and the amplitude is significantly different from zero. This seems to contradict our previous statement that the signal would not be detected if HARPS data were not included. However, as discussed by \citet[][Sect. 3.9]{gregory}, it is unwise to conclude on the significance of a signal based on the posterior probability of a given parameter instead of on the computation of the Bayes factor. Indeed, the Bayes factor includes an additional Occam penalisation for the more complex model that is not accounted for otherwise. In this case, when HARPS data are left out, the posterior probability for the one-Keplerian model is orders of magnitude higher than the corresponding two-Keplerian model probability, confirming that the 34.9-day signal is not detected without the HARPS data.

\begin{table*}[t]
{\small\center
\caption{Parameter posteriors for the HD204313 system. The epoch is BJD=2,454,993.8485.\label{table.params204313}}  

\begin{tabular}{l l c c }        
\hline\hline                 
\noalign{\smallskip}

\multicolumn{2}{c}{ Orbital parameters } 		&Planet c	&Planet b\\
\hline
\noalign{\smallskip}
Orbital period, $P^{\bullet}$ 	&[days]	 	&$34.905\pm0.012$ &$2024.1\pm3.1$\\
RV amplitude, $K^{\bullet}$ 	&[\ms]		&$3.42\pm0.22$	&$68.45\pm0.30$\\
Eccentricity, $e$$^{\bullet}$    	& 			&$0.155\pm0.071; <0.310^\dagger$ & $0.0946\pm0.0032$\\
Argument of periastron, $\omega^{\bullet}$&[deg]&$238\pm22$ 	&$292.5\pm2.7$\\
Mean longitude at epoch, $L_0^{\bullet}$ &[deg] &$46.5\pm3.5$	&$107.55\pm0.24$\\
Semi-major axis of relative orbit, $a$		&[AU] &$0.2099\pm0.0071$&$3.167\pm0.12$\\
Minimum mass, $M \sin i$ 			&[\Me] &$17.6\pm1.7$	&$1360\pm94$ \\
Minimum mass, $M \sin i$ 			&[\MJ] &$0.0553\pm0.0053$&$4.28\pm0.30$\\	
\noalign{\smallskip}
\multicolumn{2}{c}{Velocity drift$\star$} 		\\
\hline
\noalign{\smallskip}
Systemic velocity, $V_0^{\bullet}$		&[\kms]		&\multicolumn{2}{c}{$-9.73922\pm3.3\times10^{-4}$}\\
Scaling constant, $\alpha^{\bullet}$	&[\ms/dex]	&\multicolumn{2}{c}{$140\pm35$}\\
Linear$^{\bullet}$					&[dex yr$^{-1}$]	&\multicolumn{2}{c}{$-0.0100 \pm 0.0013$}\\
Quadratic$^{\bullet}$					&[dex yr$^{-2}$]	&\multicolumn{2}{c}{$(-8.1\pm2.4)\times10^{-4}$}\\
Cubic$^{\bullet}$					&[dex yr$^{-3}$]	&\multicolumn{2}{c}{$(2.97\pm0.68)\times10^{-4}$}\\
\noalign{\smallskip}
\multicolumn{2}{c}{Noise model$\ddag$} 		\\
\hline
\noalign{\smallskip}
Additional noise at \logR=-5, $\sigma_J|_{-5.0}^{\bullet}$	&[\ms]		&\multicolumn{2}{c}{$1.28 \pm 0.21$}\\
Slope, $\alpha_J^{\bullet}$						&[\ms/dex]	&\multicolumn{2}{c}{$9.6\pm6.3$}\\
$\mathrm{rms}(\mathrm{O-C})$				&[\ms]		&\multicolumn{2}{c}{$1.32\pm0.11$}\\
\noalign{\smallskip}
\multicolumn{2}{c}{Other parameters$\diamond$} 		\\
\hline
\noalign{\smallskip}
CORALIE98 systemic velocity	&[\kms]&\multicolumn{2}{c}{$9.7888 \pm 0.0016$}\\
CORALIE07 systemic velocity	&[\kms]&\multicolumn{2}{c}{$9.7754  \pm 0.0015$}\\
McDonald systemic velocity	&[\kms]&\multicolumn{2}{c}{$9.6878 \pm 0.0011$}\\
\hline
\hline
\end{tabular}
\tablefoot{

$\bullet$: MCMC jump parameter.

$\dagger$: eccentricity does not differ significantly from zero at the 99\%-level. Upper limit is reported.

$\star$: see Sect.~\ref{sect.priors40307}.

$\ddag$: the additional (stellar) noise for measurement $i$ is $\sigma_{Ji} = \sigma_J|_{-5.0} + \alpha_J \cdot (\log{(R'_{\rm HK})}_i + 5.0)$.

$\diamond$: parameters from the preliminary analysis without HARPS data (see text for details), listed here for convenience.}
}
\end{table*}

\begin{figure}
\centering
\input{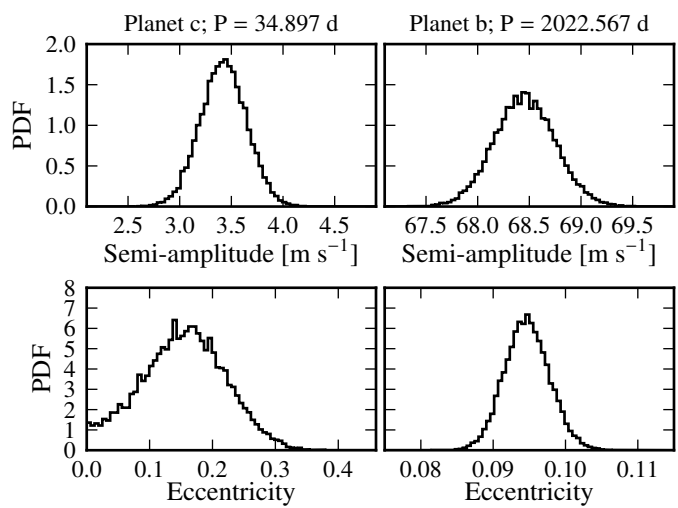}
\caption{Posterior distributions of the amplitude (top row) and eccentricity (bottom row) of the two Keplerian curves used to model the HARPS radial velocities of HD204313. Because the eccentricity distribution of planet b is much more concentrated that the corresponding distribution of planet c, the vertical scale was set to 20 times larger than the scale shown for planet c. \label{fig.HD204313_posteriors}}
\end{figure}

\begin{figure}
\centering
\input{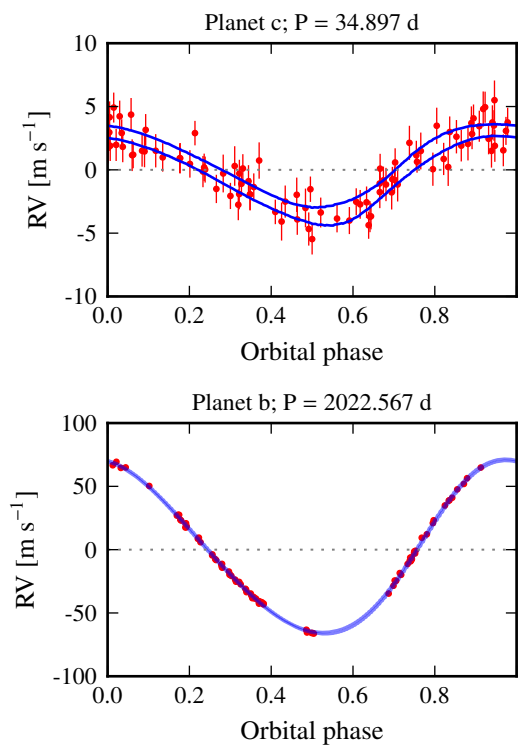}
\caption{Radial velocity data phase-folded to the best-fit period of each of the two planetary candidates in the HD204313 system after subtracting the effect of the other signal and the long-term drift. The blue lines represent the 95-\% highest density interval (HDI).\label{fig.HD204313_orbits}}
\end{figure}

The mean and 68.3\% credible intervals of the model parameters are listed in Table~\ref{table.params204313}. The posterior distributions of the amplitudes and eccentricities are presented in Fig.~\ref{fig.HD204313_posteriors} and the phase-folded orbits and the model confidence intervals are shown in Fig.~\ref{fig.HD204313_orbits}. The companion on a 34.9-day orbit is a Neptune-like candidate and lies in a nearly circular orbit ($e < 0.3$ at the 99\% level). The parameters agree with those reported previously \citep{mayor2011}.

{ The large separation between the two planetary candidates leaves little doubt as to the long-term stability of the system. We nevertheless performed numerical integrations using Mercury \citep{chambers99} for over $7\times10^{5}$ years. We proceeded as for HD1461 and HD40307. No secular evolution of the orbital parameters was observed, except for planet $c$, whose semi-major axis exhibits a fractional change of 19 parts per million, which
is of the same order as the energy conservation during the integration.}

\subsection{Search for the additional signals.}
Our analysis of the combined CORALIE, HARPS, and McDonald data does not detect the long-period companion announced by \citet{robertson2012}. Indeed, the long-term trend seen in the residuals of the one-Keplerian model is not constrained from above, unlike the peak seen in the data analysed by \citet{robertson2012} (see their Fig.~2, which shows a peak slowly decreasing for periods longer than around 10000 days). If we leave out the HARPS data set, which is the only one exhibiting the long-term drift, as mentioned above, a peak appears at a period of around 3500 days, but with insignificant power.    

The GLS of the residuals of a model with two Keplerian and a long-term drift exhibit a peak at 4.7 days with an amplitude of 70 \cms. Although the peak is well below the 0.01 level, we decided to perform a Bayesian comparison between a model with and without a Keplerian at this period. The two estimators of the evidence agree that the proposed signal is not significantly detected: the Bayes factor in favour of the simpler model is $108\pm37$ using the \Perr\ estimator, and $2180\pm190$ for the \CJ\ technique. Once again, we see a large difference between the two estimators, and as for HD40307, the \CJ\ estimate seems to punish more complex models more severely. We therefore conclude that the only two significant signals in HD204313 are those at 2020 and at 34.9 days.

\section{Summary and discussion \label{sect.discussion}}
We analysed the entire data set produced by the HARPS southern extrasolar planet search programme on HD1461, HD40307, and HD204313, three systems that each contain at least one known planet \citep{rivera2010, mayor2009, tuomi2013b, segransan2010, robertson2012}. For HD204313, we also employed CORALIE and McDonald observations to constrain the mean motion of a long-period companion. The data sets span { more than} ten years and reveal long-term variability associated with changes in the mean activity level of the star on timescales of around a decade.

The model employed contains a term that represents the effect of the evolving activity level of the target stars, as well as a statistical term to account for the short-term activity effect (the so-called jitter of the star), which is much harder to describe using a deterministic model. The long-term activity signal has been shown to correlate with the evolution of the activity proxy \logR\ \citep{lovis2011b}, and we therefore used the HARPS measurement of this activity proxy in time to provide priors for the long-term activity effect. Some assumptions are key to our model: 1) the RV signal produced by the long-term activity variability scales linearly with \logR, and 2) the stellar jitter is adequately described as an additional Gaussian noise with an amplitude that scales linearly with \logR. We have tested an alternative model with a fixed amplitude for the additional noise, but the data mildly preferred the varying-jitter model.

{ Potential periodicities were identified using a generalised Lomb-Scargle periodogram and by resorting to a genetic algorithm. After we chose the candidate signals,} we determined { their significance and the total} number of Keplerian signals present in each data set using the full machinery of Bayesian model comparison. Estimating the marginal likelihood is problematic \citep[see, for example][]{gregory2007}. Using different techniques to obtain this estimate \citep[e.g.][]{chibjeliazkov2001, perrakis2014} has permitted us to compare them, and to identify the cases needing more data or a more sophisticated model. The Bayesian model comparison is supposed to perform better than the periodogram analysis that
is usually employed by the HARPS team to detect weak signals. { One of the} reasons { involves the uncertainties of the subtracted signals and} is discussed, for example, by \citet{lovis2011} and \citet{tuomi2012}. { A more general reason is discussed, for example, by \citet{sellke2001} and involves the interpretation of the $p$-value obtained for the periodogram peak power as a false-alarm probability. The periodogram is still an essential tool for identifing periodicities in time series, but fails at providing robust estimates of the significance of each signal. In this sense, Bayesian model comparison constitutes a well-established method for computing the probability of all involved models, albeit not without technical difficulties, such as estimating the Bayesian evidence.} The periodogram analysis and the Bayesian model comparison are expected to produce equally good results for  detections with a strong and high signal-to-noise ratio, { provided that the threshold for the peak power is chosen correctly}. A detailed study of the limits of each method is needed to fully understand the techniques used to mine signals in radial velocity data. In the cases analysed here, all signals supported by the Bayesian approach would have also been obtained with the periodogram analysis using the 1\% p-value criterion.

After we established the number of signals, we resorted to other observables obtained from the HARPS spectra (CCF asymmetry and width, and an activity proxy) to conclude on the nature of the RV variations. In general, if the frequency of a RV variation is seen in any of the other observables, it casts doubt on its planetary origin. 

Finally, we conclude the following.
\begin{itemize}
\item HD1461 hosts two super-Earth planet candidates, with orbital periods $P=5.77$ days and $P=13.5$ days. { The 13.5-day signal detected in the HD1461 system is close to the first harmonic of the rotational period, which could mean that the signal is produced by activity \citep{boisse2011}. We verified that the signal is coherent over a timescale of  several years and that no sign of variability is seen at this frequency in the ancillary observables or in the activity indicator \logR.} The minimum masses are $M\sin i=6.4$ \Me\ and $M\sin i=5.6$ \Me\  for the 5.77-day and 13.5-day signals, respectively. The long-term activity signal has an amplitude of 1.5 \ms\ and a period of 9.6 years.

\item HD40307 hosts four certain planetary companions at periods of between $4.3$ and $51.6$ days and with masses of between  $M\sin i=3.6$ \Me\ and $M\sin i=8.7$ \Me. We find inconclusive evidence for an additional { companion} at $P\sim200$ days that it would be extremely interesting to confirm, as { such a companion would} orbit within the habitable zone of the star.

\item HD204313 is a system with a Neptune-mass planet on a 34.9-day orbit and a 4.3-\MJ\ candidate on an outer orbit ($P=2024$ days).
\end{itemize}

Only six systems with a massive ($M > 4$ \MJ) companion on an outer ($P > 1000$ days) orbit have been detected to date. Out of these, HD204313 has the highest mass ratio between the orbiting companions. The system most closely resembling HD204313 is arguably HD38529 \citep{fischer2001, fischer2003, wright2009}, with a 0.8-\MJ\ candidate on a 14.3-day orbit, and a 12.3-\MJ\ object on an 2140-day orbit. The mass ratio of this system is around five times lower than for HD204313, however. The system around HD74156 \citep{naef2004} consists of two companions in orbits alike to those of the  HD204313 system (a 1.8 \MJ\ object at 51.6 days and an 8.2 \MJ\ companion on a $\sim$ 2500-day orbit), but with a much lower mass ratio.

In addition to HD40307, only a handful of systems with more than three planets has been detected using RVs: $\mu$ Ara, 55 Cnc, HD10180, and GJ876. The continuing monitoring of this target has permitted these detections, and may in the future permit unveiling further planetary signals and improving the modelling of the activity effect.

HD1461 and HD40307 are two of only six solar-type planetary systems with at least two super-Earth companions whose mass is known to better than 50\%, and they will become prime targets for the follow-up mission CHEOPS. The other systems are HD20794 \citep{pepe2011}, HD7924 \citep{fulton2015}, { HD219134 \citep{motalebi2015}}, and Kepler-102 \citep{marcy2014}. Except for Kepler-102, which was first discovered as a transiting candidate system, these systems required more than 100 RV measurements to be detected. This shows the difficulty associated with detecting this type of companions. 

Additionally, as the required data sets span many years, activity cycles are omnipresent. Correctly modelling their effect on RV data is therefore necessary to unveil the full system architecture, from the short-period companions out to the habitable-zone planets. We have shown evidence that periodicities arise at the aliasing frequencies when an incomplete correction of the long-term variability is performed. A fully satisfactory solution is not at hand. In the meantime, adapting the observing strategy based on the activity level of the star seems reasonable. We have shown here that the observations of HD40307 obtained during the active period of the star contribute little to constraining the planetary system characteristics (see also Fig.~\ref{fig.HD40307_jitter}). Reducing the observing cadence during high-activity periods can save hours of telescope time.

\begin{acknowledgements}
We acknowledge Ewan Cameron (Oxford) for bringing the \citet{perrakis2014} estimator to our attention and for his help in understanding the limitations of the TPM estimator.
This work has been carried out within the frame of the National Centre for Competence in Research "PlanetS" supported by the Swiss National Science Foundation (SNSF).  It made use of the Data Analyses Center for Exoplanets - \url{http://dace.unige.ch} -  a platform of PlanetS. R.F.D. acknowledges funding from the European Union Seventh Framework Programme (FP7/2007-2013) under Grant agreement number 313014 (ETAEARTH). S.U., C.L., D.S., F.P., C.M. and A.W. acknowledge the financial support of the SNSF. CM acknowledges the support of the Swiss National Science Foundation under grant BSSIO\_155816 ``PlanetsInTime". We thank the University of Geneva for their continuous support to our planet search programmes. M. Gillon is Research Associate at the FNRS. P.F. and N.C.S. acknowledge support by Funda\c{c}\~ao para a Ci\^encia e a Tecnologia (FCT) through Investigador FCT contracts of reference IF/01037/2013 and IF/00169/2012, respectively, and POPH/FSE (EC) by FEDER funding through the program ``Programa Operacional de Factores de Competitividade - COMPETE''. P.F. further acknowledges support from Funda\c{c}\~ao para a Ci\^encia e a Tecnologia (FCT) in the form of an exploratory project of reference IF/01037/2013CP1191/CT0001. This research has made use of the SIMBAD database and of the VizieR catalogue access tool operated at CDS, France. 
\end{acknowledgements}


\begin{appendix}
\section{Parameter priors}

\begin{table*}[t]
{\small\centering
\caption{Parameter prior distributions for the HD1461 system. The epoch is BJD=2,455,155.3854 for planets $b$ and $c$ and BJD=2,455,195.8367 for the magnetic cycle.\label{table.priors1461}}            
\begin{tabular}{l l c c}        
\hline\hline                 
\noalign{\smallskip}

& 								&\multicolumn{2}{c}{Prior distribution}\\
\multicolumn{2}{c}{ Orbital parameters }	&Planet b / c	&  Magnetic cycle$\star$\\
\hline
\noalign{\smallskip}
Orbital period, $P$ 	&[days]	 		&$J(1.0, 10^4)$	&$N(3522.4, 75.4)$\\

RV amplitude, $K$ 	&[\ms]			&$U(0.0, 200)$	&$U(0.0, 200)$\\

Eccentricity, $e$    	& 			&$B(0.867, 3.03)$	&--\\

Argument of periastron, $\omega$&[deg]	&$U(0.0, 360.0)$	&--\\

$e^{1/2} \sin(\omega)$ & & -- & $N(-0.388, 0.078)$\\

$e^{1/2} \cos(\omega)$ & & -- & $N(0.129, 0.096)$\\

Mean longitude at epoch, $L_0$ &[deg] 	&$U(0.0, 360.0)$	&$N(124.0, 21.0)$\\
Systemic velocity, $V_0$		&[\kms]		&\multicolumn{2}{c}{$U(-10.061, -10.055)$}\\
\noalign{\smallskip}
\multicolumn{2}{c}{Noise model$\ddag$} 		\\
\hline
\noalign{\smallskip}
Additional noise at \logR=-5, $\sigma_J|_{-5.0}$&[\ms]		&\multicolumn{2}{c}{$U(0, 50)$}\\
Slope, $\alpha_J$						&[\ms/dex]	&\multicolumn{2}{c}{$U(-200, 200)$}\\
\hline
\hline
\end{tabular}

\tablefoot{\\
$U(x_{min};  x_{max})$: uniform distribution between $x_{min}$ and $x_{max}$.\\
$J(x_{min};  x_{max})$: Jeffreys (log-flat) distribution between $x_{min}$ and $x_{max}$.\\
$N(\mu; \sigma)$: normal distribution with mean $\mu$ and standard deviation $\sigma$.\\
$B(a, b)$: beta distribution.\\
$\star$: see Sect.~\ref{sect.priors40307}.\\
$\ddag$: the additional (stellar) noise for measurement $i$ is $\sigma_{Ji} = \sigma_J|_{-5.0} + \alpha_J \cdot (\log{(R'_{\rm HK})}_i + 5.0)$.
}

}
\end{table*}

\begin{table*}[t]
{\small\center
\caption{Prior distributions for the HD40307 system. Priors are identical for all signals in the model. The epoch is BJD=2,454,521.6791.\label{table.40307priors}}            

\begin{tabular}{l l c }        
\hline\hline                 
\noalign{\smallskip}

\multicolumn{2}{c}{ Orbital parameters } 		&Prior distribution\\
\hline
\noalign{\smallskip}
Orbital period, $P$ 	&[days]	 	&$J(1.0, 10^4)$\\
RV amplitude, $K$ 	&[\ms]		&$U(0.0, 200)$\\
Eccentricity, $e$    	& 			&$B(0.867, 3.03)$\\
Argument of periastron, $\omega$&[deg]	&$U(0, 360)$\\
Mean longitude at epoch, $L_0$ &[deg] &$U(0, 360)$\\
Systemic velocity, $V_0$		&[\kms]		&$U(28.996, 33.668)$\\
\noalign{\smallskip}
\multicolumn{2}{c}{Velocity drift (long-term activity effect)$\star$} 		\\
\hline
\noalign{\smallskip}
Scaling constant, $\alpha$	&[\ms/dex]	&$U(0, 100)$\\
Linear					&[$10^{-4}$ dex yr$^{-1}$]&$N(373.8, 8.6)$\\
Quadratic					&[$10^{-4}$ dex yr$^{-2}$]	&$N(46.3, 4.6)$\\
Cubic					&[$10^{-4}$ dex yr$^{-3}$]	&$N(-17.3, 1.0)$\\
\noalign{\smallskip}
\multicolumn{2}{c}{Noise model$\ddag$} 		\\
\hline
\noalign{\smallskip}
Additional noise at \logR=-5, $\sigma_J|_{-5.0}$&[\ms]		&$U(0, 50)$\\
Slope, $\alpha_J$						&[\ms/dex]	&$U(0, 50)$\\
\hline
\hline
\end{tabular}
\tablefoot{
The argument of periastron $\omega$ is unconstrained for these nearly-circular orbits.

$U(x_{min};  x_{max})$: uniform distribution between $x_{min}$ and $x_{max}$.\\
$J(x_{min};  x_{max})$: Jeffreys (log-flat) distribution between $x_{min}$ and $x_{max}$.\\
$N(\mu; \sigma)$: normal distribution with mean $\mu$ and standard deviation $\sigma$.\\
$B(a, b)$: beta distribution.\\
$\star$: see Sect.~\ref{sect.priors40307}.\\
$\ddag$: the additional (stellar) noise for measurement $i$ is $\sigma_{Ji} = \sigma_J|_{-5.0} + \alpha_J \cdot (\log{(R'_{\rm HK})}_i + 5.0)$.
}
}
\end{table*}
\begin{table*}[t]
{\small\center
\caption{Parameter prior distributions for the HD204313 system. The epoch is BJD=2,454,993.84858.\label{table.priors204313}}            

\begin{tabular}{l l c c }        
\hline\hline                 
\noalign{\smallskip}

& 								&\multicolumn{2}{c}{Prior distribution}\\
 \multicolumn{2}{c}{ Orbital parameters }	&Planet b	&Planet c\\
\hline
\noalign{\smallskip}
Orbital period, $P$ 	&[days]	 		&$N(2046.3, 9.1)$	&$N(34.989, 0.033)$\\
RV amplitude, $K$ 	&[\ms]			&$N(66.9, 1.3)$	&$N(4.9, 1.0)$\\
Eccentricity, $e$    	& 				&$N(0.125, 0.017)$	&$B(0.93, 5.5)$\\
Argument of periastron, $\omega$&[deg]	&$N(303.0, 8.3)$	&$U(0, 360)$\\
Mean longitude at epoch, $L_0$ &[deg] 	&$N(109.10, 0.91)$	&$N(35.6, 15.0)$\\
Systemic velocity, $V_0$		&[\kms]		&\multicolumn{2}{c}{$U(-9,762, -9.697)$}\\
\noalign{\smallskip}
\multicolumn{2}{c}{Velocity drift (long-term activity effect)$\star$} 		\\
\hline
\noalign{\smallskip}
Scaling constant, $\alpha$	&[\ms/dex]	&\multicolumn{2}{c}{$U(0, 300)$}\\
Linear			&[$10^{-2}$ dex yr$^{-1}$]	&\multicolumn{2}{c}{$N(-1.02, 0.13)$}\\
Quadratic			&[$10^{-4}$ dex yr$^{-2}$]	&\multicolumn{2}{c}{$N(-5.5, 2.5)$}\\
Cubic			&[$10^{-4}$ dex yr$^{-3}$]	&\multicolumn{2}{c}{$N(3.68, 0.92)$}\\
\noalign{\smallskip}
\multicolumn{2}{c}{Noise model$\ddag$} 		\\
\hline
\noalign{\smallskip}
Additional noise at \logR=-5, $\sigma_J|_{-5.0}$&[\ms]		&\multicolumn{2}{c}{$U(0, 50)$}\\
Slope, $\alpha_J$						&[\ms/dex]	&\multicolumn{2}{c}{$U(-200, 200)$}\\
\hline
\hline
\end{tabular}

\tablefoot{\\
$\mathrm{U}(x_{min};  x_{max})$: uniform distribution between $x_{min}$ and $x_{max}$.\\
$N(\mu; \sigma)$: normal distribution with mean $\mu$ and standard deviation $\sigma$.\\
$B(a, b)$: beta distribution.\\
$\star$: see Sect.~\ref{sect.priors40307}.\\
$\ddag$: the additional (stellar) noise for measurement $i$ is $\sigma_{Ji} = \sigma_J|_{-5.0} + \alpha_J \cdot (\log{(R'_{\rm HK})}_i + 5.0)$.
}

}
\end{table*}

\end{appendix}

\end{document}